\documentclass[
12pt,
letterpaper,
reprint,
prx,
aps,
floatfix,
showkeys,
superscriptaddress,
longbibliography,
colorlinks
]{revtex4-2}

\usepackage[utf8]{inputenc}
\usepackage{amsmath,amsthm,amssymb,amsfonts}
\theoremstyle{remark}
\usepackage{mathtools}
\usepackage{graphicx}
\graphicspath{{./figures/}}
\usepackage{booktabs}
\usepackage[x11names, table]{xcolor}
\usepackage{rotating}
\usepackage{orcidlink}
\usepackage{adjustbox}
\usepackage{array}
\usepackage[mathlines]{lineno}
\usepackage{appendix}
\usepackage{aligned-overset}
\usepackage{algorithm}
\usepackage{algpseudocode}
\usepackage{bbm}
\usepackage[normalem]{ulem}
\usepackage{cancel}
\usepackage[T1]{fontenc} %

\usepackage{pifont}
\newcommand{\cmark}{\textcolor{green!80!black}{\ding{51}}}
\newcommand{\xmark}{\textcolor{red}{\ding{55}}}

\AtBeginDocument{%
  \heavyrulewidth=.08em
  \lightrulewidth=.05em
  \cmidrulewidth=.03em
  \belowrulesep=.65ex
  \belowbottomsep=0pt
  \aboverulesep=.4ex
  \abovetopsep=0pt
  \cmidrulesep=\doublerulesep
  \cmidrulekern=.5em
  \defaultaddspace=.5em
}
\usepackage{hyperref}

\newcommand{\abs}[1]{\left\lvert#1\right\rvert} 
\newcommand{\absSmall}[1]{\lvert#1\rvert} 

\newcommand{\bra}[1]{\langle#1\rvert} 
\newcommand{\ket}[1]{\lvert#1\rangle} 
\newcommand{\proj}[1]{\ket{#1}\bra{#1}} %
\newcommand{\braket}[2]{ \langle #1 | #2 \rangle} 
\newcommand{\braopket}[3]{\langle #1 | #2 | #3\rangle} 
\newcommand{\ev}[1]{\left\langle #1 \right\rangle} 
\newcommand{\evSmall}[1]{\langle #1 \rangle} 
\newcommand{\tr}[1]{\text{Tr}\left[#1\right]}
\newcommand{\trSmall}[1]{\text{Tr}[#1]}

\newcommand{\re}[1]{\text{Re}\left[#1\right]}  
 
\newcommand{\im}[1]{\text{Im}\left[#1\right]} 
 
\newcommand{\evText}[1]{\text{E}\left[#1\right]}

\newcommand{\var}[1]{\text{Var}\left[#1\right]} 
\newcommand{\varSmall}[1]{\text{Var}[#1]} 

\newcommand{\varSubSmall}[2]{\text{Var}_{#1}[#2]}

\newcommand{\inth}[1]{\int \text{d}#1\;}
\newcommand{\intg}[3]{\int_{#1}^{#2} \text{d}#3\;}

\newcommand{\intginf}[1]{\intg{-\infty}{\infty}{#1}}

\newcommand{\intS}[1]{\int_S\text{d}#1\;}
\newcommand{\deriv}[2]{\frac{\text{d}#1}{\text{d}#2}}

\newcommand{\h}[1]{\hat{#1}}

\newcommand{\R}{\mathbb{R}} 
\newcommand{\C}{\mathbb{C}} 
\newcommand{\Z}{\mathbb{Z}}

\newcommand{\IQ}{\mathcal{I}_Q}
\newcommand{\IC}{\mathcal{I}_C}
\newcommand{\IP}{\mathcal{I}_\text{prior}}

\newcommand{\om}{\omega}
\newcommand{\Om}{\Omega}
\newcommand{\si}{\sigma}

\newcommand{\T}{\text{T}}

\newcommand{\diag}[1]{\text{diag}\left(#1\right)}

\newcommand{\order}[1]{\mathcal{O}(#1)}

\usepackage{outlines}

\begin{document}
\title{Bayesian frequency estimation at the fundamental quantum limit} 

\author{James~W.~Gardner\,\orcidlink{0000-0002-8592-1452}}
    \email{james.gardner@anu.edu.au}
    \affiliation{OzGrav-ANU, Centre for Gravitational Astrophysics, Research Schools of Physics, and of Astronomy and Astrophysics, The Australian National University, Canberra, ACT 2601, Australia}
    \affiliation{Division of Physics, Mathematics and Astronomy, California Institute of Technology, Pasadena, California 91125, USA}
\author{Tuvia~Gefen\,\orcidlink{0000-0002-3235-4917}\,}
    \affiliation{Racah Institute of Physics, The Hebrew University of Jerusalem, Jerusalem 91904, Givat Ram, Israel}
\author{Ethan~Payne\,\orcidlink{0000-0003-4507-8373}}
    \affiliation{Division of Physics, Mathematics and Astronomy, California Institute of Technology, Pasadena, California 91125, USA}
    \affiliation{LIGO Laboratory, California Institute of Technology, Pasadena, California 91125, USA}
\author{Su~Direkci\,\orcidlink{0009-0002-3867-4337}}
    \affiliation{Division of Physics, Mathematics and Astronomy, California Institute of Technology, Pasadena, California 91125, USA}
\author{Sander~M.~Vermeulen\,\orcidlink{0000-0003-4227-8214}}
    \affiliation{Division of Physics, Mathematics and Astronomy, California Institute of Technology, Pasadena, California 91125, USA}
\author{Simon~A.~Haine\,\orcidlink{0000-0003-1534-1492}}
    \affiliation{Department of Quantum Science and Technology and Department of Fundamental and Theoretical Physics, Research School of Physics, The Australian National University, Canberra, ACT 0200, Australia}
\author{Joseph~J.~Hope\,\orcidlink{0000-0002-5260-1380}}
    \affiliation{Department of Quantum Science and Technology and Department of Fundamental and Theoretical Physics, Research School of Physics, The Australian National University, Canberra, ACT 0200, Australia}
\author{Lee~McCuller\,\orcidlink{0000-0003-0851-0593}}
    \affiliation{Division of Physics, Mathematics and Astronomy, California Institute of Technology, Pasadena, California 91125, USA}
\author{Yanbei~Chen\,\orcidlink{0000-0002-9730-9463}\,}
    \affiliation{Division of Physics, Mathematics and Astronomy, California Institute of Technology, Pasadena, California 91125, USA}

\date{\today}
\begin{abstract}
Searching for a weak signal at an unknown frequency is a canonical task in experiments probing fundamental physics such as gravitational-wave observatories and ultra-light dark matter haloscopes. These state-of-the-art sensors are limited by quantum noise arising from the fundamental uncertainty about the state of the device. Classically, frequency estimation suffers from a threshold effect in the signal-to-noise ratio such that weak signals are extremely hard to localise in frequency. We show that this phenomenon persists at the fundamental quantum limit but that the classical approach, a quadrature measurement, can nevertheless be beaten by a coherent protocol of projecting onto the ``quantum whitened'' possible quantum states. Quantum whitening is a covariant measurement, and we examine it analytically in the wide-prior limit and numerically for finite-width priors. Beyond accelerating searches for unknown frequencies, quantum whitening may be used generally to sense the parameter of a unitary encoding given no prior information about the parameter.
\end{abstract}
\maketitle
\allowdisplaybreaks

\begin{figure*}[ht!]
    \centering
    \includegraphics[width=0.9\linewidth]{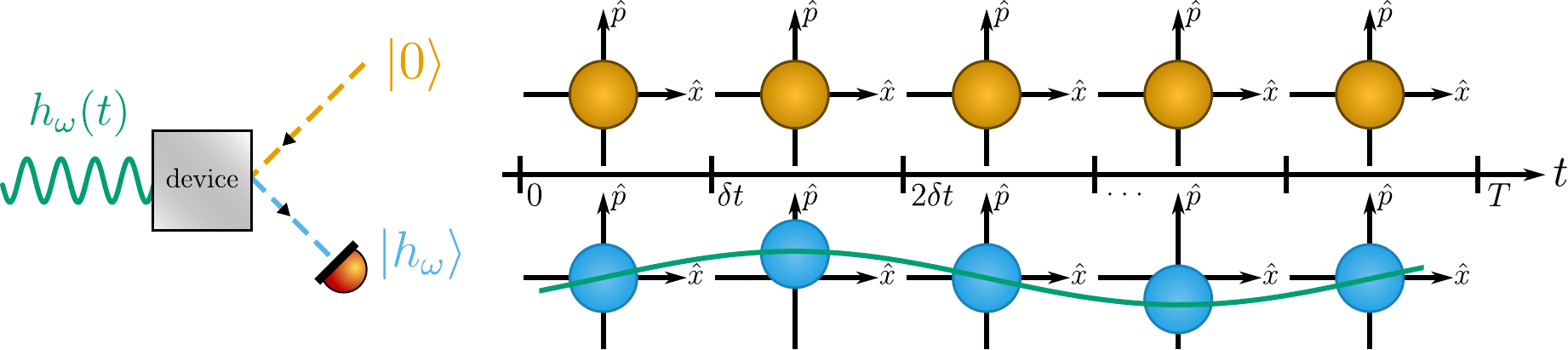}
    \caption{A sinusoidal signal $h_\omega(t)$ in time $t$ displaces a bath of vacuum modes $\ket{0}$ to an oscillating coherent state $\ket{h_\omega}$ via a linear quantum device. The ball-and-stick diagrams on the right idealise the Wigner function of the bath discretised in time in Appendix~\ref{app:Discretising the time domain}. The signal is windowed to the bounded interval $(0, T)$ which is discretised into $M=T/\delta t$ time-bins of width $\delta t$. While here we choose $M=5$ for the purposes of illustration, we take the continuum limit of $\delta t\to0$ with $T$ fixed in the text.}
    \label{fig:waveform estimation diagram}
\end{figure*}

Many promising searches for fundamental physics probe for weak signals at unknown frequencies encoded in the displacement of a bosonic mode. For example, searches using ground-based gravitational-wave observatories for the post-merger remnant of binary neutron star mergers could test our theories of exotic matter at supranuclear densities~\cite{ethanpaper,10.1088/2514-3433/ac2256ch9,sarin_lasky_2022} and searches for axions using table-top haloscopes could directly detect dark matter and resolve the strong $\mathcal{CP}$ problem~\cite{kim2010axions,choi2021recent,rosenberg2000searches,graham2015experimental}. However, the fundamental Bayesian quantum limits of the sensitivity of these searches are not well understood. Here, a Bayesian analysis is necessary because we assume that only a single event is seen and that the prior on the possible frequencies of the signal is wide. For example, the range of possible axion masses spans five orders of magnitude~\cite{Peccei2008} at the ultra-light end of the vast 90 orders of magnitude of possible masses for dark matter more generally~\cite{battaglieri2017us}. Understanding how to achieve the fundamental quantum limit of these devices using unconventional measurement protocols may lead to improved sensitivity and thus accelerate these searches for fundamental physics.

Let us now define a canonical frequency estimation problem. We want to sense the following sinusoidal signal
\begin{align}\label{eq:signal}
    h_\omega(t) = A\cos(\omega t + \phi)
\end{align}
where the amplitude $A$ (with units of square root Hertz) and phase $\phi$ are known, but the frequency $\omega$ is unknown. We assume a flat prior $\pi(\om)$ on the possible frequencies within the interval $(\omega_0-\Delta\omega/2,\omega_0+\Delta\omega/2)$ with centre frequency $\omega_0$ and a wide prior width $\Delta\omega$. Suppose that the sinusoidal signal interacts with a linear device from time $t=0$ to $t=T$, where $T$ is the integration time. If the bosonic bath of incoming modes to the sensor is in the multi-mode vacuum state $\ket0$, then the coherent quantum state of the bath of outgoing modes is as follows 
\begin{align}\label{eq:state}
    |h_\omega\rangle &= e^{i\intg{0}{T}{t} h_\omega(t) \h x(t)} |0\rangle, 
\end{align}
where the amplitude quadrature of the bath is $\h x(t)=[\h a(t) + \h a^\dagger(t)]/\sqrt{2}$ given the bath annihilation and creation operators $\h a(t)$ and $\h a^\dagger(t)$, respectively, which obey the commutator $[\h a(t), \h a^\dagger(t')]=\delta(t-t')$. Here, we set $\hbar = 1$ and have absorbed the convolution with the linear susceptibility of the sensor into the signal~\cite{KuboRPP66FluctuationdissipationTheorem,BuonannoPRD02SignalRecycled}. %
This coherent state consists of the vacuum displaced along the phase quadrature of the bath $\h p(t)=[-i\h a(t) + i \h a^\dagger(t)]/\sqrt{2}$ by the signal as shown in Fig.~\ref{fig:waveform estimation diagram}. We assume here that the direction of the displacement is known and fixed, because we want to model ground-based gravitational-wave observatories consisting of tuned resonant optical cavities and a Michelson interferometer held at dark port such that the signal only appears in the phase quadrature~\cite{bond2016interferometer}.

We want to determine the optimal measurement of the final state $|h_\omega\rangle$ in Eq.~\ref{eq:state} to best constrain the posterior distribution on the possible signal frequencies. Ref.~\cite{ethanpaper} recently showed an advantage in frequency estimation by performing a frequency-domain number-resolving measurement $\h n(\Om)$ over a time-domain quadrature measurement $\h p(t)$. In this paper, we go beyond this head-to-head comparison and find the fundamental quantum limit of frequency estimation and how to achieve it. First, in Sec.~\ref{sec:background}, we review the techniques of optimal quantum sensing. Second, in Sec.~\ref{sec:quantum whitening}, we calculate the general fundamental limit and optimal measurement. Third, in Sec.~\ref{sec:single-particle states} and Sec.~\ref{sec:coherent states}, we review the geometry of the single-particle and coherent states, respectively, to establish a waveform basis of quantum states. Finally, we apply the results to different cases of single-particle and coherent states in Sec.~\ref{sec:applications} and draw conclusions in Sec.~\ref{sec:conclusions}.

\section{Background}
\label{sec:background}
Let us review quantum parameter estimation, with Bayesian and Fisher information approaches, and classical frequency estimation.

\subsection{Review of Bayesian quantum estimation}
Let us consider Bayesian quantum estimation of a single generic parameter $\theta$ (see, e.g., Refs.~\cite{macieszczak2014bayesian, jarzyna2015true, 
rubio2019quantum,
kaubruegger2021quantum,
marciniak2022optimal,direkci2024heisenberg}). We suppose that we have a prior distribution $\pi(\theta)$ on the parameter $\theta$ that is encoded in a quantum state $\h \rho(\theta)$. We then perform some measurement described by a positive operator-valued measure (POVM) with a set of effects $\{\h E_x\}_x$ such that $E_x\geq0$ and $\inth{x}E_x=1$. We obtain the measurement outcome $x$ given $\theta$ with likelihood $L(x|\theta)=\trSmall{\h\rho(\theta)\h E_x}$. Marginalising over $\theta$, then the total probability of obtaining $x$ is given by the evidence $p(x)=\inth{\theta}\pi(\theta)L(x|\theta)$. By Bayes's rule, the posterior on $\theta$ given $x$ is then given by
\begin{align}
    \label{eq:Bayes rule}
    p(\theta|x) = \frac{\pi(\theta)L(x|\theta)}{p(x)}.
\end{align}
From the posterior given $x$, we form some estimator $\check{\theta}_x$ of $\theta$, which can be biased. The mean-square error (MSE) of this estimator at a given $\theta$ is
\begin{align}
    \label{eq:MSE}
\text{MSE}(\theta) = \int\text{d}x\; L(x|\theta) (\check{\theta}_x - \theta)^2.
\end{align}
We wish to minimise the Bayesian MSE (BMSE) defined as the MSE averaged against the prior or, equivalently by Eq.~\ref{eq:Bayes rule}, the posterior variance averaged against the evidence: %
\begin{align}
    \label{eq:BMSE}
    \text{BMSE} &= \int\text{d}\theta\; \pi(\theta)\; \text{MSE}(\theta) = \int\text{d}x\; p(x)\; V_\text{post}(x)
\end{align}
where the variance of the posterior given $x$ is
\begin{align}\label{eq:posterior variance}
    V_\text{post}(x)=\int\text{d}\theta\; p(\theta|x) (\check{\theta}_x - \theta)^2.
\end{align}
Here, the optimal estimator $\check{\theta}_x$ to minimise the BMSE for a given measurement scheme is the mean of the posterior distribution. For a given quantum state $\h\rho(\theta)$, the minimum BMSE (MBMSE) over all possible measurements is given as follows~\cite{macieszczak2014bayesian}
\begin{align}\label{eq:MBMSE}
    \Delta^2\theta = \Delta^2\theta^{(0)}-\trSmall{\hat{\bar\rho}\h L^2}
\end{align}
where $\Delta^2\theta^{(0)}=\int\text{d}\theta\;\pi(\theta)\,\theta^2$ is the variance of the prior $\pi(\theta)$. The optimal measurement basis $\ket{x}$ and estimator $\check{\theta}_x$ is captured by the Bayesian symmetric logarithmic derivative (BSLD) defined as
\begin{align}\label{eq:BSLD}
    \h L=\inth{x}\check{\theta}_x\proj{x}
\end{align}
which is the solution to the following Lyapunov equation
\begin{align}\label{eq:Lyapunov equation}
    \hat{\bar\rho}' = \frac{1}{2}\{\hat{\bar\rho}, \h L\}
\end{align}
where $\{\cdot,\cdot\}$ is the anticommutator and the mixed state is
\begin{align}
    \label{eq:mixed state}
    \hat{\bar\rho}=\inth{\theta}\pi(\theta)\h\rho(\theta), \quad %
    \hat{\bar\rho}'=\inth{\theta}\pi(\theta)\h\rho(\theta)\,\theta.
\end{align}
Here, $\hat{\bar\rho}$ captures the classical uncertainty in the final state $\h\rho(\theta)$ given the prior, and $\hat{\bar\rho}'$ which we call the ``Bayesian derivative'' acts informally like the average derivative of $\hat{\bar\rho}$ with respect to the parameter $\theta$. Indeed, in the case of a Gaussian prior then $\hat{\bar\rho}'=\Delta^2\theta^{(0)}\inth{\theta}\pi(\theta)\partial_\theta \h\rho(\theta)$~\cite{macieszczak2014bayesian}. For a given prior $\pi(\theta)$, the Lyapunov equation in Eq.~\ref{eq:Lyapunov equation} can then be solved given the spectral decomposition of $\hat{\bar\rho}=\sum_j \lambda_j \proj{\phi_j}$ as
\begin{align}\label{eq:BSLD sol}
   \hat L = \sum_{j,k}\frac{2}{\lambda_{j}+\lambda_{k}} \braopket{\phi_{j}}{\hat{\bar\rho}'}{\phi_{k}}\ket{\phi_{j}}\bra{\phi_{k}}   
\end{align}
where the sum excludes $j,k$ such that $\lambda_j+\lambda_k=0$. We can then calculate the MBMSE in Eq.~\ref{eq:MBMSE} as
\begin{align}\label{eq:MBMSE sol}
    \Delta^2\theta = \Delta^2\theta^{(0)} - \sum_{j,k}\frac{4\lambda_j}{(\lambda_j+\lambda_k)^2}\abs{\braopket{\phi_j}{\hat{\bar\rho}'}{\phi_k}}^2.
\end{align}
We assume in Eqs.~\ref{eq:MBMSE}--\ref{eq:mixed state} that the prior is centred such that $\inth{\theta}\pi(\theta)\theta=0$. For example, for sensing a frequency within $\omega\in(\omega_0-\Delta\omega/2,\omega_0+\Delta\omega/2)$ we formally estimate $\omega-\omega_0$ instead of $\omega$. However, we will ignore this technicality throughout the rest of the paper and just refer to estimating the frequency $\omega$.

\subsection{Review of Fisher information}
\begin{table}[]
\begingroup
\setlength{\tabcolsep}{6pt} %
\renewcommand{\arraystretch}{1} %
\begin{tabular}{@{}l|ll@{}}
\toprule
Method          & Bayesian                    & Fisher         \\ \midrule
Figure-of-merit & BMSE (Eq.~\ref{eq:BMSE})         & MSE (Eq.~\ref{eq:MSE})     \\
Fundamental limit & MBMSE (Eq.~\ref{eq:MBMSE}) & QCRB (Eq.~\ref{eq:QCRB}) \\
Optimal measurement & BSLD (Eq.~\ref{eq:BSLD sol}) & SLD (Eq.~\ref{eq:SLD sol})\\
Prior           & Any prior & No prior \\
Estimator       & Biased                      & Unbiased                     \\
Valid           & Everywhere                  & Asymptotically               \\
Computation     & Harder                      & Easier                       \\
\bottomrule
\end{tabular}%
\endgroup
\caption{Comparison of Bayesian and Fisher approaches to quantum parameter estimation.}
\label{tab:Bayesian vs Fisher}
\end{table}

Let us discuss the asymptotic limit of Bayesian estimation where we measure $M\gg1$ independent and identical copies of the quantum state. In this frequentist regime, the prior width is negligibly narrow and we instead study local estimation of infinitesimal deviations about a given $\theta$. Formally, this connection between the Bayesian and frequentist approaches is given by the Bernstein-von Mises theorem.

We now review the Fisher information approach for estimating a single parameter $\theta$. In our motivating problem, although this parameter determines a classical waveform in Eq.~\ref{eq:signal} which is encoded in a quantum state in Eq.~\ref{eq:state}, this should be distinguished from true waveform estimation where there are a continuum of parameters to estimate (see, e.g., Refs.~\cite{TsangPRL11FundamentalQuantum,Miao+2017,gardner2024achieving}). The residual MSE in Eq.~\ref{eq:MSE} at $\theta$ for a given measurement obeys the classical Cram\'er-Rao bound (CCRB)
\begin{align}\label{eq:CCRB}
    \text{MSE}(\theta) \geq [M\IC(\theta)]^{-1},
\end{align}
where the classical Fisher information (CFI) is given by 
\begin{align}\label{eq:CFI}
    \IC(\theta) = \sum_{x}\frac{[\partial_\theta L(x|\theta)]^2}{L(x|\theta)}.
\end{align}
The CFIs from independent and identical observations sum, which contributes the factor of $M$ in Eq.~\ref{eq:CCRB}.
Lifting from a given measurement to the quantum state, the quantum Cram\'er-Rao bound (QCRB) is
\begin{align}
    \label{eq:QCRB}
    \text{MSE}(\theta) \geq [M\IQ(\theta)]^{-1},
\end{align}
where the maximum CFI over all possible measurements is the quantum Fisher information (QFI) which is given similarly to Eq.~\ref{eq:MBMSE sol} by 
\begin{align}\label{eq:QFI}
    \IQ(\theta) = \sum_{j,k}\frac{4\lambda_j}{(\lambda_j+\lambda_k)^2}\abs{\braopket{\phi_j}{\partial_\theta\h \rho(\theta)}{\phi_k}}^2
\end{align}
where the spectral decomposition of the state is $\h\rho(\theta)=\sum_j \lambda_j \ket{\phi_j}\bra{\phi_j}$ and the sum excludes $j,k$ such that $\lambda_j+\lambda_k=0$. Similarly to how Eq.~\ref{eq:MBMSE} relates the MBMSE to the BSLD, the QFI is $\IQ(\theta)=\trSmall{\h\rho\h S^2}$ in terms of $\h S$, the symmetric logarithmic derivative (SLD). The SLD solves the Lyapunov equation $\partial_\theta\hat{\rho}(\theta) = \frac{1}{2}\{\hat{\rho}(\theta), \h S\}$ similarly to Eq.~\ref{eq:Lyapunov equation}. The solution is similar to Eq.~\ref{eq:BSLD sol} and is given by
\begin{align}\label{eq:SLD sol}
   \hat S = \sum_{j,k}\frac{2}{\lambda_{j}+\lambda_{k}} \braopket{\phi_{j}}{\partial_\theta\h \rho(\theta)}{\phi_{k}}\ket{\phi_{j}}\bra{\phi_{k}}   
\end{align}
where we exclude $\lambda_j+\lambda_k=0$ as usual. The expectation value $\trSmall{\h\rho\h S^2}$ thus recovers the QFI in Eq.~\ref{eq:QFI}. The equality between the maximum CFI and the QFI only holds in the single-parameter case, which we focus on, as the multi-parameter case requires handling non-commutativity~\cite{Holevo2011book,gardner2024achieving}.

We compare the Bayesian and Fisher methods in Table~\ref{tab:Bayesian vs Fisher}. Although the Fisher information describes the asymptotic scaling of the MSE, questions remain about how fast a given optimal measurement reaches the asymptotic limit such that we need to maximise the domain of unambiguous estimation along with the point estimate sensitivity~\cite{vasilyev2024optimal}. I.e., two different measurements can both be optimal asymptotically (their CFIs attain the QFI) but the MSE of one can converge to the quantum limit much faster than the other. This behaviour is already accounted for implicitly in the Bayesian method, however, such that we need not alter our approach.

The QFI is also related to the infinitesimal expansion of the fidelity between quantum states as follows
\begin{align}
    \mathcal{F}(\theta) &= \tr{\sqrt{\sqrt{\h\rho(\theta)}\h\rho({\theta+\delta\theta})\sqrt{\h\rho(\theta)}}\,}^2\nonumber 
    \\&= 1 - \frac{1}{4} \delta\theta^2 \IQ(\theta) + \order{\delta\theta^3} %
    \label{eq:fidelity}
    .
\end{align}
Here, we assume that the rank of $\h\rho(\theta)$ is constant in $\theta$, e.g.\ if the state is pure $\h\rho(\theta)=\proj{h_\theta}$ such that the fidelity is $\mathcal{F}(\theta) = \absSmall{\braket{h_{\theta}}{h_{\theta+\delta\theta}}}^2$.

For the Gaussian case, the Fisher approach simplifies considerably. The CFI in Eq.~\ref{eq:CFI} for a multi-variate normal distribution of $n$ variables with $n$-by-1 parameter-dependent mean vector $\vec{\mu}_n(\theta)$ and $n$-by-$n$ parameter-independent covariance matrix $\Sigma_n$ is equal to
\begin{align}
    \label{eq:CFI Gaussian, mean only}
    \IC(\theta) = \partial_\theta\vec\mu_n(\theta)^\T \Sigma_n^{-1} \partial_\theta\vec\mu_n(\theta).
\end{align}
Similarly, the QFI in Eq.~\ref{eq:QFI} for an $n$-mode Gaussian state with $2n$-by-1 parameter-dependent mean vector $\vec{\mu}_{2n}(\theta)$, since there are two quadratures for each mode, and $2n$-by-$2n$ parameter-independent covariance matrix $\Sigma_{2n}$ is 
\begin{align}
    \label{eq:QFI Gaussian, mean only}
    \IQ(\theta) = \partial_\theta\vec\mu_{2n}(\theta)^\T \Sigma_{2n}^{-1} \partial_\theta\vec\mu_{2n}(\theta)
\end{align}
which is achieved by measuring the quadrature along $\Sigma^{-1}\partial_\theta\vec\mu_{2n}(\theta)$~\cite{monras2013phase}. The Gaussian shift QFI is thus saturated by a Gaussian measurement.

We can also incorporate a priori information about $\theta$ from a prior distribution $\pi(\theta)$, thus treating the parameter as a random variable. In the case that the MSE and QFI are independent of the parameter, then van Trees's Bayesian QCRB states that~\cite{van2002detection,gill1995applications,genoni2013optimal}
\begin{align}
    \text{MSE} \geq [M(\IQ + \IP)]^{-1}
\end{align}
where the Fisher information of the prior is
\begin{align}\label{eq:prior Fisher information}
    \IP = \inth{\theta}\frac{\pi'(\theta)^2}{\pi(\theta)}.
\end{align}
Here, the integral should be restricted to the support of the prior. For example, if the prior is a Gaussian with variance $\sigma^2$, then the prior Fisher information is $\IP=1/\sigma^2$. This bound is not tight except asymptotically, however, and the full Bayesian approach is required for true incorporation of the prior.

\subsection{Review of classical frequency estimation}
\label{sec:Review of classical frequency estimation}

Let us briefly discuss the classical problem of estimating the frequency of a weak sinusoidal signal hidden in white noise. This is the scenario for our motivating problem in Eq.~\ref{eq:state} if we perform a time-domain quadrature measurement $\h p(t)$, since the quantum shot noise of the vacuum is white and the expectation values follow the signal as $\evSmall{\h p(t)}=h(t)$. The noise here comes entirely from the quantum state entering the linear device shown in Fig.~\ref{fig:waveform estimation diagram} and the classical signal is deterministic, i.e.\ $h(t)$ is not a continuous random variable (in contrast to Ref.~\cite{gardner2024stochastic,gardner2025lindblad}). It is well-known for this classical problem that the CCRB in Eq.~\ref{eq:CCRB} is not a tight bound below some signal-to-noise ratio (SNR) threshold~\cite{RifeITIT74SingleTone,SteinhardtI8IICASSP85ThresholdsFrequency,james2002characterization}. Here, the SNR is $A\sqrt{2T}$ in amplitude units where the factor of two comes from the quantum noise of the vacuum state of a harmonic oscillator which equals $\varSmall{\h p}=\frac{1}{2}$ given $[\h x, \h p]=i$. Below the SNR threshold, we need to instead use a different classical bound such as the BMSE. For example, later in Fig.~\ref{fig:MBMSE vs SNR}, we will see that the BMSE for the time-domain quadrature measurement $\h p(t)$ does not attain the CCRB except for above $A\sqrt{2T}\sim10$. Instead of the BMSE, we could use the Barankin bound but this requires an unbiasedness assumption which does not hold as generally as the Bayesian approach~\cite{KnockaertITSP97BarankinBound}.

The harmonic frequencies appearing in the Fourier series on the interval $(0, T)$ are $\omega_n=2\pi n/T$ for $n\in\Z$. One might then expect the classical frequency resolution to scale as $1/T$ in amplitude units, but the CCRB actually decreases as $T^{-3/2}$~\cite{SteinhardtI8IICASSP85ThresholdsFrequency, schmitt2017submillihertz,schmitt2021optimal}. This can be shown from Eq.~\ref{eq:CFI Gaussian, mean only}. The CFI acquires a factor of $T^2$ from the square of the derivative $-At\sin(\omega t+\phi)$ of the signal in Eq.~\ref{eq:signal} and obtains the third factor of $T$ from the integral over time.

We now want to investigate what happens for the case of quantum frequency estimation above and below the SNR threshold, and whether other measurement schemes than quadrature measurement can improve the sensitivity.

\section{Quantum whitening}
\label{sec:quantum whitening}
Before we return to our motivating problem of frequency estimation of the coherent state in Eq.~\ref{eq:state}, we will first solve a more general problem. We will consider general families of pure states $\left\{ |\psi_{\theta}\rangle\right\} _{\theta\in\mathbb{R}}$, where $\theta$ is the parameter to be estimated. We will focus on the case where this family of states has a Toeplitz symmetry, i.e.\ the inner product $\langle\psi_{\theta}|\psi_{\theta'}\rangle$ is a function of only $\theta-\theta'$. This will be useful to our motivating problem, since the geometry of the coherent states has a Toeplitz symmetry in a particular limit which means that we want to understand Toeplitz families of states. Thus, in this section, we will first review the covariant measurement of an additive unitary. Second, we will show that a Toeplitz family of states is always generated by an additive unitary, and we will interpret the covariant measurement of a continuous Toeplitz family as projecting on the ``quantum whitened'' possible states. Finally, we will calculate the fundamental quantum limit, the MBMSE, for a Toeplitz family when we have no prior information about the parameter. 

\subsection{Additive unitary case}
Let us first review the case where the parameter is encoded by an additive unitary. Let the final state be $\ket{\psi_\theta}=e^{-i\h H\theta}\ket{\psi_0}$ for some Hamiltonian $\h H$ and initial state $\ket{\psi_0}$. The unitary $\h U(\theta)=e^{-i\h H\theta}$ is additive since $\h U(\theta)\h U(\theta')=\h U(\theta+\theta')$ and $\h U^\dagger(\theta)=\h U(-\theta)$. The Hamiltonian $\h H$ is therefore the generator of the unitary group $\{ \hat{U}(\theta)\} _{\theta\in\mathbb{R}}$, which is the relevant symmetry transformation of this problem. The inner product between quantum states is thus $\braket{\psi_{\theta}}{\psi_{\theta'}} = \bra{\psi_0}e^{i\h H(\theta-\theta')}\ket{\psi_0}$. This inner product is Toeplitz, i.e.\ it is only a function of $\theta-\theta'$. As such, we refer to $\{\ket{\psi_\theta}\}_{\theta\in\mathbb{R}}$ as a Toeplitz family of states. 

Since we are ultimately interested in frequency estimation of the coherent states, we consider the case where the Hilbert space is infinite-dimensional and the Hamiltonian $\h H$ has a continuous spectrum. (We discuss the finite and discrete cases in Appendix~\ref{app:periodic case}.) Let the spectral decomposition be $\h H = \intginf{k} k \proj{k}$ for some continuous basis satisfying $\braket{k}{l}=\delta(k-l)\chi_S(k)$ where $\chi_{S}(k)$ is the indicator function for some set $S\subset\R$ to be determined. The eigenbasis of the Hamiltonian is thus $\{\ket{k}\}_{k\in S}$. We will show later that $S$ is the support of the spectral measure, i.e.\ the Fourier transform of the Toeplitz inner product between quantum states above. Since $\braket{k}{k}=0$ for $k\notin S$ such that $\ket{k}=0$, then we may write the Hamiltonian as
\begin{align}
    \label{eq:Hamiltonian}
    \h H = \intS{k} k \proj{k}.
\end{align}
In terms of the spectral decomposition of $\h H$, the unitary is $\h U(\theta)=\intS{k} e^{-ik\theta} \proj{k}$. This implies that the mapping, $\theta \mapsto \h U(\theta)$ is injective, i.e.\ for every $\theta_1 \neq \theta_2$ in $\mathbb{R}$, we have that $\h U(\theta_1) \neq \h U(\theta_2)$. The initial state is $\ket{\psi_0}=\intS{k}\psi_0(k)\ket{k}$ for some complex wavefunction $\psi_0(k)$ supported on $S$. The initial state's probability distribution in $k$ is thus $\abs{\psi_0(k)}^2$, and the final state is $\ket{\psi_\theta}=\intS{k} e^{-ik\theta}\psi_0(k) \ket{k}$. The inner product between final states, $\braket{\psi_{\theta}}{\psi_{\theta'}} = \intS{k} e^{ik(\theta-\theta')} \abs{\psi_0(k)}^2$, equals the value at $\theta-\theta'$ of the Fourier transform of the initial state's probability distribution in $k$.

\subsubsection{Covariant measurement}
\label{sec:covariant}
What is the optimal measurement of the final state to estimate the parameter $\theta$? Since the inner product is not necessarily a delta function, we cannot estimate the parameter perfectly but some measurement bases may still outperform others. If the figure-of-merit is covariant with respect to the symmetry group $\{ \hat{U}(\theta)\} _{\theta\in\mathbb{R}}$, i.e.\ if the prior $\pi(\theta)$ is flat, then it is known that the optimal measurement is also covariant with respect to this symmetry~\cite{holevo1978estimation,chiribella2005optimal, WisemanMilburn2009book}. We discuss covariant measurements more in Appendix~\ref{sec:Covariant figure-of-merit}, and we will return later to the assumption that the prior is flat. 

We thus want to find the covariant states, $\{\ket{\theta}\}_{\theta\in\mathbb{R}}$, that satisfy $\h U(\theta)\ket{\theta'}=\ket{\theta+\theta'}$. Let the states be $\ket{\theta}=\intS{k}c(\theta,k)\ket{k}$ for some $c(\theta,k)$ such that covariance implies that $c(\theta,k)=e^{-ik\theta}c(0,k)$. The inner product between states is then $\braket{\theta}{{\theta'}} = \intS{k} e^{i(\theta -\theta') k}\abs{c(0,k)}^2$ which equals $\delta(\theta-\theta')$ such that the states are orthogonal if $S=\R$ and $\abs{c(0,k)}=1/\sqrt{2\pi}$. The covariant states are thus unique up to the complex phase of $c(0,k)$ which is a gauge freedom described by the unitary $\h V=\intS{k} e^{i\vartheta_k}\proj{k}$ for a given set of a phases $\{\vartheta_k\}_{k\in S}$. Since this gauge unitary commutes with the group action, $[\h U(\theta), \h V]=0$, it can be ignored without loss of generality. Therefore, we define the covariant states as a Fourier transform of the eigenbasis of the Hamiltonian as follows
\begin{align}\label{eq:quantum whitening}
    \ket{\theta} = \frac{1}{\sqrt{2\pi}}\intS{k}e^{-i\theta k}\ket{k} 
\end{align}
where the inverse Fourier transformation is $\ket{k} = \frac{1}{\sqrt{2\pi}}\intginf{\theta}e^{i\theta k}\ket{\theta}$. The inner product between covariant states is thus
\begin{align}\label{eq:covariant inner product}
    \braket{\theta}{{\theta'}} = \frac{1}{2\pi} \intS{k} e^{i(\theta -\theta') k}.
\end{align}
It is optimal to measure the covariant states provided that the prior is flat, which we will revisit shortly. If $S=\R$ such that the covariant states are orthogonal ($\braket{ \theta}{ \theta'} = \delta(\theta - \theta')$), then we project onto them by measuring the covariant operator, $\h W = \intginf{\theta}\theta\proj{ \theta}$. This is the canonical conjugate of the Hamiltonian, $[\h W, \h H]=i$, such that $\h U^\dagger(\theta)\h W\h U(\theta)= \h W + \theta$. This is similar to the canonical quadratures which obey $[\h x, \h p]=i$ and $e^{i\theta\h p}\h xe^{-i\theta\h p}=\h x+\theta$ and whose eigenstates are related by the Fourier transform.

If instead $S\subsetneq\R$ such that the covariant states are not orthogonal, then they form a covariant non-projective POVM on the subspace spanned by the eigenbasis of the Hamiltonian. The POVM is $\{\h E_\theta\}_{\theta\in(-\pi,\pi]}$ where the effects are $\h E_\theta=\proj{\theta}$. This POVM is normalised although the covariant states may not be. In this case, we can still form the covariant operator, $\h W = \intginf{\theta}\theta\proj{ \theta}$, but now the covariant states are not its eigenbasis since they are not orthogonal. Ref.~\cite{macieszczak2014bayesian} showed that replacing the non-projective POVM $\{\h E_\theta\}_{\theta\in(-\pi,\pi]}$ with projection onto the orthogonal eigenbasis of $\h W$ cannot increase the BMSE. Since we already know that the covariant POVM is optimal, the projective measurement onto the eigenbasis of $\h W$ is also optimal (its BMSE equals the MBMSE). This means that $\h W$ equals the BSLD $\h L$ in Eq.~\ref{eq:BSLD}. Conversely, to show that a given POVM is optimal in general, it suffices to form the operator like $\h W$ (which can also be interpreted as projection onto its eigenbasis) and show that it equals the BSLD.

From the initial family of non-orthogonal quantum states $\{\ket{\psi_\theta}\}_{\theta\in\mathbb{R}}$ related by an additive unitary, we have defined two continuous families of states: the stationary states $\{\ket{k}\}_k$ and covariant states $\{\ket{\theta}\}_{\theta\in\mathbb{R}}$. These span the same subspace of possible quantum states spanned by $\{\ket{\psi_\theta}\}_{\theta\in\mathbb{R}}$. We may restrict ourselves to this subspace without loss of generality, which can significantly reduce the dimensionality of the Hilbert space as we will see later.

\subsection{Toeplitz family of states}
\label{sec:whitening}
We now consider a general Toeplitz family of states $\{\ket{\psi_\theta}\}_{\theta\in\mathbb{R}}$ and want to show that it is generated by an additive unitary. Since the geometry of the quantum states is Toeplitz, there exists some positive semi-definite function $G:\R\to\C$, which is called the symbol of the Toeplitz family, such that $\braket{\psi_\theta}{\psi_{\theta'}} = G(\theta-\theta')$ and $G(0)=1$. Since $G$ is not necessarily a delta function, the states $\{\ket{\psi_\theta}\}_{\theta\in\mathbb{R}}$ are not orthogonal and so we first want to find a basis to express them in.

Let us perform a Fourier transform and then normalise to define the following Fourier states
\begin{align}\label{eq:quantum Fourier transform}
    \ket{k} = \frac{1}{2\pi \sqrt{g(k)}} \intginf{\theta} e^{i k\theta} \ket{\psi_\theta}
\end{align}
Here, we define the spectral measure of the symbol $G(\theta) = \intginf{k} e^{i k \theta} g(k)$ as
\begin{align}\label{eq:g(k) definition}
    g(k) = \frac{1}{2\pi} \intginf{\theta} e^{-i k \theta} G(\theta).
\end{align}
If $g(k)=0$ for a given $k$, then we take $\ket{k}=0$. Let $S=\{k\in\R:g(k)>0\}$ be the support of $g$. The inner product between Fourier states in Eq.~\ref{eq:quantum Fourier transform} is thus
\begin{align}\label{eq:Fourier inner product}
    \braket{k}{l} = \delta(k - l)\chi_{S}(k).
\end{align}
On the support of $g$, the Fourier states are orthonormal, $\braket{k}{l} = \delta(k - l)$ for $k\in S$ and $l\in S$, and otherwise the inner product vanishes, $\braket{k}{l} = 0$ if $k\notin S$ or $l\notin S$. The inverse transformation between Fourier states and the original family is then
\begin{align}\label{eq:inverse QFT}
    \ket{\psi_\theta} = \intS{k} e^{-i k\theta} \sqrt{g(k)}\ket{k}
\end{align}
The function $g(k)$ is a valid probability distribution by Bochner's theorem because it is normalised ($\intS{k} g(k)=G(0)=1$) and non-negative since it is the Fourier transform of a positive semi-definite function. Similarly, the Gram matrix of inner products is Hermitian and positive semi-definite such that its eigenvalues $g(k)$ are non-negative. %
While this construction could be applied to finite-dimensional systems such as qubits or qudits, we focus on the infinite-dimensional case. We also assume here that $\theta$ ranges over all of $\R$. If instead the Hilbert space is finite dimensional or $\theta$ is continuous within a bounded interval because $G$ is periodic, then $S$ is discrete. In this latter case, we need to use the Fourier series instead and then normalise as we show in Appendix~\ref{app:periodic case}.

We can now show that, for any Toeplitz family of states, there exists an additive unitary that generates it. (We already showed that the additive unitary case is Toeplitz.) For a given Toeplitz symbol $G(\theta)$, let the Hamiltonian be defined as $\h H = \intS{k} k \proj{k}$ with eigenbasis $\{\ket{k}\}_{k\in S}$ from Eq.~\ref{eq:quantum Fourier transform}. Let the initial state be
\begin{align}
    \label{eq:initial state}
    \ket{\psi_0}=\intS{k}\sqrt{g(k)}\ket{k}
\end{align}
such that its probability distribution in $k$ is the spectral measure $g(k)$. The wavefunction in the previous section is thus $\psi_0(k)=\sqrt{g(k)}$. Then, the final state $\ket{\psi_\theta}=\h U(\theta)\ket{\psi_0}$ agrees with Eq.~\ref{eq:inverse QFT} and the Gram matrix equals the symbol $G(\theta-\theta')=\braket{\psi_{\theta}}{\psi_{\theta'}}$ as required. The additive unitary and Toeplitz family cases are thus equivalent.

If the prior is flat, then the covariant measurement is thus optimal by the result in the additive unitary case. The inner product between the original states $\{\ket{\psi_\theta}\}_{\theta\in\mathbb{R}}$ and covariant states $\{\ket{\theta}\}_{\theta\in\mathbb{R}}$ defined in Eq.~\ref{eq:quantum whitening} is %
\begin{align}\label{eq:original to whitened}
    \braket{ \theta}{\psi_{\theta'}} = \frac{1}{\sqrt{2\pi}}  \intS{k} \sqrt{g(k)} e^{i(\theta - \theta')k}.
\end{align}
The likelihood of measuring $\theta$ from quantum whitening ($\h E_\theta=\proj{\theta}$) for a given parameter $\theta'$ and state $\h\rho(\theta')=\proj{\psi_{\theta'}}$ is $L(\theta|\theta')=\trSmall{\h\rho(\theta')\h E_\theta} = \absSmall{\braket{ \theta}{\psi_{\theta'}}}^2$ which equals
\begin{align}\label{eq:likelihood}
    L(\theta|\theta') 
    &= \frac{1}{2\pi}  \intS{k}\intS{l} \sqrt{g(k)g(l)} e^{i(\theta - \theta')(k-l)}.
\end{align}
For the Toeplitz case, and in particular the coherent states that we consider later, we call the covariant states $\{\ket{\theta}\}_{\theta\in\mathbb{R}}$ the ``quantum whitened states'' and the covariant operator $\h W$ the ``quantum whitening measurement operator''. This is in analogy to the classical whitening process, where a continuous random variable with a stationary two-point correlation function is mapped to a white noise random variable by a similar procedure of Fourier transforming, dividing by the square root of the eigenvalues, and then inverse Fourier transforming.

Similarly to the previous section, from the initial non-orthogonal Toeplitz family $\{\ket{\psi_\theta}\}_{\theta\in\mathbb{R}}$, we have defined the families of the Fourier states (stationary states) $\{\ket{k}\}_{k\in\mathbb{R}}$ and quantum whitened states (covariant states) $\{\ket{\theta}\}_{\theta\in\mathbb{R}}$. The Fourier states are orthogonal on $S$ and the quantum whitened states are orthogonal if $S=\R$.

\subsection{Fundamental quantum limit}
\label{sec:waveform basis}
We now calculate the MBMSE analytically in the general case of a Toeplitz family with the uninformative and uniform prior.

Let us start with just the Toeplitz assumption and assuming nothing about the prior. For a given Toeplitz family of states $\{\ket{\psi_\theta}\}_{\theta\in\mathbb{R}}$ and prior $\pi(\theta)$, the mixed state and Bayesian derivative in Eq.~\ref{eq:mixed state} in the Fourier basis in Eq.~\ref{eq:quantum Fourier transform} are
\begin{align}\label{eq:mixed state, Fourier basis}
    &   \hat{\bar\rho} = \intS{k} \intS{l} \tilde \pi(k - l) \sqrt{g(k)g(l)}\ket{k}\bra{l},
    \\& \hat{\bar\rho}' = \intS{k} \intS{l} i \tilde \pi'(k - l)  \sqrt{g(k)g(l)} \ket{k}\bra{l}\nonumber
\end{align}
where we use $\h\rho(\theta)=\proj{\psi_\theta}$ and Eq.~\ref{eq:inverse QFT}. The Fourier transform of the prior is $\tilde \pi(k) = \intginf{\theta}  e^{-i k \theta} \pi(\theta)$ such that $i \tilde \pi'(k) = \intginf{\theta} \theta e^{-i k \theta} \pi(\theta)$ and the inverse Fourier transform is $\pi(\theta) = \frac{1}{2\pi}\intginf{k} e^{i k \theta} \tilde \pi(k)$. We show this in Appendix~\ref{app:proof of whitening}. The next step in finding the MBMSE is to calculate the BSLD in Eq.~\ref{eq:BSLD sol} from the spectral decomposition of the mixed state $\hat{\bar\rho}$, however, the eigenbasis is not known for a general prior. We will show below that the mixed state is diagonal in the Fourier basis for a flat prior. Otherwise, we can diagonalise the mixed state numerically to calculate the BSLD as we do in Appendix.~\ref{app:Numerics}. Similarly, a covariant measurement is only necessarily optimal if the prior on $\theta$ is flat such that the figure-of-merit is covariant. For arbitrary priors, it may not be optimal to project onto the quantum whitened states.

\subsubsection{Uninformative, uniform prior on $\R$}
\label{sec:Improper prior}
Let us now additionally assume that the prior is flat such that we can apply the covariant measurement result. This needs to be done with care since we assumed that $\theta$ and $k$ were continuous on $\R$. This prior is the limit of a sequence of wider and wider normalised priors such as rectangular or Gaussian functions. In principle, we would want to calculate the MBMSE for each finite prior and then take the limit of the sequence of MBMSEs as the prior width goes to infinity. However, as discussed above, we cannot diagonalise the mixed state $\h{\bar\rho}$ in Eq.~\ref{eq:mixed state, Fourier basis} for an arbitrary prior analytically. We conjecture that the MBMSE and BMSE of quantum whitening for any such sequence will converge to the value that we obtain below. We show in Appendix~\ref{sec:Wide-prior limit} that this limit works in the case of amplitude estimation. Numerically, we show later in Sec.~\ref{sec:Frequency estimation case} that this also works for frequency estimation.

Here, we instead calculate the MBMSE directly for a prior that is normalised and uniform on $\R$. These two properties are all that the following proof needs to assume, i.e.\ that $\partial_\theta\pi(\theta)=0$ and $\intginf{\theta}\pi(\theta)=1$. The uniformity means that $\tilde\pi(k-l)\propto\delta(k-l)$ such that the mixed state in Eq.~\ref{eq:mixed state, Fourier basis} is diagonal in the Fourier basis. The normalisation of the prior is necessary, else the non-unit norm of the prior will appear in the MBMSE as we discuss in Appendix~\ref{sec:Wide-prior limit}.

We need to be careful assigning an exact value to the prior. We do so nevertheless so that we can write the following equations explicitly, but we emphasise that our conclusions only require that the prior be normalised and flat on $\R$. Suppose that $\pi(\theta)\equiv\varepsilon$ for some infinitesimal $0<\varepsilon\ll1$ such that $\intginf{\theta}\pi(\theta) = 1$ and $\tilde \pi(k-l) = \varepsilon2\pi\delta(k-l)$. Informally, we choose that $\varepsilon=1/[2\pi\delta(0)]$ in the following equations such that $\tilde \pi(k-l) = \delta(k-l)/\delta(0)$ but this division by $\delta(0)$ needs to be handled with the upmost care. The only property that we use is that $\delta(0)/\delta(0)=1$ such that the flat prior and evidence will cancel in Bayes's rule in Eq.~\ref{eq:Bayes rule} below.

\subsubsection{BMSE of quantum whitening}
\label{sec:BMSE of quantum whitening}
The prior being flat implies that the covariant measurement, i.e.\ quantum whitening, is optimal. The MBMSE is thus the BMSE of quantum whitening $\h W$. Let us find the posterior $p(\theta'|\theta)$ on the parameter $\theta'$ for a given measurement result $\theta$. Given the likelihood $L(\theta|\theta')$ in Eq.~\ref{eq:likelihood}, the evidence is $p(\theta)\equiv1/[2\pi\delta(0)]$ which is normalised like the prior ($\intginf{\theta} p(\theta)=1$). The posterior thus equals the likelihood ($p(\theta'|\theta)=L(\theta|\theta')$) by Bayes's rule in Eq.~\ref{eq:Bayes rule}. The posterior is normalised ($\intginf{\theta'}p(\theta'|\theta)=1$) and centred on the measurement result ($\check\theta_{\theta} = \theta$). The MBMSE, which equals the BMSE in Eq.~\ref{eq:BMSE} for $\h W$, is thus
\begin{align}\label{eq:MBMSE, improper prior}
    \Delta^2\theta = \mathcal{V} := \frac{1}{4} \intS{k} \frac{g'(k)^2}{g(k)}. %
\end{align}
Here, the MSE in Eq.~\ref{eq:MSE} and variance of the posterior in Eq.~\ref{eq:posterior variance} are both constant and equal to the MBMSE 
\begin{align}
    \text{MSE}(\theta') = V_\text{post}(\theta) \equiv \mathcal{V}
\end{align}
and the prior and evidence are normalised. We prove this result in Appendix~\ref{app:proof of whitening, 2}.

The value of $\mathcal{V}$ in Eq.~\ref{eq:MBMSE, improper prior} has a familiar functional form. It equals a quarter of the Fisher information in Eq.~\ref{eq:prior Fisher information} were we to treat $g(k)$ in Eq.~\ref{eq:g(k) definition} as a prior on $k$ for some hypothetical dual estimation problem. We emphasise that this hypothetical dual problem of estimating $k$ and its classical Fisher information $4\mathcal{V}$ are not the same as our actual problem of estimating $\theta$ and the quantum Fisher information which we discuss below. In particular, we wish to minimise the MBMSE with respect to $\theta$ and thus the Fisher information of $g(k)$. Informally, since $\theta$ and $k$ are related by a Fourier transform, this scenario is similar to the classical time-frequency uncertainty principle known as the Gabor limit in signal processing. This intuition will become clearer when we discuss the Gaussian case in Sec.~\ref{sec:Gaussian case}.

\subsubsection{Comparison to the QFI}
Let us now calculate the actual QFI for this estimation problem, which we distinguish from the intuition about the functional form of the MBMSE provided above. The value of $\mathcal{V}$ in Eq.~\ref{eq:MBMSE, improper prior} for the MBMSE is in contrast with the QFI which will not be reached except for at high SNR. Using the additive unitary representation with the Hamiltonian $\h H$ in Eq.~\ref{eq:Hamiltonian} and initial state $\ket{\psi_0}$ in Eq.~\ref{eq:initial state}, the QFI with respect to $\theta$ is $\IQ(\theta)=4\varSubSmall{\ket{\psi_0}}{\h H}$ which equals
\begin{align}\label{eq:QFI, 4Var[g(k)]}
    \IQ(\theta)=4\left(\intS{k} k^2 g(k)-\left[\intS{k} k g(k)\right]^2\right)
\end{align}
which is four times the variance of $g(k)$. In contrast to the QCRB, the MBMSE in Eq.~\ref{eq:MBMSE, improper prior} in principle depends on more than just the second moment of $g(k)$, although we will show that it is equal in the Gaussian case below. In particular, we will show cases in Sec.~\ref{sec:applications} where the QFI is finite and the MBMSE diverges.

The value for the QFI in Eq.~\ref{eq:QFI, 4Var[g(k)]} can also be shown without using the additive unitary representation. Let us Taylor expand the Toeplitz symbol for a small parameter difference $\delta\theta$ as
\begin{align}
    G(\delta\theta)\approx 1+\delta\theta G'(0)+\frac{1}{2}\delta\theta^2G''(0)
\end{align}
where $\re{G'(0)}=0$ holds such that the states are normalised. The derivatives of the symbol at zero are related to the moments of the spectral measure as $G'(0) = i \intS{k} k g(k)$ and $G''(0) = - \intS{k} k^2 g(k)$. The QFI from the fidelity in Eq.~\ref{eq:fidelity} thus equals the value in Eq.~\ref{eq:QFI, 4Var[g(k)]}. 

\subsubsection{Optimality of quantum whitening}
\label{sec:separate proof}
We can also prove that quantum whitening is optimal without using the result that covariant measurement is optimal. We include this alternative proof here for completeness and provide further details in Appendix~\ref{app:proof of whitening, 3}.

For the uninformative, uniform prior ($\tilde \pi(k-l) = \delta(k-l)/\delta(0)$), the mixed state in Eq.~\ref{eq:mixed state, Fourier basis} is normalised ($\trSmall{\hat{\bar\rho}} = 1$) and diagonal in the Fourier basis since
\begin{align}\label{eq:rho, improper prior}
    \hat{\bar\rho} = [1/\delta(0)] \intS{k} g(k) \ket{k}\bra{k}.
\end{align}
This was the point of choosing the flat prior: we now know that the eigenbasis of the mixed state is the Fourier basis and thus can calculate the BSLD. The matrix coefficients of the Bayesian derivative in Eq.~\ref{eq:mixed state, Fourier basis} are
\begin{align}\label{eq:rho', improper prior}
    \braopket{k}{\hat{\bar\rho}'}{l} = i [\delta'(k - l)/\delta(0)]\sqrt{g(k)g(l)}
\end{align}
such that the coefficients of the BSLD in Eq.~\ref{eq:BSLD sol} are
\begin{align}\label{eq:BSLD, improper prior, unsimplified}
   \braopket{k}{\h L}{l} = i\delta'(k - l) \frac{2\sqrt{g(k)g(l)}}{g(k)+g(l)}\chi_S(k)\chi_S(l).
\end{align}
The distributions $\delta'(k-l)f_1(k,l)$ and $\delta'(k-l)f_2(k,l)$ are equal if the functions $f_1$ and $f_2$ and their partial derivatives are equal at $k=l$, i.e.\ if the following conditions are satisfied %
\begin{align}\label{eq:conditions on delta derivative}
    f_1(k,k)&=f_2(k,k),
    \\\partial_x f_1(x+l,l)|_{x=0}&=\partial_x f_2(x+l,l)|_{x=0},\nonumber
    \\\partial_x f_1(k,k-x)|_{x=0}&=\partial_x f_2(k,k-x)|_{x=0}.\nonumber
\end{align}
This can be shown using integration by parts against an arbitrary test function. Eq.~\ref{eq:BSLD, improper prior, unsimplified} thus implies that here 
\begin{align}\label{eq:BSLD, improper prior}
    \braopket{k}{\h L}{l} = i\delta'(k - l)\chi_S(k)\chi_S(l).
\end{align}
In comparison, the matrix coefficients of the quantum whitening measurement operator are
\begin{align}\label{eq:W coeffs, improper prior}
    \braopket{k}{\h W}{l}=i\delta'(k-l)\chi_S(k)\chi_S(l)
\end{align}
which equal those of the BSLD $\h L$ in Eq.~\ref{eq:BSLD, improper prior}. This can be verified by showing that $\h W$ satisfies the Lyapunov equation in Eq.~\ref{eq:Lyapunov equation}. In particular, the matrix coefficients of the anticommutator are
\begin{align}\label{eq:anticommutator, improper prior}
    \braopket{k}{\frac{1}{2}\{\hat{\bar\rho}, \hat W\}}{l} &= \frac{1}{2}i[\delta'(k-l)/\delta(0)][g(k) + g(l)]
\end{align}
which equal the matrix coefficients of the Bayesian derivative in Eq.~\ref{eq:rho', improper prior} since the conditions in Eq.~\ref{eq:conditions on delta derivative} hold. This means that it is optimal to measure the quantum whitened states $\ket{\theta'}$, either via the non-projective POVM if $S\subsetneq\R$ or by projection if $S=\R$. Either way, the optimal estimator for $\theta$ is simply the measurement result $\theta'$. 

We have now determined the fundamental quantum limit and optimal measurement for an arbitrary Toeplitz family given no prior information about the parameter. We will now apply these results to various families of single-particle and coherent states. However, first we must establish the inner product structure, i.e.\ the geometry, of single-particle and coherent states.

\section{Geometry of single-particle states}
\label{sec:single-particle states}
Before we return to our problem of determining the fundamental quantum limit on frequency estimation of the coherent state in Eq.~\ref{eq:state}, it will be informative to first study the simpler problem of estimating the parameter governing the waveform of a single-particle state, i.e.\ the multi-mode Fock state with a single particle in a given temporal mode. This single-particle state will provide a good introduction to the techniques that we use later for the coherent state. Although the inner product structure of the multi-mode single-particle and coherent states is well-known, we review it here for completeness as we will build upon it later in our Bayesian analysis. %

Let us provide a basic physical model for the single-particle state. It may be imagined as the result of coupling an optical field to a two-level atomic transition with the field initially in vacuum and the qubit excited. Suppose that after some energy-conserving unitary dynamics the final state of the qubit is the ground state, then the final state of the optical field will be a single-particle state of exactly one photon. The temporal mode of the state is determined by the dynamics and the unknown signal.

We will demonstrate the following key idea: that we may restrict ourselves to the ``waveform subspace'' spanned by the possible quantum states which significantly reduces the dimension of the Hilbert space as will be shown. To find a basis for this waveform subspace, we need to first understand how the geometry of the quantum states depends on the corresponding geometry of the classical waveforms, i.e.\ temporal modes. We will show that these geometries are the same for single-particle states but different for coherent states which is what makes our motivating problem about coherent states interesting.

\subsection{General single-particle state case}
Let us generalise the classical waveforms beyond Eq.~\ref{eq:signal} to now allow for any family parameterised by $\theta$ of $L^2(\R)$ functions, i.e.\ square-integrable complex waveforms:
\begin{align}\label{eq:generalised signal}
    \left\{h_\theta \,\bigg|\, \theta\in\R,\, h_\theta: \R\to\C,\, \intginf{t}\abs{h_\theta(t)}^2<\infty\right\}
\end{align}
We want to estimate the classical waveform--governing parameter $\theta$ from measurements of the quantum state that encodes it. There are many possible ways for a classical waveform to be encoded in a quantum state, but we focus on the single-particle and coherent state cases. For a given classical waveform and encoding scheme, we may refer to the corresponding quantum state as a ``quantum waveform''. 

Let the single-particle quantum state corresponding to a given classical waveform be defined as $\ket{x_\theta}=\h b_{\theta}^\dag\ket{0}$ where the temporal mode annihilation operator is defined as 
\begin{align}\label{eq:temporal mode operator}
    \h b_\theta = \intginf{t} x_\theta^*(t) \h a(t)
\end{align}
and we decompose the classical waveform as $h_\theta(t)=-i\sqrt{2}\alpha_\theta x_\theta(t)$ where $\alpha_\theta\in\C$ and $\intginf{t} \abs{x_\theta(t)}^2 = 1$. Here, the normalisation $\alpha_\theta$ is discarded since we assume that the state only has one particle but later this will be the complex amplitude of the corresponding coherent state. Thus, without loss of generality, let us assume that $\alpha_\theta\equiv1$ for the remainder of this section. The inner product between single-particle states is then related to the commutator between different temporal modes as
\begin{align}\label{eq:single-particle inner product}
    \braket{x_\theta}{x_{\theta'}} = [\h b_{\theta}, \h b_{\theta'}^\dag] = \intginf{t} x_{\theta}^*(t) x_{\theta'}(t) 
\end{align}
which is the $L^2(\R)$ inner product of the corresponding classical waveforms which are normalised. 

The classical and quantum geometries are thus the same since their inner products are the same: There is no added structure or quantum noise from measuring the single-particle states compared to the classical waveforms. This means that quantum whitening the single-particle states is the same as classically whitening their corresponding classical waveforms, i.e.\ temporal modes, as we show in Appendix~\ref{app:Quantum whitening of single-particle states}. We will see that these geometries are not the same in the coherent state case.

\subsection{QFI of single-particle states}
To determine the QFI, suppose that $\theta'=\theta+\delta\theta$ with $\delta\theta$ small such that we can Taylor expand the normalised waveform as
\begin{align}\label{eq:normalised waveform expansion}
    x_{\theta+\delta\theta}(t)\approx x_{\theta}(t)+\delta\theta\partial_\theta x_{\theta}(t)+\frac{1}{2}\delta\theta^2\partial_\theta^2 x_{\theta}(t)
\end{align}
where we keep only terms $\order{\delta\theta^2}$ henceforth. The inner product in Eq.~\ref{eq:single-particle inner product} is thus
\begin{align}
    \braket{x_\theta}{x_{\theta+\delta\theta}} &\approx 1 + \delta\theta \intginf{t} x_{\theta}^*(t) \partial_\theta x_{\theta}(t) \\&+ \frac{1}{2}\delta\theta^2 \intginf{t} x_{\theta}^*(t) \partial_\theta^2 x_{\theta}(t).\nonumber
\end{align}
Since the waveforms $x_\theta(t)$ are normalised, by computing the norm of both sides of Eq.~\ref{eq:normalised waveform expansion}, we can show that from the linear term in $\delta \theta$ that
\begin{align}
    \intginf{t} \re{x_{\theta}(t)\partial_\theta x_{\theta}^*(t)} = 0
\end{align}
and from the quadratic term in $\delta \theta^2$ that
\begin{align}
    \intginf{t} \re{x_{\theta}(t)\partial_\theta^2 x_{\theta}^*(t)} = - \intginf{t} \abs{\partial_\theta x_{\theta}(t)}^2.
\end{align}
This means that the infinitesimal fidelity between single-particle states is
\begin{align}
    \mathcal{F}(\theta) 
    &= \absSmall{\braket{x_{\theta}}{x_{\theta+\delta\theta}}}^2
    \\&\approx 1 - \delta\theta^2 \intginf{t} \abs{\partial_\theta x_{\theta}(t)}^2 \nonumber
    \\&+ \delta\theta^2 \left(\intginf{t} \im{x_{\theta}^*(t) \partial_\theta x_{\theta}(t)}\right)^2\nonumber
\end{align}
such that by Eq.~\ref{eq:fidelity} the QFI is
\begin{align}\label{eq:single-particle QFI}
    \IQ(\theta) &= 4 \intginf{t} \abs{\partial_\theta x_{\theta}(t)}^2 \\&- 4\left(\intginf{t} \im{x_{\theta}^*(t) \partial_\theta x_{\theta}(t)}\right)^2.\nonumber
\end{align}
This agrees with the QFI in Eq.~\ref{eq:QFI, 4Var[g(k)]} if the family of waveforms is Toeplitz. We will compare this below to the QFI of a coherent state with an average particle number of one. 

\section{Geometry of coherent states}
\label{sec:coherent states}

We now finally return to our problem of estimating the frequency $\omega$ from the coherent state in Eq.~\ref{eq:state}. We address the toy problem of estimating the displacement of a single harmonic oscillator in Appendix~\ref{app:Bayesian quantum estimation of displacement}. Here, we need to find the optimal measurement across many displaced harmonic oscillators in time. The total Hilbert space is too large to simulate numerically, e.g., the dimension is $d^M$ if we discretise the total time $T$ into $M=T/\delta t$ oscillators in time-bins of width $\delta t$ and truncate each oscillator to dimension $d$ in the Fock basis, see Appendix~\ref{app:Discretising the time domain}. However, the waveform subspace $\text{span}\{\ket{h_\omega}\}_{\omega\in\mathbb{R}}$ spanned by the possible quantum states can be much smaller as we will see later. Intuitively, this is because the inner product between neighbouring coherent states in the family $\{\ket{h_\omega}\}_{\omega\in\mathbb{R}}$ may not decay quickly.

We want to find a basis for the waveform subspace and then calculate the MBMSE in Eq.~\ref{eq:MBMSE} from computing $\hat{\bar\rho}$ and $\hat{\bar\rho}'$ in that basis. To find a basis, we need to first calculate the Gram matrix $G_{\omega,\omega'}=\braket{h_{\omega}}{h_{\omega'}}$ of inner products between different coherent states to understand the structure of the waveform subspace. We will calculate the Gram matrix for coherent states generally and then consider the case of frequency estimation and other examples summarised in Table~\ref{tab:Toeplitz}.

\subsection{General coherent state case}
Let us now generalise the classical waveforms beyond Eq.~\ref{eq:signal} and coherent states beyond Eq.~\ref{eq:state} to now any family of classical waveforms in Eq.~\ref{eq:generalised signal} and coherent states defined as follows
\begin{align}\label{eq:generalised state}
    \ket{h_\theta} &= e^{\frac{i}{\sqrt2}\intginf{t} \left[ h_\theta(t) \h a^\dagger(t) + h_\theta^*(t) \h a(t)\right]} \ket{0}.
\end{align}
Here, we choose a convention such that the expectation values of the quadratures follow the classical waveform~as 
\begin{align}\label{eq:quadrature EVs}
    \braopket{h_\theta}{\h x(t)}{h_\theta}&=-\im{h_\theta(t)},
    \\\braopket{h_\theta}{\h p(t)}{h_\theta}&=\re{h_\theta(t)}.\nonumber
\end{align}
We now want to relate the Hilbert space structure of the classical waveforms to that of the coherent quantum states like we did above for the single-particle states. The inner product between the coherent states is 
\begin{align}
    \label{eq:inner product of coherent states}
    \braket{h_{\theta}}{h_{\theta'}} = e^{-\frac{1}{4} \intginf{t}\left[\abs{h_\theta(t)}^2+\abs{h_{\theta'}(t)}^2-2h_{\theta}^*(t) h_{\theta'}(t)\right]},
\end{align}
where the first two terms in the exponent are the $L^2(\R)$ norms squared of the classical waveforms $h_\theta$ and $h_{\theta'}$, and the third term is twice the $L^2(\R)$ inner product between them. We prove this result in Appendix~\ref{app:inner product}. This inner product is not necessarily Toeplitz but we will see examples where it will be.

Even if the classical waveforms are orthogonal, the inner product between the corresponding quantum states is nonzero for finite SNR because of the quantum noise of the coherent state, unlike the single-particle states. If the $L^2(\R)$ inner product $\intginf{t} h_{\theta}^*(t) h_{\theta'}(t)$ is real, e.g.\ if the classical waveforms are real, then the inner product between quantum states in Eq.~\ref{eq:inner product of coherent states} becomes
\begin{align}
    \label{eq:inner product of states, real L2 inner product}
    \braket{h_{\theta}}{h_{\theta'}} = e^{-\frac{1}{4} \intginf{t}\abs{h_\theta(t) - h_{\theta'}(t)}^2}
\end{align}
where the exponent is the $L^2(\R)$ distance between the classical waveforms $h_\theta$ and $h_{\theta'}$. 

The $L^2(\R)$ norm squared of the classical waveform $h_\theta$ which appears in the exponent of Eq.~\ref{eq:inner product of coherent states} is equal to twice the average number of particles in the corresponding coherent state since
\begin{align}
    \label{eq:average total number}
    \bar N_\theta = \braopket{h_\theta}{\h N}{h_\theta} = \frac{1}{2}\intginf{t} \abs{h_\theta(t)}^2
\end{align}
where $\h N = \intginf{t}\h a^\dagger(t)\h a(t)$ is the total number operator and the units of $h_\theta(t)$ are square root Hertz. This is implied by the coherent state being an eigenstate of the annihilation operator $\hat a(t)\ket{h_\theta} = \frac{i}{\sqrt2}h_\theta(t)\ket{h_\theta}$.

\subsection{QFI of coherent states}
Let us consider how distinguishable parameter values are when they are infinitesimally far apart. Suppose that $\theta'=\theta+\delta\theta$ with $\delta\theta$ small such that we can Taylor expand the signal as
\begin{align}
    h_{\theta'}(t)\approx h_{\theta}(t)+\delta\theta\partial_\theta h_{\theta}(t)+\frac{1}{2}\delta\theta^2\partial_\theta^2 h_{\theta}(t)
\end{align}
and we drop $\order{\delta\theta^3}$ terms henceforth. The inner product in Eq.~\ref{eq:inner product of coherent states} is then
\begin{align}
    \label{eq:inner product, infinitesimal}
    \braket{h_{\theta}}{h_{\theta+\delta\theta}} &\approx e^{-\frac{1}{4} \delta\theta^2 \intginf{t}\abs{\partial_\theta h_{\theta}(t)}^2+i\varphi}
\end{align}
where the phase is
\begin{align}
    \varphi=\frac{1}{2}\delta\theta\intginf{t}\im{h_{\theta}^*(t)\left(\partial_\theta h_{\theta}(t)+ \frac{1}{2}\delta\theta \partial_\theta^2 h_{\theta}(t)\right)}
\end{align}
which vanishes if the $L^2(\R)$ inner product between waveforms $\intginf{t} h_{\theta}^*(t) h_{\theta+\delta\theta}(t)$ is real. The fidelity between these states is thus
\begin{align}
    \mathcal{F}(\theta) = \absSmall{\braket{h_{\theta}}{h_{\theta+\delta\theta}}}^2 &\approx e^{-\frac{1}{2} \delta\theta^2 \intginf{t}\abs{\partial_\theta h_{\theta}(t)}^2}
\end{align}
such that by Eq.~\ref{eq:fidelity} the QFI is
\begin{align}\label{eq:QFI, general coherent}
    \IQ(\theta) = 2 \intginf{t}\abs{\partial_\theta h_{\theta}(t)}^2
\end{align}
which is the $L^2(\R)$ norm squared of the partial derivative of the classical waveform $\partial_\theta h_{\theta}$. This value for the QFI can also be shown from the Gaussian formula in Eq.~\ref{eq:QFI Gaussian, mean only}. It can furthermore be recovered from the QFI in Eq.~\ref{eq:QFI, 4Var[g(k)]} if the family is Toeplitz.

The CFI for a time-varying quadrature measurement $\h x_{\vartheta(t)}(t)=\cos[\vartheta(t)]\h x(t)+\sin[\vartheta(t)]\h p(t)$ saturates the QFI, where $\vartheta(t)=\arg[\partial_\theta h_\theta(t)]+\frac{\pi}{2}$ such that $\partial_\theta\braopket{h_\theta}{\h x_{\vartheta(t)}(t)}{h_\theta}=\abs{\partial_\theta h_\theta(t)}$ by Eq.~\ref{eq:quadrature EVs}. (Here, $\h x_{\vartheta(t)}(t)$ should be distinguished from the normalised temporal mode $x_\theta(t)$ in Eq.~\ref{eq:temporal mode operator}.) This can be shown, e.g., by discretising the state in time following Appendix~\ref{app:Discretising the time domain}, calculating the CFI in Eq.~\ref{eq:CFI Gaussian, mean only}, and then taking the continuum limit. This Fisher information approach, however, is only tight asymptotically and so we need to calculate the MBMSE instead to understand the fundamental quantum limit for finite SNR.

\subsubsection{Comparison to the QFI of single-particle states}
Suppose that we decompose the classical waveform as $h_\theta(t)=-i\sqrt{2}\alpha_\theta x_\theta(t)$ into its norm $\alpha_\theta=\sqrt{\bar N_\theta}$ and $L^2(\R)$-normalised waveform $x_\theta(t)$. Here, we assume that $\alpha_\theta>0$ without loss of generality, since any overall complex phase of $h_\theta(t)$ can be absorbed into $x_\theta(t)$. Then, the QFI in Eq.~\ref{eq:QFI, general coherent} decomposes as follows
\begin{align}\label{eq:QFI, coherent, decomposed}
    \IQ(\theta) = 4 (\partial_\theta\alpha_\theta)^2+4 \alpha_\theta^2\intginf{t}\abs{\partial_\theta x_\theta(t)}^2
\end{align}
where the first term comes from changes in the norm of the classical waveform (average number of particles) and the second term comes from changes in the normalised waveform (temporal mode function). 

The QFI for a coherent state with an average particle number of one ($\alpha_\theta\equiv1$) is thus
\begin{align}\label{eq:QFI, coherent state, nbar=1}
    \IQ(\theta) = 4 \intginf{t}\abs{\partial_\theta x_\theta(t)}^2.
\end{align}
Let us compare this $\bar N_\theta=1$ coherent state $\ket{h_\theta} = e^{\h b_\theta^\dagger-1/2} \ket{0}$ from Eq.~\ref{eq:coherent state, operator definition} to the single-particle state $\ket{x_\theta}=\h b_{\theta}^\dag\ket{0}$ with the same temporal mode operator $\h b_\theta$ in Eq.~\ref{eq:temporal mode operator} given by the same normalised waveform $x_\theta(t)$. The coherent state QFI in Eq.~\ref{eq:QFI, coherent state, nbar=1} is an upper bound on the single-particle state QFI in Eq.~\ref{eq:single-particle QFI}. The difference in QFI is the second term in Eq.~\ref{eq:single-particle QFI} which equals
\begin{align}
    \IQ^{\ket{h_\theta}}(\theta) - \IQ^{\ket{x_\theta}}(\theta) 
    &= 4\left(\intginf{t} \abs{x_{\theta}(t)}^2\partial_\theta\arg[x_{\theta}(t)] \right)^2.
\end{align}
This difference vanishes, e.g., (1) if $x_\theta(t)$ is real or (2) if $x_\theta(t)$ is complex but the parameter does not affect its phase such that $\partial_\theta\arg[x_{\theta}(t)]=0$. This is because non-vacuum coherent states are sensitive to phase changes but Fock states are not since Fock states are rotationally symmetric in the phase plane. 

For example, suppose that we want to sense a rotation of phase space by $\h U(\theta)=e^{-i\theta\h N}$. A given coherent state $\ket{h_0}$ with $h_0(t)=-i\sqrt{2}\alpha_0 x_0(t)$ simply rotates around the origin in the phase plane under this unitary as $\ket{h_\theta}=\ket{h_0e^{-i\theta}}$ such that the QFI is $\IQ(\theta)=4\alpha_0^2$. In particle, if the average particle number is one, then the coherent state QFI is 4. In comparison, a given single-particle state $\ket{x_0}$ is invariant up to a global phase as $\h U(\theta)\ket{x_0}=e^{-i\theta}\ket{x_0}$ such that the QFI vanishes. 

\subsection{Frequency estimation of a windowed sinusoid}

\begin{table*}[]
\begingroup
\setlength{\tabcolsep}{6pt} %
\renewcommand{\arraystretch}{1} %
\begin{tabular}{@{}llllll@{}}
\toprule
Parameter & Signal & Gram matrix & Toeplitz & Circulant & QFI \\ \midrule
General case, $\theta$ & Eq.~\ref{eq:generalised signal}: family of square-integrable functions & Eq.~\ref{eq:inner product of coherent states} & \hspace{0.5cm}---        & \hspace{0.5cm}--- & Eq.~\ref{eq:QFI, general coherent} \\\arrayrulecolor{black!30}\midrule
Frequency, $\omega$ & Eq.~\ref{eq:signal windowed (0, T)}: sinusoidal & Eq.~\ref{eq:cosine inner product of states, (0, T)}$^{\star}$ & \hspace{0.5cm}\cmark$^{\star}$        & \hspace{0.5cm}\xmark   & Eq.~\ref{eq:QFI, sinusoid, (0, T)}$^{\star}$     \\
 & Eq.~\ref{eq:complex exponential signal, (0, T)}: complex exponential & Eq.~\ref{eq:complex exponential inner product of states, (0, T)} & \hspace{0.5cm}\cmark        & \hspace{0.5cm}\xmark   & Eq.~\ref{eq:QFI, complex exponential, (0, T)}       \\
 \arrayrulecolor{black!30}\midrule
Amplitude, $A$ & Eq.~\ref{eq:general signal for amplitude estimation}: general scale factor family & Eq.~\ref{eq:inner product, amplitude estimation} & \hspace{0.5cm}\cmark         & \hspace{0.5cm}\xmark   & Eq.~\ref{eq:QFI, amplitude}     \\
 & Eq.~\ref{eq:signal windowed (0, T)}: sinusoidal & Eq.~\ref{eq:Gram matrix, amplitude estimation}$^{\star}$ & \hspace{0.5cm}\cmark         & \hspace{0.5cm}\xmark    & Eq.~\ref{eq:QFI, amplitude, sinusoid}$^{\star}$    \\
 & Eq.~\ref{eq:complex exponential signal, (0, T)}: complex exponential & Eq.~\ref{eq:Gram matrix, complex case, amplitude estimation} & \hspace{0.5cm}\cmark         & \hspace{0.5cm}\xmark    & Eq.~\ref{eq:QFI, amplitude, complex exponential}    \\
 \arrayrulecolor{black!30}\midrule
Phase, $\phi$ & Eq.~\ref{eq:signal windowed (0, T)}: sinusoidal & Eq.~\ref{eq:phase estimation, Gram matrix}$^{\star}$ & \hspace{0.5cm}\cmark$^{\star}$         & \hspace{0.5cm}\cmark$^{\star}$    & Eq.~\ref{eq:QFI, sinusoidal phase}$^{\star}$    \\
 & Eq.~\ref{eq:complex exponential signal, (0, T)}: complex exponential & Eq.~\ref{eq:phase estimation, Gram matrix, complex case} & \hspace{0.5cm}\cmark        & \hspace{0.5cm}\cmark    & Eq.~\ref{eq:QFI, complex phase}     \\
\arrayrulecolor{black}\bottomrule
\end{tabular}%
\endgroup
\caption{Summary of the properties and QFI of the Gram matrix $G_{\theta,\theta'}=\braket{h_{\theta}}{h_{\theta'}}$ of inner products between coherent states for estimating different parameters from different signals. The Gram matrix is Toeplitz if $G_{\theta,\theta'}=G_{\theta-\theta'}$ and circulant if $G_{\theta,\theta'}=G_{\theta-\theta'+n P}$ for $n\in\Z$ and some finite period $P$. We reference the equations for the signal defined on the time interval $(0, T)$ but the $(-T/2,T/2)$ case is similar. ${}^{\star}$This property or equation holds only in the many-cycles limit of $T\gg1/\min(\omega,\omega')$.} %
\label{tab:Toeplitz}
\end{table*}

Let us examine this inner product structure for the case of the sinusoidal signal. The general state in Eq.~\ref{eq:generalised state} recovers the sinusoidal state in Eq.~\ref{eq:state} by windowing the signal in Eq.~\ref{eq:signal} to the time interval $(0, T)$ as 
\begin{align}\label{eq:signal windowed (0, T)}
    h_\omega(t) = A\cos(\omega t + \phi) \chi_{(0, T)}(t)
\end{align}
where $\chi_{(0, T)}(t)$ is the indicator function on $(0, T)$. This choice of window function and time interval is only a convention, as we discuss in Appendices~\ref{app:(-T/2, T/2)}--\ref{app:complex exponential}. The average number of particles in Eq.~\ref{eq:average total number} is thus
\begin{align}
    \label{eq:average total number, cosine case}
    \bar N_\omega &= \frac{A^2 T}{4}+\frac{A^2}{8 \omega }\left(\sin [2 (\omega T +\phi )]-\sin [2 \phi]\right)
\end{align}
which equals $\bar N_\omega\approx A^2T/4$ in the relevant many-cycles limit of $T\gg1/\omega$. In this regime, the average number of particles thus equals an eighth of the SNR squared, since the SNR is $A\sqrt{2T}$. Meanwhile, the $L^2(\R)$ inner product between classical waveforms is 
\begin{align}
    \intginf{t}h_{\omega}^*(t) h_{\omega'}(t) 
    &= \frac{1}{2}A^2T\,\text{sinc}[(\omega-\omega')T]
    \\&+ \frac{A^2\sin[(\omega+\omega')T+2 \phi]-A^2\sin (2 \phi )}{2(\omega+\omega')}\nonumber
\end{align}
where $\text{sinc}(0)=1$ and $\text{sinc}(x)=\sin(x)/x$ for $x\neq0$. This equals $\frac{1}{2}A^2T\,\text{sinc}[(\omega-\omega')T]$ in the many-cycles limit of $T\gg1/(\omega+\omega')$. The inner product between quantum states in Eq.~\ref{eq:inner product of coherent states}, or Eq.~\ref{eq:inner product of states, real L2 inner product} since the waveform is real, is thus
\begin{align}
    \label{eq:cosine inner product of states, (0, T), exact}
    \braket{h_{\omega}}{h_{\omega'}} &= e^{-\frac{1}{4}A^2 T g(\omega,\omega')}
\end{align}
where the exponent is defined as
\begin{align}
    g(\omega,\omega') &= 1 - \text{sinc}[(\omega-\omega')T]
    \\&+\frac{1}{4 \omega T}\left(\sin [2 (\omega T +\phi )]-\sin [2 \phi]\right)\nonumber
    \\&+ \frac{1}{4 \omega' T}\left(\sin [2 (\omega' T +\phi )]-\sin [2 \phi]\right)\nonumber
    \\&- \frac{\sin[(\omega+\omega')T+2 \phi]-\sin (2 \phi )}{(\omega+\omega')T}\nonumber
\end{align}
such that in the many-cycles limit of $T\gg1/\min(\omega,\omega')$ the Gram matrix of inner products becomes 
\begin{align}
    \label{eq:cosine inner product of states, (0, T)}
    \braket{h_{\omega}}{h_{\omega'}} &\approx e^{-\frac{1}{4}A^2 T \left(1 - \text{sinc}[(\omega-\omega')T]\right)}
\end{align}
In this regime, the Gram matrix is a Toeplitz matrix since it is purely a function of $\omega-\omega'$, whereas in general the Gram matrix in Eq.~\ref{eq:cosine inner product of states, (0, T), exact} is not Toeplitz. The many-cycles limit is relevant because we consider long integration times and want to know how the sensing performance scales with time. 

The sinc function in the exponent of the inner product of states in Eq.~\ref{eq:cosine inner product of states, (0, T)} comes from the asymptotic value of the $L^2(\R)$ inner product of classical waveforms $\intginf{t}h_{\omega}^*(t) h_{\omega'}(t)$ in the many-cycles limit. This sinc function vanishes for frequencies that are far apart $(\omega-\omega')T\gg1$ or separated by a harmonic frequency of $\omega-\omega'=n\pi/T$ for $(0, T)$ in Eq.~\ref{eq:cosine inner product of states, (0, T)}. When the sinc function vanishes such that the classical waveforms are orthogonal, however, the inner product between the waveform states is nonzero and equals $\braket{h_{\omega}}{h_{\omega'}}\approx e^{-A^2T/4}$ for finite SNR because of the quantum noise of the coherent state.

The QFI in Eq.~\ref{eq:QFI, general coherent} for frequency estimation of the signal defined on $(0, T)$ in Eq.~\ref{eq:signal windowed (0, T)} is
\begin{align}
\label{eq:QFI, sinusoid, (0, T)}
    \IQ(\om)\approx\frac{1}{3} A^2T^3
\end{align}
where we take the many-cycles limit of $\omega T\gg1$. The QCRB in Eq.~\ref{eq:QCRB} in amplitude units is thus
\begin{align}
\label{eq:QCRB, frequency estimation}
    \text{MSE}(\om)^{-1/2}\approx\sqrt{3}A^{-1}T^{-3/2}
\end{align}
which scales as $1/T^{3/2}$ instead of the na\"ive harmonic frequency spacing of $1/T$, as discussed in Sec.~\ref{sec:Review of classical frequency estimation}. To reach a given frequency resolution $\delta\omega$, this means that we require only a time $T$ proportional to $1/\delta\om^{2/3}$ rather than $1/\delta\omega$ (cf.\ Ref.~\cite{schmitt2021optimal}). Heuristically, this scaling can be understood as the number of states that are approximately orthogonal. Suppose that we require that $\abs{\braket{h_{\omega}}{h_{\omega+\delta\om}}} \leq \varepsilon$ for some $\varepsilon>0$ such that the probability $\varepsilon^2$ of measuring $\ket{h_{\omega+\delta\om}}$ if the signal is actually $\ket{h_{\omega}}$ is small. For two close frequencies satisfying $\delta\om\, T\ll1$, the inner product in Eq.~\ref{eq:inner product, infinitesimal} and Eq.~\ref{eq:cosine inner product of states, (0, T)} is
\begin{align}\label{eq:inner product, infinitesimal, cosine}
    \braket{h_{\omega}}{h_{\omega+\delta\om}} \approx e^{-\frac{1}{24}\delta\om^2A^2T^3}
\end{align}
such that the frequencies need to be at least as far apart as
\begin{align}
    \label{eq:frequency bin width}
    \delta\om \geq 2A^{-1}T^{-3/2}\sqrt{6\log(1/\varepsilon)}.
\end{align}
This means that the maximum number of approximately orthogonal waveform states $N=\Delta\omega/\delta\omega$ in a frequency interval of width $\Delta\omega$ is
\begin{align}
    \label{eq:number of approximately orthogonal states}
    N=\frac{\Delta\omega AT^{3/2}}{2\sqrt{6\log(1/\varepsilon)}}.
\end{align}
Suppose that $\varepsilon$ is small enough that we can well-determine which of these $N$ waveform states the measured state is closest in frequency to. Then, the remaining error in our estimate of $\omega$ simply comes from the resolution of the frequency bins of width $\delta\omega$. If the prior is flat, then we may use that the standard deviation of the continuous uniform distribution is $(b-a)/(2\sqrt{3})$ on the interval $(a,b)$ to conclude that the error in estimating the frequency given Eq.~\ref{eq:frequency bin width} is
\begin{align}
    \label{eq:frequency estimation error}
    \sigma_\omega = A^{-1}T^{-3/2}\sqrt{2\log(1/\varepsilon)}
\end{align}
which has the same scaling with $A^{-1}T^{-3/2}$ as the QCRB in Eq.~\ref{eq:QCRB, frequency estimation}. This provides an intuitive explanation of the scaling of the QCRB. We assumed above that $\delta \omega\,T\ll1$ such that we require that the SNR $A\sqrt{2T}$ satisfies
\begin{align}
    \label{eq:approximately orthogonal condition}
    A\sqrt{T} \gg 2\sqrt{6\log(1/\varepsilon)}.
\end{align}
This illustrates that the QFI is only valid for high SNRs since $\varepsilon$ is small. At low SNR, the QCRB cannot be achieved by any measurement which motivates our use of Bayesian techniques.

We discuss frequency estimation in the case of a complex exponential signal instead of the real sinusoidal one above in Appendix~\ref{app:complex exponential}, including for different window functions than the rectangular one above. We also discuss instead estimating the amplitude with the frequency and phase known in Appendix~\ref{app:Amplitude estimation} and estimating the phase with the amplitude and frequency known in Appendix~\ref{app:Phase estimation}. We summarise our results in Table~\ref{tab:Toeplitz} including whether the Gram matrix is Toeplitz and circulant (i.e.\ periodic). 

\section{Applications}
\label{sec:applications}
Let us now examine some examples of the MBMSE in Eq.~\ref{eq:MBMSE, improper prior} for different Toeplitz families of states $\{\ket{\psi_\theta}\}_{\theta\in\mathbb{R}}$ with different symbols $G(\theta)$ and the uninformative, uniform prior on the parameter $\theta$. We will then present our numerical results for frequency estimation of a sinusoidal signal in a coherent states where the Gram matrix is only approximately Toeplitz and the prior is uniform on a bounded interval.

\subsection{Gaussian case}
\label{sec:Gaussian case}
Suppose that the symbol is $G(\theta) = e^{-\frac{1}{2}\varsigma^2 \theta^2}$ which is proportional to a Gaussian distribution with variance $1/\varsigma^2$. Then, the spectral measure $g(k)=e^{-k^2/(2\varsigma^2)}/(\varsigma\sqrt{2\pi})$ is a Gaussian probability distribution with variance $\varsigma^2$. The likelihood in Eq.~\ref{eq:original to whitened} is $L(\theta|\theta') = \frac{2\varsigma }{{\sqrt{2\pi}}} e^{-2\varsigma^2(\theta-\theta')^2}$ which is Gaussian with mean $\theta'$ and variance $1/(4\varsigma^2)$. The MBMSE in Eq.~\ref{eq:MBMSE, improper prior} is thus $\mathcal{V} = 1/(4\varsigma^2)$ from the MSE. This can also be shown from the dual estimation problem since the Fisher information of $g(k)$ with respect to $k$ is $1/\varsigma^2$. The MBMSE decreases as $\varsigma$ increases since $G(\theta)$ becomes narrower such that neighbouring parameters can be more easily distinguished. Thus we want to minimise the Fisher information of $g(k)$. This value of $\mathcal{V}$ also equals the QCRB in Eq.~\ref{eq:QFI, 4Var[g(k)]} (cf.\ Eq.~\ref{eq:QFI, amplitude}) because the QCRB is tight for a finite number of measurements in the Gaussian case.

For example, the Gram matrix for amplitude estimation of a coherent state in Eq.~\ref{eq:inner product, amplitude estimation} is Gaussian with $\varsigma^2=\bar N_1$. The MBMSE in Eq.~\ref{eq:MBMSE Gaussian prior, amplitude estimation} is $1/(4\bar N_1)$ in the limit of wider and wider priors. This result agrees with the value of $\mathcal{V}$ for the MBMSE from Eq.~\ref{eq:MBMSE, improper prior} above. 

For another example, consider estimating the centre of a single-particle state corresponding to a Gaussian pulse. Suppose that the normalised classical waveform in Eq.~\ref{eq:temporal mode operator} is
\begin{align}\label{eq:x(t) Gaussian}
    x_\theta(t) = \frac{1}{(2\pi)^{1/4}\sqrt{\sigma}}e^{-(t-\theta)^2/(4\sigma^2)}
\end{align}
The Gram matrix in Eq.~\ref{eq:single-particle inner product} is thus $G(\theta)=e^{-\theta^2/(8 \sigma ^2)}$ and so the QCRB and MBMSE are both $\sigma^2$. In comparison, as we show in Appendix~\ref{sec:centre estimation}, the coherent state case driven by $h_\theta(t)=-i\sqrt{2}\alpha x_\theta(t)$ has the same QCRB of $\sigma^2$ if the average number of particles is one ($\alpha=1$) but has divergent MBMSE since the Gram matrix is non-Gaussian and asymptotically nonzero.

\subsubsection{Constant case}
Suppose that $G(\theta)\equiv 1$ such that the state is independent of the parameter and the posterior equals the prior since no information is gained from any measurement. Then, $g(k)=\delta(k)$ and the MBMSE in Eq.~\ref{eq:MBMSE, improper prior} diverges as it equals the prior variance. Similarly, the QFI in Eq.~\ref{eq:QFI, 4Var[g(k)]} vanishes such that the QCRB diverges. This is the wide limit $\varsigma\to0$ of the Gaussian case.

\subsubsection{Singular case}
Let us now consider the opposite case. Suppose that $G(\theta)=\delta(\theta)/\delta(0)$ such that the states are all orthogonal, e.g.\ the position eigenstates, then $g(k)\equiv1/[2\pi\delta(0)]$ which is the uninformative, uniform probability distribution on $\R$ similarly to the prior. The MBMSE in Eq.~\ref{eq:MBMSE, improper prior} vanishes as the parameter can be exactly determined from a single measurement projecting onto the states themselves. Similarly, the QFI in Eq.~\ref{eq:QFI, 4Var[g(k)]} diverges such that the QCRB vanishes. This is the narrow limit $\varsigma\to\infty$ of the Gaussian case.

\subsection{Frequency estimation of single-particle states}
\label{sec:sinc case}
Suppose that the symbol is $G(\theta)=\text{sinc}(a\theta)$ for some $a>0$. We give an example of this below for frequency estimation of a single-particle state. The spectral measure is then $g(k)=\frac{1}{2a}\chi_{(-a, a)}(k)$ which is the uniform probability distribution on the bounded interval $(-a, a)$. This rectangular function is discontinuous at $k=\pm a$ such that the MBMSE in Eq.~\ref{eq:MBMSE, improper prior} diverges. The proof of the posterior variance in Eq.~\ref{eq:MBMSE, improper prior} given in Appendix~\ref{app:proof of whitening, 2} assumes that $g$ is smooth which it is not here. Nevertheless, we can independently verify that the MBMSE diverges since Eq.~\ref{eq:original to whitened} implies that the posterior of quantum whitening is $p(\theta|\theta')=(a/\pi)\,\text{sinc}^2[a(\theta' - \theta)]$ which has unbounded variance. This is because the sinc function $\text{sinc}(\theta)$ falls off as $1/\theta$, which is too slow for the integral in the posterior variance in Eq.~\ref{eq:posterior variance} to converge. Since the symbol $G(\theta)$ is not constant, however, we do learn something from the quantum whitening measurement such that the posterior is better than the prior in a sense, but their variances are both undefined. This highlights that the BMSE can be a poor figure-of-merit for the uninformative, uniform prior on $\R$. In contrast, the QFI in Eq.~\ref{eq:QFI, 4Var[g(k)]} is $\IQ(\theta)=4a^2/3$ such that the QCRB in Eq.~\ref{eq:QCRB} is finite.

A more robust metric would be to compute the limit of the ratio between the posterior and prior variances for a sequence of, e.g., uniform priors on wider and wider bounded intervals that converge to the uninformative, uniform prior on $\R$. Analytically, we do not know the MBMSE for finite prior width. Numerically, however, we see that the variance of the uniform prior on $\theta\in(-\Delta/2,\Delta/2)$ scales as $\Delta^2$ and the quantum whitening posterior variance scales as $\Delta$ as expected. This means that the ratio between the posterior and prior variances scales as $1/\Delta$ such that the relative gain improves for wider and wider priors. We proved above that quantum whitening is optimal in the limit of wide priors such that this relative gain proportional to $1/\Delta$ is the best that we can achieve.

For example, suppose that we want to estimate the frequency of a single-particle state corresponding to the normalised classical waveform $x_\omega(t) = -i\frac{1}{\sqrt{T}}e^{i(\omega t + \phi)} \chi_{(-T/2, T/2)}(t)$ in Eq.~\ref{eq:temporal mode operator} (cf. Eq.~\ref{eq:complex exponential signal, (-T/2, T/2)}). The Gram matrix is thus $G(\omega)=\text{sinc}(\omega T/2)$ such that $g(k)=\frac{1}{T}\chi_{(-T/2, T/2)}(k)$ and the QFI is $\IQ(\omega)=T^2/3$ which is finite unlike the MBMSE. For the real sinusoid case, the same result holds for $x_\omega(t) = \sqrt{\frac{2}{T}}\cos(\omega t + \phi) \chi_{(-T/2, T/2)}(t)$ (cf. Eq.~\ref{eq:signal windowed (-T/2, T/2)}) in the many-cycles limit since then the Gram matrix is the same. We address the coherent state case of frequency estimation below.

\subsection{Frequency estimation of coherent states}
\label{sec:Frequency estimation case}
Let us now finally return to the problem of frequency estimation of the coherent states. For the sinusoidal signal defined on $(0, T)$ in Eq.~\ref{eq:signal windowed (0, T)}, the exact Gram matrix in Eq.~\ref{eq:cosine inner product of states, (0, T), exact} is not Toeplitz but the Gram matrix in Eq.~\ref{eq:cosine inner product of states, (0, T)} in the many-cycles limit is Toeplitz. If we wanted to avoid the many-cycles limit, we could instead study the complex exponential signal defined on $(-T/2, T/2)$ in Eq.~\ref{eq:complex exponential signal, (-T/2, T/2)} whose exact Gram matrix in Eq.~\ref{eq:complex exponential inner product of states, (-T/2, T/2)} is Toeplitz and has the same functional form as Eq.~\ref{eq:cosine inner product of states, (0, T)}. %

\begin{figure}
    \centering
    \includegraphics[width=\columnwidth]{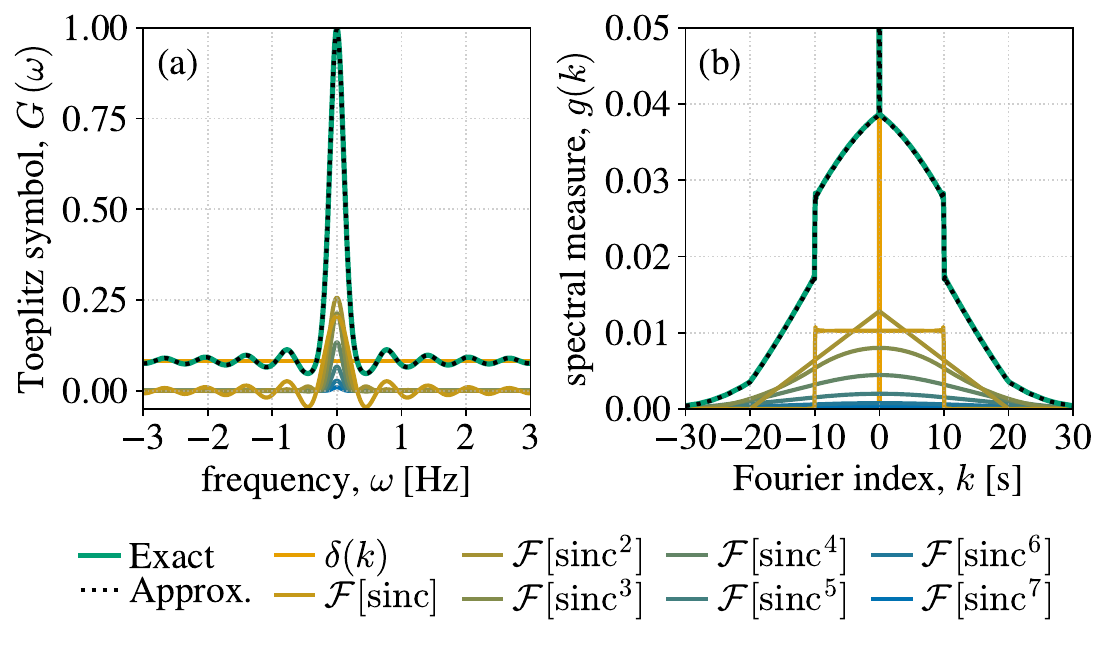}
    \caption{Exact and approximate (truncated at 7th order) values for the Taylor expansion of the (a) symbol $G(\omega)$ and (b) spectral measure $g(k)$ for the frequency estimation Gram matrix in Eq.~\ref{eq:cosine inner product of states, (0, T)} with $A=1$~Hz$^{-0.5}$ and $T=10$~s such that the SNR is $A\sqrt{2T}\approx4.5$. A divergence at DC and discontinuities at $\pm T$ are introduced in $g(k)$ from the Fourier transform of the constant and sinc parts of $G(\omega)$, respectively.}
    \label{fig:exp_sinc_Taylor}
\end{figure}
Suppose that the symbol is $G(\omega) = e^{-\frac{1}{4}A^2 T \left[1 - \text{sinc}(\omega T)\right]}$ whose Fourier transform, the spectral measure $g(k)$, we cannot compute analytically. However, as shown in Fig.~\ref{fig:exp_sinc_Taylor}, we observe numerically that $g(k)$ is discontinuous at the same points $k=\pm T$ as the Fourier transform $\frac{1}{2T}\chi_{(-T, T)}(k)$ of $\text{sinc}(\omega T)$ such that the MBMSE in Eq.~\ref{eq:MBMSE, improper prior} again diverges. This is in comparison to the QCRB in Eq.~\ref{eq:QCRB, frequency estimation} which is finite but unattainable. There is also a singularity at DC due to the $e^{-\frac{1}{4}A^2 T}$ asymptotic value of $G(\omega)$ for large $\omega$. 

\subsubsection{Low-SNR expansion}
These divergences can be explained analytically in the low-SNR limit $A^2 T\ll1$ since then the symbol is
\begin{align}
    G(\omega) 
    &\approx e^{-\frac{1}{4}A^2 T} \bigg[1 + \frac{A^{2}T}{4} \text{sinc}(\omega T) + \frac{A^{4}T^2}{32} \text{sinc}^2(\omega T)\bigg]
\end{align} %
where we drop terms of order $\order{A^6T^3/384}$. Each term here corresponds to one additional particle: the first term to zero particles, the second to one particle, the third to two particles etc. The spectral measure is then
\begin{align}
    g(k) \approx e^{-\frac{1}{4}A^2 T} \bigg[\delta(k) &+ \frac{A^{2}}{8} \chi_{(-T, T)}(k) \\&+ \frac{A^{4}}{128}\max(0,2T-\abs{k})\bigg].\nonumber
\end{align}
Here, the $n$th term in the symbol $G(\omega)$ is proportional to $\text{sinc}^n(\omega T)$ and thus goes as $1/\omega^n$ and its Fourier transform, i.e.\ the corresponding term in the spectral measure $g(k)$, is a piecewise degree-$(n-1)$ polynomial with support $(-nT,nT)$ and discontinuities in the $n$th derivative if $n\geq1$. For example, as shown above, the $n=0$ term goes to the Dirac delta function at $k=0$, the $n=1$ term goes to a rectangular function on $(-T, T)$, the $n=2$ term goes to a triangular function on $(-2T,2T)$, the $n=3$ term goes to a piecewise quadratic function on $(-3T,3T)$, etc. This expansion is only valid in the low SNR--limit but we see numerically the contributions to $g(k)$ from these different terms and the appearance of the discontinuities at $k=\pm T$ and divergence at DC for finite SNR too as shown in Fig.~\ref{fig:exp_sinc_Taylor}.

\subsubsection{Limiting behaviour for wider and wider priors}
Similarly to the discussion in the sinc case above, this means that the BMSE is a poor figure-of-merit for frequency estimation with the uninformative, uniform prior on $\R$ and that we want to instead study the ratio between the MBMSE, i.e.\ the minimum average posterior variance, and the prior variance for a sequence of wider and wider priors. Unlike the sinc case above, however, the symbol $G(\omega)$ here asymptotes to a nonzero value $e^{-\frac{1}{4}A^2 T}$ for large $\omega$. This means that the posterior variance scales as $\Delta\omega^2$ like the variance of the uniform prior on $(-\Delta\omega/2,\Delta\omega/2)$ for large $\Delta\omega$. We give another example in Eq.~\ref{eq:Gram matrix, Gaussian pulse centre} where the Gram matrix has a nonzero constant component and thus the posterior variance diverges quadratically.

We observe numerically that the gap between the prior and posterior variances closes, albeit quite slowly, for increasing $\Delta\omega$ and fixed SNR. This is expected as the effect on the variance from the central peak due to the other terms in the expansion of $G(\omega)$ is slowly suppressed for wider and wider priors, even though it may still contain a significant probability mass for a given finite prior. For example, for SNR 4.5, the numerical ratio of posterior to prior variance is approximately 0.06 for a prior spanning two orders of magnitude and only rises to 0.08 for a prior spanning four orders of magnitude. This means that the decrease in variance of the posterior compared to the prior is significant for wide but finite priors, even if we expect it to diminish in the limit of an infinitely wide prior.

The comparison above is between the prior and posterior distribution from quantum whitening. More precisely, we approximate the posterior distribution with the likelihood below Eq.~\ref{eq:original to whitened} which only equals the posterior in the limit of an infinitely wide prior. Is quantum whitening optimal for finite priors? We have only shown that quantum whitening is optimal for an infinite, flat prior. We calculate the MBMSE numerically from Eq.~\ref{eq:MBMSE sol} using the Toeplitz symbol above and methods in Sec.~\ref{app:Numerics}. The results for the MBMSE are similar to the posterior variance but not identical. The small residual difference between the MBMSE and posterior variance could be because (1) we use a finite numerical resolution, (2) we use the likelihood instead of the true posterior, and/or (3) quantum whitening is close to but not fully optimal for a wide but finite prior. We defer better understanding the finite prior case to future work, including what happens with non-uniform priors.

For the coherent state case, the quantum whitened states in Eq.~\ref{eq:quantum whitening} are not necessarily coherent states themselves since they are particular superpositions of coherent states. The coherent states form an overcomplete basis such that any state in the subspace is some superposition of them. This is in contrast to the single-particle state case where the quantum whitened states are themselves single-particle states with the classically whitened waveforms/temporal modes. The quantum whitened coherent states and the coherent states with classically whitened waveforms are thus not necessarily the same.

\begin{figure}
    \centering
    \includegraphics[width=\columnwidth]{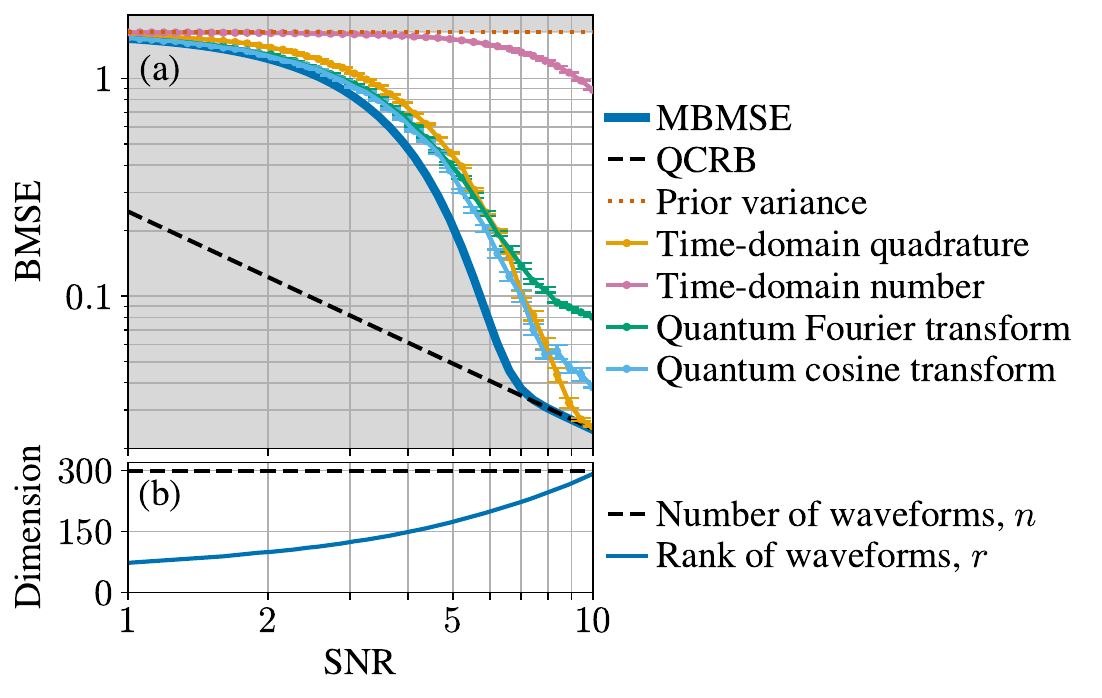}
    \caption{(a) BMSE in amplitude units for different measurement schemes versus the SNR in amplitude units $A\sqrt{2T}$ for different signal amplitudes $A$ and fixed total time $T$ compared to the MBMSE, QCRB, and prior variance for the parameters given in Table~\ref{tab:parameters}. The shaded grey region is inaccessible. (b)~Rank of the Gram matrix versus the SNR compared to the number of waveforms simulated numerically.}
    \label{fig:MBMSE vs SNR}
\end{figure}

\subsubsection{Non-Toeplitz case}
Let us now examine what happens if we do not assume that the Gram matrix is Toeplitz by not taking the many-cycles limit. Using the numerical approach in Appendix~\ref{app:Numerics}, we can study our motivating example of frequency estimation of the sinusoidal signal in Eq.~\ref{eq:signal windowed (0, T)} with known amplitude and phase and the exact non-Toeplitz Gram matrix in Eq.~\ref{eq:cosine inner product of states, (0, T), exact}. Let us fix the parameters given in Table~\ref{tab:parameters} such as the total time $T$ and prior width $\Delta \omega$ and vary the known amplitude $A$ to reach different values of the SNR $A\sqrt{2T}$. As the SNR increases, the MBMSE exhibits an SNR threshold as it goes from the prior variance to the QCRB as shown in Fig.~\hyperref[fig:MBMSE vs SNR]{\ref*{fig:MBMSE vs SNR}a}. For intermediate values of the SNR, the MBMSE is not achieved by quadrature measurement in time, which is the classical measurement discussed in Sec.~\ref{sec:Review of classical frequency estimation}. The MBMSE is also not achieved by performing a number-resolving measurement in time, e.g.\ illuminating a single-photon detector in the case of an optical device. Additionally, it is not saturated by first performing a discrete quantum Fourier or cosine transform to the bath discretised in the time-domain as in Appendix~\ref{app:Discretising the time domain} and then performing a number-resolving measurement to each of the resulting modes. This is the filter-based photon-counting strategy recently compared in Ref.~\cite{ethanpaper} to the time-domain quadrature measurement, although the advantage that we see here between these two strategies is smaller than that seen in Ref.~\cite{ethanpaper} likely due to the simplifications that we have made here to the signal and filter models and in assuming that all other parameters are known. This quantum Fourier transform and number-resolving measurement should be distinguished from the Fourier transform in Eq.~\ref{eq:quantum Fourier transform}. The optimal measurement that achieves the MBMSE here is not quantum whitening, because we have not taken the limit of wider and wider priors, but is to project onto analogous superpositions of the coherent waveform states. The rank $r$ of the Gram matrix remains stable under increasing the frequency resolution as discussed in Appendix~\ref{app:Numerics} but increases with the SNR for a fixed number of waveforms $n$ as shown in Fig.~\hyperref[fig:MBMSE vs SNR]{\ref*{fig:MBMSE vs SNR}b}. This means that the Fourier transform in Eq.~\ref{eq:quantum Fourier transform} is an efficient representation of the waveform subspace.

\begin{table}[]
\begingroup
\setlength{\tabcolsep}{6pt} %
\renewcommand{\arraystretch}{1} %
\begin{tabular}{@{}ll@{}}
\toprule
Parameter                                       & Value                     \\ \midrule
Known phase, $\phi$                             & 0                         \\
Total time, $T$                                 & 10~s                 \\
Harmonic frequency spacing, $2\pi/T$            & $2\pi\times0.1$~Hz   \\
Prior width, $\Delta \omega$                & $2\pi\times0.9$~Hz   \\
Centre frequency, $\omega_0$                    & $2\pi\times0.5$~Hz   \\
Centre number of cycles, $\omega_0 T/(2\pi)$ & $0.5$             \\
Minimum frequency, $\omega_\text{min}:=\omega_0-\Delta\omega/2$    & $2\pi\times0.05$~Hz  \\
Minimum number of cycles, $\omega_\text{min} T/(2\pi)$ & $5$             \\ %
Maximum frequency, $\omega_\text{max}:=\omega_0+\Delta\omega/2$    & $2\pi\times0.95$~Hz  \\
Maximum number of cycles, $\omega_\text{max} T/(2\pi)$ & $9.5$             \\
\arrayrulecolor{black!30}\midrule
Number of oscillators, $M$                      & 101                       \\
Time resolution, $\delta t$                     & 0.1~s                \\
Nyquist frequency, $\pi/\delta t$               & $2\pi\times5$~Hz      \\
Number of waveforms, $n$                        & 300                       \\
Frequency resolution, $\Delta \omega/n$         & $2\pi\times 3$~mHz \\
Phase difference per time bin, $\Delta\omega\delta t$ & $0.18\pi$                 \\
\arrayrulecolor{black!100}\bottomrule
\end{tabular}%
\endgroup
\caption{Parameters for the BMSE versus SNR results shown in Fig.~\ref{fig:MBMSE vs SNR}. The first group of parameters are physically meaningful in the continuum limit whereas the second group of parameters, below the grey line, only arise from discretising the time domain numerically as done in Appendix~\ref{app:Discretising the time domain}.}
\label{tab:parameters}
\end{table}

\section{Outlook}
\label{sec:conclusions}
In this paper, we have determined the fundamental Bayesian quantum limit of frequency estimation of a bosonic system. This Bayesian perspective yields an understanding of phenomena, such as the SNR threshold effect, which are hidden from the traditional Fisher information approach. The key is to understand the Hilbert space geometry of the multi-mode coherent states, and how it relates to their corresponding classical waveforms. For example, the coherent states all overlap for finite SNR, even if their classical waveforms are orthogonal. In the many-cycles limit, such that the inner product becomes Toeplitz, and if we have no prior information about the frequency, we prove that quantum whitening the states is optimal. This means that we project onto particular superpositions of the coherent states. This result applies beyond frequency estimation to any Toeplitz family of states. A divergence in the Bayesian limit arises, however, in the case of frequency estimation with an infinitely wide prior. We handle this numerically by using finite-width priors and taking the singular value decomposition of the waveform basis, which works even when the inner product is not Toeplitz. We observe that, although the SNR threshold appears just as it does classically, the classical limit can nevertheless be surpassed by a measurement similar to quantum whitening. These results may accelerate future searches for fundamental physics, such as gravitational-wave probes of neutron-star physics and the direct detection of ultra-light dark matter, as well as improve emerging quantum technologies using nonstandard, nonclassical techniques.

Many avenues of future work are promising such as determining how to experimentally realise quantum whitening, studying other parameterised families of waveforms such as Lorentzians, and considering covariant measurements for estimating the phase. We are also curious as to what happens if we marginalise over the amplitude and phase in the case of frequency estimation rather than treating them as known. Ref.~\cite{ethanpaper} shows that the head-to-head comparison between quadrature and number-resolving measurements changes dramatically whether or not we marginalise. We may want to simultaneously estimate the amplitude and phase rather than marginalise over them. However, tight general Bayesian multiparameter bounds are unknown. Additionally, the Bayesian limit is not known if we optimise over the intervening quantum control and preparation of the initial quantum state, potentially allowing for entanglement with noiseless ancillae (such as in Ref.~\cite{supaper}). Much therefore remains to be understood about Bayesian waveform estimation, especially compared to the Fisher information regime.

\begin{acknowledgements}
We thank the following people for their advice provided during this research: Katerina Chatziioannou, Rana Adhikari and the Caltech quantum control group, the Caltech Chen quantum group, and the ANU CGA squeezer group. This research is supported by the Australian Research Council Centre of Excellence for Gravitational Wave Discovery (Project No. CE170100004 and CE230100016). J.W.G. and this research are supported by an Australian Government Research Training Program Scholarship and also partially supported by the US NSF grant PHY-2011968. In addition, Y.C. acknowledges the support by the Simons Foundation (Award Number 568762) and the NSF grant PHY-2309231. T.G. acknowledges funding provided by  the Quantum Science and Technology early-career fellowship of the Israel Council for Higher Education. S.A.H. acknowledges support through an Australian Research Council Future Fellowship grant FT210100809. 
\end{acknowledgements}

\section*{Data availability}
The code and data that supports this work is openly available online~\cite{repo}.

\appendix
\section{Circulant family and flat prior}
\label{app:periodic case}
Suppose that we have a circulant family of non-orthogonal states $\{\ket{\psi_\theta}\}_{\theta\in\mathbb{R}}$ such that $\braket{\psi_{\theta}}{\psi_{\theta'}} = G(\theta-\theta')$ for some periodic function $G$ and the prior is uniform on the finite period. Let the period of the symbol $G$ be $2\pi$ without loss of generality such that the prior is $\pi(\theta)=[1/(2\pi)]\chi_{(-\pi,\pi]}(\theta)$ and we may restrict the family to $\{\ket{\psi_\theta}\}_{\theta\in(-\pi,\pi]}$.

Similarly to the Toeplitz case in Sec.~\ref{sec:whitening}, we can define the Fourier $\{\ket{\varphi_k}\}_{k\in\Z}$ and whitened $\{\ket{\phi_\theta}\}_{\theta\in(-\pi,\pi]}$ families of states. Many of these formulae are directly analogous to those in the main text, but we provide them in full here regardless for completeness. Instead of the Fourier transform in Eq.~\ref{eq:quantum Fourier transform} which has $\theta\in\R$ and thus $k\in\R$, we now calculate the Fourier series for $\theta\in(-\pi,\pi]$ and $k\in\Z$ as
\begin{align}
    \ket{\varphi_k} = \frac{1}{2\pi \sqrt{g_k}} \intg{-\pi}{\pi}{\theta} e^{i k\theta} \ket{\psi_\theta}
\end{align} %
By analogy to the continuous case in Eq.~\ref{eq:g(k) definition}, we define the discrete spectral measure as
\begin{align}\label{eq:discrete spectral measure}
    g_k = \frac{1}{2\pi} \intg{-\pi}{\pi}{\theta} e^{-i k \theta} G(\theta).
\end{align}
Here, we take $\ket{\varphi_k}=0$ if $g_k=0$ for a given $k$. Let $S=\{k\in\Z:g_k>0\}$ be the support of $g$. Similarly to Eq.~\ref{eq:Fourier inner product}, the inner product between Fourier states is thus
\begin{align}
    \braket{\varphi_k}{\varphi_l} = \delta_{kl}\chi_{S}(k).
\end{align}
Instead of Eq.~\ref{eq:inverse QFT}, the inverse transformation between states is now
\begin{align}\label{eq:QFS inverse}
    \ket{\psi_\theta} = \sum_{k\in S} e^{-i k\theta} \sqrt{g_k}\ket{\varphi_k}
\end{align}
and the symbol is $G(\theta) = \sum_{k\in S} e^{i k \theta} g_k$ such that $\ket{\psi_\theta}$ and $G(\theta)$ both have period $2\pi$. Similarly to the continuous case, the discrete spectral measure $g_k$ is a valid probability distribution since it is normalised ($\sum_{k\in S} g_k=1$) and non-negative due to the positive semi-definiteness of the inner product.

Similarly to the Toeplitz case, for any circulant family of states, there exists an additive unitary and initial state that generates it. The additive unitary $\h U(\theta)=e^{-i\h H\theta}$ for this circulant family is generated by the Hamiltonian $\h H = \sum_{k\in S} k \proj{\varphi_k}$ with the discrete eigenbasis $\{\ket{\varphi_k}\}_{k\in S}$. If the support of $g$ is partial, $S\subsetneq \Z$, then, e.g., the eigenbasis could only be for $k\geq0$ if $S=\mathbb{N}_0$ or be finite dimensional if $S$ is finite (we consider both of these cases below). If the initial state is $\ket{\psi_0}=\sum_{k\in S}\sqrt{g_k}\ket{\varphi_k}$, then the final state $\ket{\psi_\theta}=\h U(\theta)\ket{\psi_0}$ agrees with Eq.~\ref{eq:QFS inverse} and the Gram matrix recovers the periodic symbol $G(\theta-\theta')=\braket{\psi_{\theta}}{\psi_{\theta'}}$. The QFI is thus
\begin{align}
    \IQ(\theta)=4\left(\sum_{k\in S} k^2 g_k-\left[\sum_{k\in S} k g_k\right]^2\right)
\end{align}
which is four times the variance of $g_k$ and similar to Eq.~\ref{eq:QFI, 4Var[g(k)]}.

Instead of Eq.~\ref{eq:quantum whitening}, the quantum whitened states are now defined for $\theta\in(-\pi,\pi]$ as
\begin{align}
    \ket{\phi_\theta} = \frac{1}{\sqrt{2\pi}}\sum_{k\in S} e^{-i\theta k}\ket{\varphi_k} %
\end{align}
such that the inverse transformation is $\ket{\varphi_k} = \frac{1}{\sqrt{2\pi}}\intg{-\pi}{\pi}{\theta}e^{i\theta k}\ket{\phi_\theta}$. Although these states are not necessarily normalised, the effects $\h E_\theta=\proj{\phi_\theta}$ of the quantum whitening POVM $\{\h E_\theta\}_{\theta\in(-\pi,\pi]}$ are normalised since $\intg{-\pi}{\pi}{\theta}\h E_\theta=\h \Pi_S$, where $\h \Pi_S$ is the projector onto the subspace spanned by the eigenbasis of the Hamiltonian $\{\ket{\varphi_k}\}_{k\in S}$. The quantum whitened states are also covariant $\h U(\theta)\ket{\phi_{\theta'}}=\ket{\phi_{\theta+\theta'}}$ as before. Furthermore, if $S=\Z$, then these are the unique covariant and orthogonal states, up to a gauge unitary that can be ignored. The proof of this result is similar to the $S=\R$ case given in Sec.~\ref{sec:covariant} just with integrals replaced with sums.

Instead of Eq.~\ref{eq:covariant inner product}, the inner product between quantum whitened states is
\begin{align}\label{eq:covariant inner product, circulant}
    \braket{\phi_\theta}{\phi_{\theta'}} &= \frac{1}{2\pi} \sum_{k\in S} e^{i(\theta -\theta') k}
\end{align} %
If $S=\Z$, then the quantum whitened states are orthogonal ($\braket{\phi_\theta}{\phi_{\theta'}} = \delta(\theta - \theta')$) and we may construct a projective measurement operator describing quantum whitening, $\h W = \intg{-\pi}{\pi}{\theta}\theta\proj{\phi_\theta}$. (We can form the same operator in the non-projective case below, but its eigenbasis may be different.) For example, if estimating the rotation of a particle constrained to a one-dimensional ring, then $S=\Z$ since the angular momentum eigenstates are indexed by $\Z$ and quantum whitening is a projective measurement. 

If $S\subsetneq\Z$, however, then the quantum whitened states are not orthogonal and form a non-projective POVM. For example, consider the phase estimation of a coherent state in Appendix~\ref{app:Phase estimation}. There, $S=\mathbb{N}_0$ since the Fock states are indexed by $\mathbb{N}_0$ such that the whitened states are not orthogonal ($\braket{\phi_\theta}{\phi_{\theta'}} =\frac{1}{2\pi} [1-e^{i(\theta - \theta')}]^{-1}$). The normalisation of the POVM is then $\intg{-\pi}{\pi}{\theta}\h E_\theta=\h \Pi_S\neq1$.

For another example where $S\subsetneq\Z$, the relevant covariant measurement of the rotation of a qudit is well known~\cite{Holevo2011book}. Let the Hamiltonian be $\h H = \h J_z$ where $\{\ket{\varphi_k}\}_{k\in S}$ are the $\h J_z$ eigenstates for finite $S=\{-j,\mathellipsis,j\}$, the spin is $j=(d-1)/2$, and the dimension is $d$. The final state is thus $\ket{\psi_\theta} = \sum_{k=-j}^j e^{-i k\theta} \sqrt{g_k}\ket{\varphi_k}$ for some $g_k$ determined by the initial state. We assume that the dimension $d$ is odd and the spin $j$ is integer such that the symbol $G(\theta)=\sum_{k=-j}^{j} e^{i k \theta} g_k$ is $2\pi$ periodic. (If the dimension $d$ is instead even and the spin $j$ is half-integer, then the system is only $4\pi$ periodic and we need to appropriately modify the formalism.) The covariant states $\ket{\phi_\theta} = \frac{1}{\sqrt{2\pi}}\sum_{k=-j}^j e^{-i\theta k}\ket{\varphi_k}$ are sometimes called the ``quantum Fourier transform'' states. Their inner product in Eq.~\ref{eq:covariant inner product, circulant} is the following Dirichlet kernel
\begin{align*}
    \braket{\phi_\theta}{\phi_{\theta'}} = \frac{\sin[(\theta -\theta')(j + \frac{1}{2})]}{2\pi \sin(\frac{\theta -\theta'}{2})}.
\end{align*}
This converges to the Dirac comb $\sum_{n\in\Z}\delta(\theta-\theta'+2\pi n)$ in the limit of $j\to\infty$ corresponding to $S=\Z$. Although the continuous set $\{\ket{\phi_\theta}\}_{\theta\in(-\pi,\pi]}$ is not orthogonal, there is a $d$-dimensional orthogonal subset given by $\theta'$ satisfying $\theta-\theta'=n\pi/(j+\frac{1}{2})$ for $n=-j,\mathellipsis,j$ and any starting starting value of $\theta$. Up to this continuous degree-of-freedom of choosing $\theta$ and the normalisation of the covariant states, this $d$-dimensional orthogonal subset may be called the ``quantum Fourier transform'' basis. We saw a similar distinction between a continuum of non-orthogonal vectors and a discrete orthogonal subset of them in Sec.~\ref{sec:Review of classical frequency estimation} between the classical sinusoids indexed by a continuum of possible frequencies versus by just the harmonic frequencies.

For any $S\subset\Z$, the inner product between the original and quantum whitened states is similar to Eq.~\ref{eq:original to whitened} as
\begin{align}\label{eq:original to whitened, covariant}
    \braket{\phi_\theta}{\psi_{\theta'}} &= \frac{1}{\sqrt{2\pi}}\sum_{k\in S} \sqrt{g_k} e^{i(\theta-\theta') k}.
\end{align}
The likelihood of measuring $\theta$ for a given $\theta'$ with quantum whitening is $L(\theta|\theta')=\abs{\braket{\phi_\theta}{\psi_{\theta'}}}^2$, regardless of whether it is a projective measurement. If $S=\Z$, then Eq.~\ref{eq:original to whitened, covariant} is the function on $(-\pi,\pi]$ whose Fourier series is $\sqrt{g_k/(2\pi)}$ evaluated at $\theta-\theta'\bmod (-\pi,\pi]$.

The covariant measurement operator $\h W$ is off-diagonal in the eigenbasis of the Hamiltonian with the following matrix coefficients
\begin{align}\label{eq:W coeffs, circulant}
    \braopket{\varphi_{k}}{\hat W}{\varphi_{l}} = i\frac{(-1)^{k-l}}{k-l}\chi_S(k)\chi_S(l).
\end{align}
Unlike the $k\in\R$ case, however, the commutator between the quantum whitening operator and Hamiltonian is not canonical, $[\h W, \h H]\neq i$, as the commutator is instead
\begin{align}
    [\h W, \h H] &= i \sum_{k,l\in S;\,k\neq l} \ket{\varphi_k}\bra{\varphi_l} (-1)^{k-l}.
\end{align}
This highlights some of the common issues with defining a Hermitian phase operator.  

The continuous and discrete $k$ cases have different delta functions since $\intginf{\theta} e^{ik\theta} = 2\pi\delta(k)$ for the continuous case and $\intg{-\pi}{\pi}{\theta} e^{ik\theta} = 2\pi\delta_{k0}$ for the discrete case. One difference between the cases is then that $\intginf{\theta} \theta e^{ik\theta} = -i2\pi\delta'(k)$ for the continuous case, but $\intg{-\pi}{\pi}{\theta} \theta e^{ik\theta} = i2\pi(-1)^k/k$ for $k\neq0$ and zero otherwise for the discrete case. For example, this leads to the difference in commutator above and also the difference in $\h W$ Fourier coefficients between Eq.~\ref{eq:W coeffs, improper prior} for the continuous $k$ case and Eq.~\ref{eq:W coeffs, circulant} for the discrete $k$ case.

Although much of the structure of the circulant case with $\theta\in(-\pi,\pi]$ and $k\in\Z$ is similar to the Toeplitz case with $\theta,k\in\R$, two key results are different. First, the quantum whitening operator and Hamiltonian are not canonical conjugates in the full support case. Second, quantum whitening is not necessarily optimal with respect to the BMSE but is optimal for covariant figures-of-merit, which we will show below.

\subsection{Non-covariant figure-of-merit}
To show that quantum whitening is not optimal, we need to show that $\h W$ does not solve the Lyapunov equation for the BSLD $\h L$ in Eq.~\ref{eq:Lyapunov equation}. The BMSE in Eq.~\ref{eq:BMSE} is not covariant in this case since the MSE is not periodic. This means that although quantum whitening is covariant it need not be optimal here, unlike the $k\in\R$ case, since the covariant measurement result discussed in Sec.~\ref{sec:covariant} requires the figure-of-merit to be covariant. We will discuss this further in Appendix~\ref{sec:Covariant figure-of-merit}.

For a general prior, instead of Eq.~\ref{eq:mixed state, Fourier basis}, the mixed state and Bayesian derivative in Eq.~\ref{eq:mixed state} for the circulant case are
\begin{align}
    &   \hat{\bar\rho} = \sum_{k\in S} \sum_{l\in S} \tilde \pi_{k-l} \sqrt{g_k g_l}\ket{\varphi_k}\bra{\varphi_l},
    \\& \hat{\bar\rho}' = \sum_{k\in S} \sum_{l\in S} \tilde\varpi_{k-l} \sqrt{g_k g_l}\ket{\varphi_k}\bra{\varphi_l}\nonumber
\end{align}
where the Fourier series of the prior is $\tilde \pi_k = \intg{-\pi}{\pi}{\theta} e^{-i k \theta} \pi(\theta)$, the inverse transformation is $\pi(\theta) = \frac{1}{2\pi}\sum_{k=-\infty}^\infty e^{i k \theta} \tilde \pi_k$, and we define $\tilde\varpi_{k-l}:=\intg{-\pi}{\pi}{\theta}\pi(\theta)e^{-i\theta(k-l)}\theta$. %

For the flat prior $\pi(\theta)\equiv1/(2\pi)$, we have that $\tilde \pi_{k-l}=\delta_{kl}$ and $\tilde\varpi_{k-l}=i(-1)^{k-l}/(k-l)$ for $k\neq l$ and $\tilde\varpi_{0}=0$. Similarly to Eq.~\ref{eq:rho, improper prior}, the mixed state is thus diagonal in the Fourier basis (the eigenbasis of the Hamiltonian) since
\begin{align}
    \hat{\bar\rho} = \sum_{k\in S} g_k \proj{\varphi_k}.
\end{align}
And, instead of Eq.~\ref{eq:rho', improper prior}, the Bayesian derivative is now
\begin{align}
    \hat{\bar\rho}' = i\sum_{k,l\in S;\,k\neq l} \frac{(-1)^{k-l}}{k-l} \sqrt{g_k g_l}\ket{\varphi_k}\bra{\varphi_l}.
\end{align}
Instead of Eq.~\ref{eq:BSLD, improper prior, unsimplified} or Eq.~\ref{eq:BSLD, improper prior}, the BSLD in Eq.~\ref{eq:BSLD sol} for $k\neq l$ is thus
\begin{align}\label{eq:BSLD, circulant}
   \braopket{\varphi_{k}}{\hat L}{\varphi_{l}} &= i \frac{(-1)^{k-l}}{k-l} \frac{2\sqrt{g_k g_l}}{g_{k}+g_{l}} \chi_S(k)\chi_S(l)
\end{align}
where the diagonal coefficients of $\h L$ vanish like $\hat{\bar\rho}'$. Unlike the $k\in\R$ case, however, we do not have an analogy of the conditions in Eq.~\ref{eq:conditions on delta derivative} for simplifying expressions involving a function times the ``derivative'' of a delta function, where in the discrete $k$ case we mean functions times $\tilde\varpi$. This means that the coefficients of $\h W$ in Eq.~\ref{eq:W coeffs, circulant} and those of $\h L$ in Eq.~\ref{eq:BSLD, circulant} are not necessarily equal and quantum whitening is not optimal for the BMSE. This result does not indicate how far the covariant measurement may be from optimal, however, such that it may be close to optimal in practice.

\subsection{Covariant figure-of-merit}
\label{sec:Covariant figure-of-merit}
The issue above is not with the quantum whitening measurement, which is covariant, but with the cost function, which is not. The periodic symmetry of a circulant family of states $\{\ket{\psi_\theta}\}_{\theta\in(-\pi,\pi]}$ means that we should switch from using the MSE in Eq.~\ref{eq:MSE} to some periodic cost function.

In the periodic case, the additive unitary is generated by the Hamiltonian $\h H=\sum_{k=-\infty}^\infty k\proj{\varphi_k}$. The relevant symmetry group is $\{\h  U({\theta})\} _{\theta\in(-\pi,\pi]}$ where $\h U({\theta})=e^{-i \h H\theta}$. This unitary is additive such that $\h U({\theta_{1}})\h U({\theta_{2}})=\h U(\theta_{1}+\theta_{2}\bmod(-\pi,\pi])$. Previously, we considered the BMSE in Eq.~\ref{eq:BMSE} which is
\begin{align}
    \text{BMSE} &= \int\text{d}\theta \int\text{d}x\; \pi(\theta) L(x|\theta) (\check{\theta}_x - \theta)^2.
\end{align}
This is one example of the more general figure-of-merit $\bar C$ for a given cost function $C(\theta,\check{\theta}_x)$
\begin{align}
    \bar C &= \int\text{d}\theta \int\text{d}x\; \pi(\theta) L(x|\theta) C(\theta,\check{\theta}_x).
\end{align}
The BMSE uses the quadratic cost function, $C(\theta,\check{\theta}_x)=(\check{\theta}_x - \theta)^2$, which is not covariant, but other cost functions can be. An estimation problem is covariant with respect to the symmetry group $\{\h  U({\theta})\} _{\theta\in(-\pi,\pi]}$ if the following two conditions hold: %
\begin{enumerate}
    \item The cost function, $C(\theta_{1},\theta_{2})$, is group invariant:
    \begin{align}
        C(\theta_{1},\theta_{2})=C(\theta_{1}+x\bmod(-\pi,\pi],\theta_{2}+x\bmod(-\pi,\pi]).
    \end{align}
    \item The prior distribution is group invariant, $\pi(\theta+x\bmod(-\pi,\pi])=\pi(\theta).$
\end{enumerate}
An estimation problem that satisfies these conditions is covariant. It is then well-established that the optimal measurement is also covariant with respect to this group action even if it is a non-projective POVM~\cite{demkowicz2015quantum,Holevo2011book,bartlett2007reference,chiribella2005optimal}. 

In our case, for the problem to be covariant, the prior must be uniform, $\pi(\theta)=1/(2\pi)$. Regarding the cost function, it was shown that any $C(\theta_{1},\theta_{2})=\sum_{k=-\infty}^\infty a_{k}e^{ik(\theta_{1}-\theta_{2})}$ with $a_{k}\leq0$ is a valid covariant cost function~\cite{chiribella2005optimal,Holevo2011book}. For example, the Holevo variance, $C(\theta_{1},\theta_{2})=4\sin^{2}(\frac{\theta_{1}-\theta_{2}}{2}),$ is commonly used since it retrieves the standard variance in the limit of $|\theta_1-\theta_2| \ll1$. Therefore, for any circulant family and periodic cost function, covariant measurements are optimal given a flat prior. For example, this means that quantum whitening is optimal for the phase estimation of the coherent states in Appendix~\ref{app:Phase estimation}.

\section{Proof of the fundamental quantum limit in the Toeplitz case}
\label{app:proof of whitening}
Here, we prove the results in Sec.~\ref{sec:waveform basis} for the fundamental quantum limit.

Let us start by showing that Eq.~\ref{eq:mixed state, Fourier basis} holds for the mixed state and Bayesian derivative in the Toeplitz case with general prior. Using Eqs.~\ref{eq:mixed state} and~\ref{eq:inverse QFT}, $\h\rho(\theta)=\proj{\psi_\theta}$, and the definition of $\tilde\pi$, we can show that
\begin{align}
    \hat{\bar\rho}
    &=\intginf{\theta}\pi(\theta)\h\rho(\theta)
    \\&=\intginf{\theta}\pi(\theta)\intS{k} e^{-i k\theta} \sqrt{g(k)}\ket{k}\intS{l} e^{i l\theta} \sqrt{g(l)}\bra{l} \nonumber
    \\&=\intS{k} \intS{l} \sqrt{g(k)g(l)}\ket{k}\bra{l}\intginf{\theta}\pi(\theta)e^{-i (k-l)\theta} \nonumber
    \\&=\intS{k} \intS{l} \tilde \pi(k - l) \sqrt{g(k)g(l)}\ket{k}\bra{l}. \nonumber
\end{align}
This shows Eq.~\ref{eq:mixed state, Fourier basis}. The proof for $\hat{\bar\rho}'$ is similar but uses the expression for $i\tilde\pi'$.

\subsection{BMSE of quantum whitening}
\label{app:proof of whitening, 2}
Here, we show that Eq.~\ref{eq:MBMSE, improper prior} holds for the MBMSE given the uninformative, uniform prior. We also show the associated results in Sec.~\ref{sec:BMSE of quantum whitening}. The covariant measurement is optimal such that we need to calculate the BMSE of quantum whitening. First, the likelihood in Eq.~\ref{eq:likelihood} implies that the evidence $p(\theta)=\braopket{\theta}{\hat{\bar\rho}}{\theta}$ is
\begin{align}
    &p(\theta)
    \\&=\intginf{\theta'}\pi(\theta')L(\theta|\theta') \nonumber
    \\&=\frac{1}{(2\pi)^2\delta(0)}\intginf{\theta'}  \intS{k}\intS{l} \sqrt{g(k)g(l)} e^{i(\theta - \theta')(k-l)} \nonumber
    \\&=\frac{1}{2\pi\delta(0)}  \intS{k}\intS{l} \sqrt{g(k)g(l)} \delta(k-l) \nonumber
    \\&=\frac{1}{2\pi\delta(0)}  \intS{k} g(k) \nonumber
    \\&=\frac{1}{2\pi\delta(0)}  \nonumber
\end{align}
which is the same as the prior. Bayes's rule in Eq.~\ref{eq:Bayes rule} then implies that the posterior equals the likelihood, i.e.\ $p(\theta'|\theta) = L(\theta|\theta')$. Second, the optimal estimator is given by the mean of the posterior $\check\theta_{\theta} = 2\pi\delta(0)\braopket{ \theta}{\hat{\bar\rho}'}{ \theta}$ which is
\begin{align}
    &\check\theta_{\theta} 
    \\&= \intginf{\theta'}\theta' p(\theta'|\theta)  \nonumber
    \\&= \intginf{\theta'}\theta' \frac{1}{2\pi}  \intS{k}\intS{l} \sqrt{g(k)g(l)} e^{i(\theta - \theta')(k-l)} \nonumber
    \\&= \frac{1}{2\pi}  \intS{k}\intS{l} \sqrt{g(k)g(l)} e^{i\theta(k-l)}  \intginf{\theta'} \theta' e^{-i\theta'(k-l)}  \nonumber
    \\&= \frac{1}{2\pi}  \intS{k}\intS{l} \sqrt{g(k)g(l)} e^{i\theta(k-l)}  [-i2\pi\delta'(l-k)] \nonumber
    \\&= i \intS{l}\intS{x} \sqrt{g(x+l)g(l)} e^{i\theta x} \delta'(x) \nonumber
    \\&= -i \intS{l} \partial_x\left[\sqrt{g(x+l)g(l)} e^{i\theta x}\right]_{x=0} \nonumber
    \\&= -i \intS{l} \left[i\theta g(l) + \frac{1}{2} g'(l)\right] \nonumber
    \\&= \theta \nonumber.
\end{align}
Here, we assume that $g$ is differentiable and goes to zero on the boundary of $S$ such that $\intS{l} g'(l)=0$.
Finally, we calculate the MSE as a function of the parameter $\theta'$ similarly as
\begin{align}
    &\text{MSE}(\theta') 
    \\&= \intginf{\theta} L(\theta|\theta') (\check\theta_{\theta} - \theta')^2 \nonumber
    \\&= \intginf{\theta} \frac{1}{2\pi} \intS{k} \intS{l} \sqrt{g(k)g(l)} e^{i (\theta - \theta') (k-l)} (\theta - \theta')^2 \nonumber
    \\&= \frac{1}{2\pi}  \intS{k} \intS{l} \sqrt{g(k)g(l)} \intginf{\theta} e^{i (\theta - \theta') (k-l)} (\theta - \theta')^2 \nonumber
    \\&= \frac{1}{2\pi}  \intS{k} \intS{l} \sqrt{g(k)g(l)} [-2\pi\delta''(k-l)] \nonumber
    \\&= -  \intS{l} \intS{x} \sqrt{g(x+l)g(l)} \delta''(x) \nonumber
    \\&= - \intS{l} \partial^2_x\left[\sqrt{g(x+l)g(l)}\right]_{x=0} \nonumber
    \\&= - \intS{l} \left(\frac{1}{2} g''(l)-\frac{g'(l)^2}{4g(l)}\right) \nonumber
    \\&=  \frac{1}{4} \intS{l} \frac{g'(l)^2}{g(l)}.\nonumber
\end{align} 
Here, we now also assume that $g$ is twice differentiable and its derivative goes to zero on the boundary of $S$ such that $\intS{l} g''(l)=0$. These assumptions break for some of the applications that we study which is why the MBMSE diverges there. Since the MSE is independent of the parameter, it is equal to the BMSE and MBMSE which proves Eq.~\ref{eq:MBMSE, improper prior}.

\subsection{Optimality of quantum whitening}
\label{app:proof of whitening, 3}
Here, we provide further details for the independent proof of the optimality of quantum whitening in Sec.~\ref{sec:separate proof}. The mixed state $\hat{\bar\rho}$ and its Bayesian derivative $\hat{\bar\rho}'$ in Eqs.~\ref{eq:rho, improper prior} and~\ref{eq:rho', improper prior} come directly from applying the result for a general prior in Eq.~\ref{eq:mixed state, Fourier basis} to the flat prior ($\tilde \pi(k-l) = \delta(k-l)/\delta(0)$). By the inner product in Eq.~\ref{eq:Fourier inner product}, the matrix coefficients of the mixed state in Eq.~\ref{eq:rho, improper prior} are $\braopket{k}{\hat{\bar\rho}}{l} = [1/\delta(0)] g(k) \delta(k-l)$. The BSLD $\h L$ in Eq.~\ref{eq:BSLD, improper prior, unsimplified} then comes directly from Eq.~\ref{eq:BSLD sol} since the mixed state is diagonal in the Fourier basis, taking care to avoid indices where the eigenvalues are zero.

The conditions in Eq.~\ref{eq:conditions on delta derivative} determine when two functions multiplied by the derivative of the Dirac delta function are equal in the weak sense. They can be shown by integrating by parts against a smooth test function $\xi(k,l)$ as follows
\begin{align}
    & \intginf{k} \intginf{l} \delta'(k - l) f_1(k, l)\xi(k,l) 
    \\&= \intginf{l}\intginf{x}  \delta'(x) f_1(x+l,l) \xi(x+l,l) \nonumber %
    \\&= -\intginf{l} \partial_x\left[f_1(x+l,l)\xi(x+l,l)  \right]_{x=0}\nonumber 
    \\&= -\intginf{l} [\partial_x f_1(x+l,l)|_{x=0}  \xi(l,l) \nonumber\\&\hspace{2cm}+  f_1(l,l) \partial_x \xi(x+l,l)|_{x=0} ]\nonumber 
\end{align}
such that this equals $\intginf{k} \intginf{l} \delta'(k - l) f_2(k, l)\xi(k,l) $ for arbitrary $\xi(k,l)$ if the first two conditions in Eq.~\ref{eq:conditions on delta derivative} hold. The third condition on the partial derivative with respect to the second slot can be shown similarly by substituting $l=k-x$ instead of $k=x+l$. Direct computation of the partial derivatives of the coefficients of the BSLD in Eq.~\ref{eq:BSLD, improper prior, unsimplified} then shows that the BSLD is given by Eq.~\ref{eq:BSLD, improper prior}.

We want to show that the quantum whitening operator $\h W$ equals the BSLD and satisfies the Lyapunov equation. We can always form the covariant operator, $\h W = \intginf{\theta}\theta\proj{ \theta}$, regardless of $S$ and whether the covariant states are orthogonal as discussed in Sec.~\ref{sec:covariant}. First, let us show that the coefficients of $\h W$ are given by Eq.~\ref{eq:W coeffs, improper prior}. The inner product between the covariant and Fourier states is $\braket{k}{\theta} = \frac{1}{\sqrt{2\pi}}e^{-i\theta k}\chi_S(k)$ by Eq.~\ref{eq:quantum whitening}. The coefficients of $\h W$ are thus
\begin{align}
    \braopket{k}{\h W}{l}
    &= \bra{k}\intginf{\theta}\theta\ket{ \theta}\braket{\theta}{l}
    \\&= \intginf{\theta}\theta \frac{1}{2\pi}e^{-i\theta (k-l)} \chi_S(k)\chi_S(l)\nonumber
    \\&= \frac{1}{2\pi} \chi_S(k)\chi_S(l)\intginf{\theta}\theta e^{-i\theta (k-l)}\nonumber
    \\&= i\delta'(k-l)\chi_S(k)\chi_S(l).\nonumber
\end{align}
This shows Eq.~\ref{eq:W coeffs, improper prior}. Second, let us show that $\h W$ satisfies the Lyapunov equation by proving that Eq.~\ref{eq:anticommutator, improper prior} holds for the anticommutator $\frac{1}{2}\{\hat{\bar\rho}, \hat W\}$. We start by computing the coefficients for the product $\hat{\bar\rho}\hat W$ using Eqs.~\ref{eq:Fourier inner product} and~\ref{eq:rho, improper prior} as
\begin{align}
    &\braopket{k}{\hat{\bar\rho}\hat W}{l} 
    \\&= \bra{k}[1/\delta(0)] \intS{j} g(j) \proj{j} \intginf{\theta}\theta\ket{ \theta}\braket{\theta}{l}\nonumber
    \\&= [1/\delta(0)] g(k) \chi_S(k) \bra{k} \intginf{\theta}\theta\ket{ \theta}\braket{\theta}{l}\nonumber
    \\&= [1/\delta(0)] g(k) \chi_S(k) \braopket{k}{\h W}{l}\nonumber
    \\&= i[\delta'(k-l)/\delta(0)] g(k) \chi_S(k) \chi_S(l)\nonumber.
\end{align}
The conjugate product $\hat W\hat{\bar\rho}$ is thus
\begin{align}
    \braopket{k}{\hat W\hat{\bar\rho}}{l} &= i[\delta'(k-l)/\delta(0)] g(l) \chi_S(k) \chi_S(l)
\end{align}
such that the coefficients of the anticommutator $\frac{1}{2}\{\hat{\bar\rho}, \hat W\}$ are
\begin{align}
    \braopket{k}{\frac{1}{2}\{\hat{\bar\rho}, \hat W\}}{l} &= \frac{1}{2}i[\delta'(k-l)/\delta(0)][g(k) + g(l)] \chi_S(k) \chi_S(l)
\end{align}
which proves Eq.~\ref{eq:anticommutator, improper prior} since the characteristic functions $\chi_S(k) \chi_S(l)$ are redundant. The Lyapunov equation in Eq.~\ref{eq:Lyapunov equation} is thus satisfied by $\h W$ since the anticommutator $\frac{1}{2}\{\hat{\bar\rho}, \hat W\}$ equals the Bayesian derivative $\hat{\bar\rho}'$ in Eq.~\ref{eq:rho', improper prior} by Eq.~\ref{eq:conditions on delta derivative}.

\section{Quantum whitening of single-particle states}
\label{app:Quantum whitening of single-particle states}
In the single-particle state case, quantum whitening is the same as classically whitening the corresponding classical waveforms. We show this now. Suppose that the common classical and quantum geometry in Eq.~\ref{eq:single-particle inner product} is Toeplitz such that there exists some function $G:\R\to\C$ such that 
\begin{align}
    &  \braket{x_\theta}{x_{\theta'}} =  \intginf{t} x_{\theta}^*(t) x_{\theta'}(t) = G(\theta-\theta')
\end{align} %
where $G$ is positive semi-definite and $G(0)=1$. For the single-particle state case, the Fourier states in Eq.~\ref{eq:quantum Fourier transform} are $\ket{k}=\h c_k^\dagger\ket{0}$ where the Fourier annihilation operator is defined as $\h c_k = \intginf{t} y_k^*(t) \h a(t)$ and the Fourier waveform is 
\begin{align}\label{eq:Fourier waveform}
    & y_k(t) =\frac{1}{2\pi \sqrt{g(k)}}\intginf{\theta} e^{i k\theta} x_\theta(t)
\end{align}
such that the inverse transformation is $x_{\theta}(t) =\intginf{k} e^{-i k\theta} \sqrt{g(k)}y_k(t)$. Since the Fourier states are orthogonal, we have that 
\begin{align}
    \braket{k}{l} = [\h c_k, \h c_l^\dagger] = \intginf{t} y_k^*(t) y_l(t) = \delta(k-l).
\end{align}
The Fourier states are not normalised single-particle states, however, since their norm is $\intginf{t} \abs{y_k(t)}^2=\delta(0)$ whereas we require that $\intginf{t} \abs{x_\theta(t)}^2 = 1$. Informally, we could consider the normalised single-particle state corresponding to $y_k(t)/\sqrt{\delta(0)}$ but this division needs to be treated with the upmost mathematical care. The fact that we cannot have a continuous set of quantum state or classical waveforms that are orthogonal and have unit norm is a property of the continuum and thus is expected here.

Similarly, for the single-particle state case, the quantum whitened states in Eq.~\ref{eq:quantum whitening} are $\ket{\theta}=\hat{d}_{\theta}^\dagger\ket{0}$ where the whitened annihilation operator is defined as $\hat{d}_{\theta}=\intginf{t} z_{\theta}^*(t)\h a(t)$ and the classically whitened waveform is
\begin{align}
    &   z_{\theta}(t) = \frac{1}{\sqrt{2\pi}}\intginf{k}e^{-i\theta k} y_k(t)
\end{align}
such that the inverse transformation is $y_k(t) = \frac{1}{\sqrt{2\pi}}\intginf{\theta}e^{i\theta k} z_{\theta}(t)$. The whitened state are also orthogonal such that
\begin{align}
    & \braket{ \theta}{ \theta'} = [\hat{d}_{\theta}, \hat{d}_{\theta'}^\dagger] = \intginf{t} z_{\theta}^*(t) z_{\theta'}(t) = \delta(\theta - \theta').
\end{align}
The classical inner product between the original and whitened waveforms is the same as the quantum inner product in Eq.~\ref{eq:original to whitened} between the original and whitened states such that
\begin{align}
    \intginf{t} z_{\theta}^*(t) x_{\theta'}(t) = \frac{1}{\sqrt{2\pi}}\intginf{k}\sqrt{g(k)}e^{i(\theta - \theta')k}.
\end{align}
The quantum whitened states are also not single-photon states in the sense of Eq.~\ref{eq:temporal mode operator} since they are not normalised since $\intginf{t} \absSmall{z_{\theta}(t)}^2=\delta(0)$. 
We have shown that classical and quantum whitening are the same for the single-particle state case.

From the original classical waveforms $\{x_{\theta}:\R\to\C\}_{\theta\in\mathbb{R}}$, we have here defined two continuous families of orthogonal waveforms: the Fourier waveforms $\{y_k:\R\to\C\}_{k\in\mathbb{R}}$ and classically whitened waveforms $\{z_{\theta}:\R\to\C\}_{\theta\in\mathbb{R}}$. Just as we defined two bases for the subspace of possible quantum states, these are bases for the $L^2(\R)$ subspace of possible classical waveforms. 

\section{Bayesian estimation of displacement}
\label{app:Bayesian quantum estimation of displacement}

Let us consider the Bayesian problem of estimating the displacement $\mu$ of a single harmonic oscillator along a known direction. Without loss of generality, the final state is a coherent state with coherent amplitude $\alpha=i\mu/\sqrt{2}$ such that $\ev{\h x}=0$ and $\ev{\h p}=\mu$ where $[\h x, \h p]=i$. We now calculate the MBMSE, the BMSEs for quadrature and number-resolving measurements, the difference between amplitude and power esitmation, and the wide-prior limit.

\subsection{MBMSE}
In the case of a Gaussian prior, $\pi(\mu)=e^{-\mu^2/(2\sigma^2)}/(\sigma\sqrt{2\pi})$ with variance $\sigma^2$, then the MBMSE in Eq.~\ref{eq:MBMSE} becomes~\cite{macieszczak2014bayesian} 
\begin{align}
    \label{eq:MBMSE Gaussian prior}
    \Delta^2\mu = \sigma^2-\sigma^4\IQ^{\hat{\varrho}}(\mu)
\end{align}
where $\IQ^{\h\varrho}(\mu)$ is the QFI with respect to $\mu$ of the displaced mixed state $\h\varrho=e^{i\mu\h x}\hat{\bar\rho}e^{-i\mu\h x}$. This should be distinguished from the QFI of $\h\rho(\mu)$ for a given $\mu$ which is $\IQ^{\h\rho(\mu)}(\mu)=2$. The mixed state $\hat{\bar\rho}$ in Eq.~\ref{eq:mixed state} is a Gaussian state with zero mean and covariance matrix $\Sigma=\diag{\frac{1}{2}, \frac{1}{2} +\sigma^2}$ such that the displaced mixed state $\h\varrho$ is a Gaussian state with mean vector $\vec\mu=(0,\mu)^\T$ and the same covariance matrix $\Sigma$. Thus, the QFI in Eq.~\ref{eq:QFI Gaussian, mean only} is $\IQ^{\h\varrho}(\mu)=\frac{2}{1+2\sigma^2}$, which is independent of $\mu$ as expected, and the MBMSE in Eq.~\ref{eq:MBMSE Gaussian prior} is 
\begin{align}
    \label{eq:MBMSE displacement}
    \Delta^2\mu = \frac{\sigma^2}{1+2\sigma^2}.
\end{align}
In the wide-prior limit of $\sigma^2\gg\frac{1}{2}$, the MBMSE is $\Delta^2\mu \approx \frac{1}{2}$ whereas the MBMSE is $\Delta^2\mu \approx \sigma^2$ in the narrow-prior limit of $\sigma^2\ll\frac{1}{2}$.

\subsection{Quadrature measurement}
Suppose that we perform a quadrature measurement $\h p$ on $\h\rho(\mu)$ and obtain a measurement result $\mathfrak{p}$ with a likelihood of $L(\mathfrak{p}|\mu)\sim\mathcal{N}(\mu,\frac{1}{\sqrt2})$. Then, the evidence is also Gaussian $p(\mathfrak{p})\sim\mathcal{N}(0,\varsigma)$ with zero mean and variance $\varsigma^2=\frac{1}{2}+\sigma^2$, and the posterior is Gaussian $p(\mu|\mathfrak{p})\sim\mathcal{N}(\tilde\mu_\mathfrak{p},\sqrt{ V_\text{post}})$ with mean $\tilde\mu_\mathfrak{p}=\frac{2\sigma^2}{1+2\sigma^2}\mathfrak{p}$ and variance $V_\text{post}=\frac{\sigma^2}{1+2 \sigma ^2}$ from Eq.~\ref{eq:posterior variance} which is independent of $\mathfrak{p}$. The BMSE in Eq.~\ref{eq:BMSE} of quadrature measurement is then just the posterior variance $V_\text{post}$ which agrees with marginalising $\text{MSE}(\mu)=\frac{\mu ^2+2 \sigma ^4}{\left(2 \sigma ^2+1\right)^2}$ over the prior $\pi(\mu)\sim\mathcal{N}(0,\sigma)$. This value equals the MBMSE in Eq.~\ref{eq:MBMSE displacement} meaning that quadrature measurement is optimal. This measurement is biased because of the prior since, for a given $\mu$ and $\sigma$, we obtain $\mathfrak{p}$ with likelihood $L(\mathfrak{p}|\mu)$ and estimate $\check\mu_\mathfrak{p}$ whose expected value is $\check\mu_\mu=\frac{2\mu\sigma^2}{1+2\sigma^2}\neq\mu$. In the wide-prior limit of $\sigma^2\gg\frac{1}{2}$, the posterior distribution is simply the likelihood, $p(\mu|\mathfrak{p})=L(\mathfrak{p}|\mu)$, and the bias vanishes since $\check\mu_\mu=\mu$. 

\subsection{Number-resolving measurement}
In comparison, suppose that we perform a number-resolving measurement $\h n$ and obtain a measurement result $\mathfrak{n}$ with a Poissonian likelihood of $L(\mathfrak{n}|\mu)=e^{-\frac{\mu^2}{2}} \frac{\mu^{2\mathfrak{n}}}{2^\mathfrak{n} \mathfrak{n}!}$ such that the posterior distribution is 
\begin{align}
    p(\mu|\mathfrak{n}) &= \frac{ \mu ^{2 \mathfrak{n}} \left(\sigma ^2+1\right)^{\mathfrak{n}+\frac{1}{2}}}{2^{\mathfrak{n}+\frac{1}{2}} \Gamma \left(\mathfrak{n}+\frac{1}{2}\right) \sigma ^{2 \mathfrak{n}+1}} e^{-\frac{\mu ^2 \left(\sigma ^2+1\right)}{2 \sigma ^2}}
\end{align}
which has zero mean independent of the measurement result $\mathfrak{n}$. This means that the optimal estimator $\tilde\mu_\mathfrak{n}=0$ contains no additional information about the displacement $\mu$ and thus the BMSE equals the prior variance $\sigma^2$. This is because the likelihood is even in $\mu$, i.e.\ $L(\mathfrak{n}|\mu)=L(\mathfrak{n}|{-}\mu)$, such that we cannot distinguish between $\mu$ and $-\mu$ with a number-resolving measurement $\h n$.

\subsection{Power and magnitude estimation}
To handle this issue with the sign of the displacement $\mu$, suppose that we instead wish to estimate the power $u=\mu^2$ or magnitude $v=\abs{\mu}$ of the displacement. The Gaussian prior on $\mu$ implies that the power follows a scaled chi-squared distribution, $\pi(u) = \frac{1}{\si\sqrt{2\pi u}}e^{-u/(2\si^2)}$ with expected value $\evText{u}=\sigma^2$ and variance $\var{u}=2\sigma^4$, and the magnitude follows a half-Gaussian distribution, $\pi(v) = \frac{2}{\si\sqrt{2\pi}}e^{-v^2/(2\si^2)}$ with expected value $\evText{v}=\sigma\sqrt{2/\pi}$ and variance $\var{v}=\sigma^2(1-2/\pi)$. We estimate the deviation of each parameter from the expected value such that the prior is centred. These prior distributions are non-Gaussian such that Eq.~\ref{eq:MBMSE Gaussian prior} does not apply and we must instead solve Eq.~\ref{eq:MBMSE} which we have been unable to do analytically. However, a numerical analysis suggests that quadrature measurement $\h p$ does not attain the MBMSE for estimating the power $\mu^2$ or magnitude $\abs{\mu}$ despite attaining the MBMSE with respect to $\mu$. We also see that a number-resolving measurement $\h n$ attains the MBMSE for estimating the power $\mu^2$ but not the magnitude $\abs{\mu}$ where a different measurement from either $\h p$ or $\h n$ is required. This is different to the Fisher limit where a change of variables from $\theta$ to $\theta'$ only introduces a chain rule factor of $(\deriv{\theta}{\theta'})^2$ in the CFI and QFI and does not change the attainability of bounds. Furthermore, we observe that quadrature measurement does not attain the MBMSE with respect to $\mu$ in the case of a flat prior rather than a Gaussian one such that a different measurement from $\h p$ or $\h n$ is again required. We defer understanding these different measurements and the behaviour of the BMSE and MBMSE under changes of variables more generally to future work.

\subsection{Wide-prior limit}
\label{sec:Wide-prior limit}
Returning to the case of estimating $\mu$, we now demonstrate that the wide-prior limit of $\sigma^2\gg\frac{1}{2}$ mentioned briefly above needs to be handled carefully. This limit represents the uninformative, uniform prior on $\R$ that is nevertheless normalised, i.e.\ $\partial_\mu\pi(\mu)=0$ and $\intginf{\mu}\pi(\mu)=1$. For the quadrature measurement likelihood of $L(\mathfrak{p}|\mu)\sim\mathcal{N}(\mu,\frac{1}{\sqrt2})$, this is the Jeffreys prior for the mean $\mu$~\cite{jeffreys1946invariant}. In the direct calculation of the BMSE for quadrature measurement above, or for quantum whitening in Sec.~\ref{sec:Improper prior}, we find that the posterior is normalised and can have finite variance in some cases despite the prior. This value of the posterior variance and MBMSE can be calculated by just assuming that the prior is constant on $\R$ and normalised, e.g.\ $\pi(\mu)\equiv\varepsilon$ for some $\varepsilon>0$. Since the Fourier transform of the prior is $\tilde \pi(k)=e^{-\frac{1}{2}\sigma^2 k^2}$ for finite $\sigma$ which limits to $\tilde \pi(k)=\delta(k)/\delta(0)=\delta_{k0}$, then we could assign the prior the infinitesimal value of $\varepsilon=1/[2\pi\delta(0)]$ but we emphasise that the exact value does not matter, just that the prior is constant and normalised. 

Alternatively, we could use an improper, i.e.\ unnormalised, prior such as $\pi(\mu)\equiv1$. Then, we find that the posterior is still normalised and has constant variance equal to the MSE, independent of the measurement result. The issue is then that the BMSE in Eq.~\ref{eq:BMSE} is proportional to the prior's normalisation which diverges. Similarly, the MSE can be finite but the evidence is improper. This divergence is artificial, however, since it simply comes from how we have choose to represent a total lack of knowledge about the parameter $\mu$ as an uninformative prior. As such, we prefer the first method where the prior is chosen to be constant on $\R$ and normalised from the start.

We have shown above that the BMSE of quadrature measurement achieves the MBMSE of $\frac{1}{2}$ in the wide-prior limit and that the quadrature measurement is unbiased. If we na\"ively calculated the MBMSE from Eq.~\ref{eq:MBMSE} instead, however, then we could have ran into an issue if we did not Taylor expand to sufficiently high order. Here, we assume that we have already independently shown that the quadrature measurement is optimal and wish to verify the MBMSE. For finite $\sigma$, the BSLD in Eq.~\ref{eq:BSLD} is $\h L=\frac{2\sigma^2}{1+2\sigma^2}\h p$ with $\trSmall{\h{\bar\rho}\h L^2}=\frac{2\sigma^4}{1+2\sigma^2}$ such that the MBMSE in Eq.~\ref{eq:MBMSE} agrees with Eq.~\ref{eq:MBMSE displacement}. If we Taylor expand these expressions to first order in large $\sigma^2\gg\frac{1}{2}$, then we get that $\h L\approx[1-1/(2\sigma^2)]\h p$ such that $\trSmall{\h{\bar\rho}\h L^2}\approx\sigma^2-\frac{1}{2}-1/(2\sigma^2)\approx\sigma^2-\frac{1}{2}$ and the MBMSE in Eq.~\ref{eq:MBMSE} is $\Delta^2\mu\approx\frac{1}{2}$ which agrees with Eq.~\ref{eq:MBMSE displacement}. If we instead na\"ively only Taylor expand the expressions to zeroth order in large $\sigma$, then get that $\h L\approx\h p$ such that $\trSmall{\h{\bar\rho}\h L^2}\approx\sigma^2+\frac{1}{2}\approx\sigma^2$ and the MBMSE in Eq.~\ref{eq:MBMSE} is zero. The MBMSE similarly vanishes if we expand $\trSmall{\h{\bar\rho}\h L^2}=\frac{2\sigma^4}{1+2\sigma^2}\approx\sigma^2$ to zeroth order. These results are erroneous, however, since they come from mishandling the indeterminate form $\infty-\infty$, i.e.\ we have not expanded the expressions to sufficient order and expanded them too early to safely cancel the prior variance of $\Delta^2\mu^{(0)}=\si^2$ in Eq.~\ref{eq:MBMSE}. For this toy problem, therefore, we have shown that either we need to calculate the BMSE directly from Eq.~\ref{eq:BMSE} or that we need to take a sufficiently high-order expansion in the prior variance to calculate the MBMSE from Eq.~\ref{eq:MBMSE}.

These results are relevant to our general calculation in Sec.~\ref{sec:Improper prior} where we cannot calculate the MBMSE for finite prior variance and then take the wide-prior limit like we did above for the toy problem. Instead, we work exclusively in the wide-prior limit and directly calculate the BMSE for the optimal measurement in Eq.~\ref{eq:BMSE} rather than na\"ively apply Eq.~\ref{eq:MBMSE}. We have motivated this here since the former is always well-defined but the latter has the indeterminate form $\infty-\infty$ unless the limit is properly taken.

\section{Inner product of coherent states}
\label{app:inner product}
We have two independent ways of showing Eq.~\ref{eq:inner product of coherent states} for the inner product between multi-mode coherent states. The first way is to extract the temporal mode which is excited for each coherent state. The second way is to discretise the time domain like we might do numerically. We provide both proofs here for completeness.

\subsection{Extracting the temporal mode}
Let us start with the direct calculation without discretisation. We use the annihilation operator $\h b_\theta$ from Eq.~\ref{eq:temporal mode operator} to express the coherent state in Eq.~\ref{eq:generalised state} as follows
\begin{align}\label{eq:coherent state, operator definition}
    \ket{h_\theta} &= e^{\alpha_\theta \h b_\theta^\dagger - \alpha_\theta^* \h b_\theta} \ket{0}
    \\&= e^{-\frac{1}{2}\abs{\alpha_\theta}^2}e^{\alpha_\theta \h b_\theta^\dagger} e^{- \alpha_\theta^* \h b_\theta} \ket{0}\nonumber
    \\&= e^{-\frac{1}{2}\abs{\alpha_\theta}^2}e^{\alpha_\theta \h b_\theta^\dagger} \ket{0}.\nonumber
\end{align}
The average number of particles in this coherent state is
\begin{align}
    \bar N_\theta &= \braopket{h_\theta}{\h N}{h_\theta} 
    \\&= \intginf{t} \braopket{h_\theta}{\h a^\dagger(t)\h a(t)}{h_\theta}\nonumber
    \\&= e^{-\abs{\alpha_\theta}^2} \intginf{t} \bra{0}e^{\alpha_\theta^* \h b_\theta} \h a^\dagger(t)\h a(t) e^{\alpha_\theta \h b_\theta^\dagger} \ket{0}\nonumber
    \\&= e^{-\abs{\alpha_\theta}^2} \abs{\alpha_\theta}^2 \bra{0} e^{\alpha_\theta^* \h b_\theta} e^{\alpha_\theta \h b_\theta^\dagger} \ket{0}\nonumber
    \\&= \abs{\alpha_\theta}^2\nonumber
    \\&= \frac{1}{2}\intginf{t} \abs{h_\theta(t)}^2\nonumber
\end{align}
where we used that $e^{\h A} \h B e^{-\h A} = \h B + [\h A, \h B]$ for central $[\h A, \h B]$. This result is expected since $\bar N_\theta=\abs{\alpha_\theta}^2$ is the average number of particles for a single-mode coherent state with amplitude $\alpha_\theta$. Similarly, we can show that $\hat a(t)\ket{h_\theta} = \frac{i}{\sqrt2}h_\theta(t)\ket{h_\theta}$ and that Eq.~\ref{eq:quadrature EVs} holds for the expectation values of the quadratures. The inner product between two coherent states is thus
\begin{align}
    \braket{h_{\theta}}{h_{\theta'}} &= e^{-\frac{1}{2}\left(\abs{\alpha_\theta}^2 + \abs{\alpha_{\theta'}}^2\right)} \bra{0}e^{\alpha_\theta^* \h b_\theta} e^{\alpha_{\theta'} \h b_{\theta'}^\dagger} \ket{0}
    \\&= e^{-\frac{1}{2}\left(\abs{\alpha_\theta}^2 + \abs{\alpha_{\theta'}}^2\right) + \alpha_\theta^* \alpha_{\theta'} [\h b_\theta, \h b_{\theta'}^\dagger]} \bra{0} e^{\alpha_{\theta'} \h b_{\theta'}^\dagger} e^{\alpha_\theta^* \h b_\theta} \ket{0}\nonumber
    \\&= e^{-\frac{1}{2}\left(\abs{\alpha_\theta}^2 + \abs{\alpha_{\theta'}}^2\right) + \alpha_\theta^* \alpha_{\theta'} \intginf{t} f_{\theta}^*(t) f_{\theta'}(t)}\nonumber
    \\&= e^{-\frac{1}{2}\left[\bar N_\theta + \bar N_{\theta'} - \intginf{t} h_{\theta}^*(t) h_{\theta'}(t) \right]}\nonumber
\end{align}
where we used that $e^{\h A} e^{\h B} = e^{[\h A,\h B]} e^{\h B} e^{\h A}$ for central $[\h A,\h B]$. This completes the proof of Eq.~\ref{eq:inner product of coherent states}.

We evaluated the diagonal matrix coefficients of the number operator above in the overcomplete coherent state basis for the subspace. Meanwhile, the off-diagonal matrix coefficients of the number operator are
\begin{align}
\braopket{h_\theta}{\h N}{h_{\theta'}} = \frac{1}{2}\braket{h_\theta}{h_{\theta'}}\intginf{t} h_\theta^*(t)h_{\theta'}(t)
\end{align}
which is half of the product of the quantum and classical inner products.

\subsection{Discretising the time domain}
\label{app:Discretising the time domain}
Let us now instead discretise the bath of outgoing modes by coarse-graining them according to a short sampling time $\delta t$ such that the time domain $\R$ is broken into time bins $(j\delta t, (j+t)\delta t)$ indexed by $j\in\Z$. Let the annihilation operator of the $j$th time bin be
\begin{align}
    \h c_j &= \frac{1}{\sqrt{\delta t}} \intg{j\delta t}{(j+1)\delta t}{t} \h a(t) %
\end{align}
where $[\h c_j, \h a^\dag(t)] = \frac{1}{\sqrt{\delta t}} \chi_{(j\delta t, (j+1)\delta t)}(t)$ and $[\h c_j, \h c_k^\dag]=\delta_{jk}$ for $j,k\in\Z$. (These should be distinguished from the annihilation operator of the Fourier modes $ \h c_k$ defined above Eq.~\ref{eq:Fourier waveform}.) We now coarse-grain the outgoing state $|h_\theta\rangle$ in Eq.~\ref{eq:generalised state} as follows
\begin{align}
    \ket{h_\theta}  &= \bigotimes_{j=-\infty}^{\infty} e^{\frac{i}{\sqrt2}\intg{j\delta t}{(j+1)\delta t}{t} \left[ h_\theta(t) \h a^\dagger(t) + h_\theta^*(t) \h a(t)\right]} |0\rangle 
    \\&\approx \bigotimes_{j=-\infty}^{\infty} e^{\alpha_j \h c_j^\dagger - \alpha_j^* \h c_j} |0\rangle \nonumber
    \\&= \bigotimes_{j=-\infty}^{\infty} e^{-\frac{1}{2}\abs{\alpha_j}^2}e^{\alpha_j \h c_j^\dagger} \ket{0}.\nonumber
\end{align}
where $\alpha_j:=\frac{i}{\sqrt2}\sqrt{\delta t}\,h_\theta(j\delta t)$ is the coherent amplitude of the $j$th time bin and is dimensionless since $h_\theta(t)$ has units of square root Hertz. Here, we have discretised the signal $h_\theta(t)$ with its values at the left endpoints $h_\theta(j\delta t)$ which is a good approximation assuming that the sampling rate is fast compared to the signal. 

The coherent state is an eigenstate of the annihilation operator which can shown as follows
\begin{align}
    \h a(t) \ket{h_\theta} 
    &\approx \h a(t) \bigotimes_{j=-\infty}^{\infty} e^{-\frac{1}{2}\abs{\alpha_j}^2}e^{\alpha_j \h c_j^\dagger} \ket{0}
    \\&= \left[\frac{1}{\sqrt{\delta t}} \sum_{j=-\infty}^{\infty} \alpha_j \chi_{(j\delta t, (j+1)\delta t)}(t)\right]  \ket{h_\theta}\nonumber
\end{align}
where we again used that $e^{\h A} \h B e^{-\h A} = \h B + [\h A, \h B]$ for central $[\h A, \h B]$. The average number of particles is thus
\begin{align}
    \bar N_\theta &= \braopket{h_\theta}{\h N}{h_\theta} 
    \\&= \intginf{t} \braopket{h_\theta}{\h a^\dagger(t)\h a(t)}{h_\theta}\nonumber
    \\&\approx \intginf{t} \frac{1}{\delta t} \sum_{j=-\infty}^{\infty} \abs{\alpha_j}^2 \chi_{(j\delta t, (j+1)\delta t)}(t)\nonumber
    \\&= \sum_{j=-\infty}^{\infty} \abs{\alpha_j}^2\nonumber
    \\&\xrightarrow[\delta t\to0]{} \frac{1}{2}\intginf{t} \abs{h_\theta(t)}^2.\nonumber
\end{align}
The inner product between coherent states is thus
\begin{align}
    \braket{h_{\theta}}{h_{\theta'}}
    &\approx \prod_{j=-\infty}^{\infty} \braket{\alpha_j(\theta)}{\alpha_j(\theta')}
    \\&= e^{-\frac{1}{2}\sum_{j=-\infty}^{\infty} [|\alpha_j(\theta)|^2+|\alpha_j(\theta')|^2-2\alpha_j(\theta)^*\alpha_j(\theta')]}\nonumber
    \\&\xrightarrow[\delta t\to0]{} e^{-\frac{1}{2}\left[\bar N_\theta + \bar N_{\theta'} - \intginf{t} h_{\theta}^*(t) h_{\theta'}(t) \right]}\nonumber
\end{align}
such that we obtain Eq.~\ref{eq:inner product of coherent states} in the continuum limit.

The mean values of the quadratures of the $j$th time bin are $\evSmall{\h x_j}=\sqrt{2}\,\re{\alpha_j}$ and $\evSmall{\h p_j}=\sqrt{2}\,\im{\alpha_j}$ where $\h x_j = (\h c_j + \h c_j^\dagger)/\sqrt{2}$ and $\h p_j = (-i\h c_j + i\h c_j^\dagger)/\sqrt{2}$ such that $[\h x_j, \h p_k]=i\delta_{jk}$. This shows Eq.~\ref{eq:quadrature EVs} in the continuum limit. For example, for the sinusoidal signal in Eq.~\ref{eq:signal windowed (0, T)}, $\evSmall{\h p_j}=A\sqrt{\delta t}\,\cos(\omega j\delta t+\phi)$ oscillates in time as shown in Fig.~\ref{fig:waveform estimation diagram} such that the individual SNR of the $j$th oscillator is $A\sqrt{2\delta t}$ since $\varSmall{\h p_j}=\frac{1}{2}$ and the integrated SNR from the $M=T/\delta t$ oscillators within $(0, T)$ is $A\sqrt{2T}$.

Let us briefly comment on discretising the time domain in the case of frequency estimation. Similarly to how the long integration time $T$ sets the separation between the harmonic frequencies $\omega_{n+1}-\omega_n=2\pi/T$, the sampling frequency $f_s=1/\delta t$ sets the Nyquist frequency $f_N=f_s/2=1/(2\delta t)$ as the highest possible frequency to estimate without aliasing. This is also the meaning of assuming above that the sampling rate is fast compared to the signal. To calculate the inner product $\braket{h_{\omega}}{h_{\omega'}}$ between two different frequencies $\omega$ and $\omega'$ we require that $\max(\omega,\omega') \ll 1/(2\delta t)$ to avoid aliasing. For example, if we want to estimate a frequency $\omega$ within the interval $(\omega_0-\Delta\omega/2,\omega_0+\Delta\omega/2)$, then we require that $\omega_0+\Delta\omega/2 < 2\pi f_N=\pi/\delta t$. Since we also want to avoid negative frequencies, then we also require that $\omega_0-\Delta\omega/2>0$ such that we have that $\Delta\omega \delta t < \pi$. This condition is a constraint on the maximum phase difference accumulated over a single time bin between two different frequencies. Provided that this condition is satisfied, then we are safe to discretise the outgoing bath in time. For example, in Fig.~\ref{fig:MBMSE vs SNR}, we have that $\Delta\omega \delta t=0.18\pi$.

\section{Shifted window for sinusoid frequency estimation}
\label{app:(-T/2, T/2)}
The fact that we windowed the sinusoidal signal in Eq.~\ref{eq:signal windowed (0, T)} to the particular time interval $(0, T)$ with a square pulse is only a choice of convention. We could instead use a Gaussian window of width $T$ and the results should be similar as we show later for the complex exponential case. We can also shift the time interval, for example, to $(-T/2, T/2)$ such that the signal becomes
\begin{align}\label{eq:signal windowed (-T/2, T/2)}
    h_\omega(t) = A\cos(\omega t + \phi) \chi_{(-T/2, T/2)}(t),
\end{align}
then the inner product between states becomes
\begin{align}
    \label{eq:cosine inner product of states, (-T/2, T/2)}
    \braket{h_{\omega}}{h_{\omega'}} \approx e^{-\frac{1}{4}A^2 T(1 - \text{sinc}[(\omega-\omega')T/2])}
\end{align}
where we again assume the many-cycles limit. Whereas the sinc function in the exponent of Eq.~\ref{eq:cosine inner product of states, (0, T)} vanished for harmonic frequency spacing of $\omega-\omega'=n\pi/T$ for $(0, T)$, here the frequency spacing is $\omega-\omega'=2n\pi/T$ for $(-T/2, T/2)$.

Although the Gram matrices $G_{\omega,\omega'}=\braket{h_{\omega}}{h_{\omega'}}$ in Eq.~\ref{eq:cosine inner product of states, (0, T)} and Eq.~\ref{eq:cosine inner product of states, (-T/2, T/2)} for $(0, T)$ and $(-T/2, T/2)$, respectively, are not identical, they have a similar functional form. In particular, they are generally functions of $\omega$ and $\omega'$ but become Toeplitz in the many-cycles limit.

The QFI for frequency estimation of the signal defined on $(-T/2, T/2)$ in Eq.~\ref{eq:signal windowed (-T/2, T/2)} is
\begin{align}
\label{eq:QFI, sinusoid, (-T/2, T/2)}
    \IQ(\om)\approx\frac{1}{12} A^2T^3
\end{align}
where we take the many-cycles limit of $\omega T\gg1$. This simply differs by a factor of four from the QFI defined on $(0, T)$ in Eq.~\ref{eq:QFI, sinusoid, (0, T)}.

\section{Frequency estimation of a complex exponential}
\label{app:complex exponential}
Let us now consider the following example of a complex waveform which corresponds to a rotating coherent state
\begin{align}
    \label{eq:complex exponential signal, (0, T)}
    h_\omega(t) &= -iAe^{i(\omega t + \phi)} \chi_{(0, T)}(t)
\end{align}
such that the average number in Eq.~\ref{eq:average total number} is $\bar N_\omega = A^2 T/2$ and the quadrature expectation values in Eq.~\ref{eq:quadrature EVs} are
\begin{align}
    \braopket{h_\theta}{\h x(t)}{h_\theta}&=A\cos(\omega t + \phi)\chi_{(0, T)}(t),
    \\\braopket{h_\theta}{\h p(t)}{h_\theta}&=A\sin(\omega t + \phi)\chi_{(0, T)}(t).\nonumber
\end{align}
The $L^2(\R)$ inner product between classical waveforms is $\intginf{t}h_{\omega}^*(t) h_{\omega'}(t)=A^2\delta_T(\omega'-\om)$ where $\delta_T(x)=(e^{ixT}-1)/(ix)$ is a finite time approximation to the Dirac delta function obeying $\delta_T(0)=T$ and $\delta_T(x)=\intg{0}{T}{t}e^{ixt}$. The inner product between states in Eq.~\ref{eq:inner product of coherent states} is thus
\begin{align}
    \label{eq:complex exponential inner product of states, (0, T)} 
    \braket{h_{\omega}}{h_{\omega'}} = e^{-\frac{1}{2}A^2 \left[T - \delta_T(\omega'-\om)\right]}.
\end{align}
This Gram matrix is Hermitian but not real. In general, the Gram matrix is always Hermitian and positive semi-definite such that its eigenvalues are real and non-negative. If we shift the time interval to $(-T/2, T/2)$ instead of $(0, T)$, then the signal becomes
\begin{align}
    \label{eq:complex exponential signal, (-T/2, T/2)}
    h_\omega(t) &= -iAe^{i(\omega t + \phi)} \chi_{(-T/2, T/2)}(t)
\end{align}
such that the inner product is instead
\begin{align}
    \label{eq:complex exponential inner product of states, (-T/2, T/2)} 
    \braket{h_{\omega}}{h_{\omega'}}=e^{-\frac{1}{2}A^2T(1 - \text{sinc}\left[(\omega-\omega')T/2\right])}.
\end{align}
This is the same as the sinusoidal case in Eq.~\ref{eq:cosine inner product of states, (-T/2, T/2)} except for an overall factor of two in the exponent. This is likely because the complex exponential is composed of sinusoidal real and imaginary parts. Unlike the sinusoidal case, however, the Gram matrix is Toeplitz here even outside of the many-cycles limit.

The QFI in Eq.~\ref{eq:QFI, general coherent} for frequency estimation of the complex exponential signal defined on $(0, T)$ in Eq.~\ref{eq:complex exponential inner product of states, (0, T)} is
\begin{align}
\label{eq:QFI, complex exponential, (0, T)}
    \IQ(\om)=\frac{2}{3} A^2T^3
\end{align}
and for the signal defined on $(-T/2, T/2)$ in Eq.~\ref{eq:complex exponential signal, (-T/2, T/2)} is
\begin{align}
\label{eq:QFI, complex exponential, (-T/2, T/2)}
    \IQ(\om)=\frac{1}{6} A^2T^3.
\end{align}
These are twice the sinusoidal QFI in Eq.~\ref{eq:QFI, sinusoid, (0, T)} and Eq.~\ref{eq:QFI, sinusoid, (-T/2, T/2)}, respectively, because of the sinusoidal real and imaginary parts here.

\subsection{Alternate window functions}
\label{sec:alternate windows}
We assumed a rectangular window function above but the results are similar with other window functions. Let us show this now for Gaussian and double-exponential window functions. Although we do not do this for the later amplitude and phase estimation cases, the results should again be similar to the results there with the rectangular window.

\subsubsection{Frequency estimation of a sine-Gaussian}
Consider the following complex sine-Gaussian signal
\begin{align}
    h_\omega(t) &= Ae^{-(t-t_0)^2/(2\sigma^2)+i\omega(t-t_0)+i\phi}
\end{align}
where $A$ is the amplitude, $\sigma$ is the standard deviation of the Gaussian in units of time, $t_0$ is the centre time, and $\phi$ is the phase. We assume that these four parameters are known and we just want to estimate the frequency $\omega$. Suppose that this signal drives a coherent state such that the average number in Eq.~\ref{eq:average total number} is $\bar N_\omega = \sqrt{\pi}A^2\sigma/2$. The inner product between coherent states in Eq.~\ref{eq:inner product of states, real L2 inner product} is then
\begin{align}\label{eq:sine-Gaussian IP}
    \braket{h_{\omega}}{h_{\omega'}} = e^{-\frac{1}{2}\sqrt{\pi } A^2 \sigma  \left(1-e^{-\frac{1}{4} \sigma ^2 (\omega-\omega')^2}\right)}
\end{align}
which is asymptotically $e^{-\bar N_\omega}$ for large frequency differences. The QFI in Eq.~\ref{eq:QFI, general coherent} is thus
\begin{align}
    \IQ(\omega) = \sqrt{\pi}A^2\sigma^3 = 2\bar N_\omega \sigma^2.
\end{align}
This has the same scaling as Eqs.~\ref{eq:QFI, complex exponential, (0, T)}--\ref{eq:QFI, complex exponential, (-T/2, T/2)} given $\sigma\sim T$.

\subsubsection{Frequency estimation of a Lorentzian}
Let us now consider a double-exponential window function:
\begin{align}
    h_\omega(t) &= Ae^{-\gamma\abs{t-t_0}+i\omega(t-t_0)+i\phi}
\end{align}
where $\gamma$ is the decay rate which is known. The Fourier transform of this signal is a Lorentzian since
\begin{align}
    H_\omega(\Omega) = \intginf{t} e^{-i\Omega t} h_\omega(t) = \frac{2A\gamma e^{-i\Omega t_0+i\phi}}{\gamma^2+(\Omega-\omega)^2}.
\end{align}
This is similar to how fact that the Laplace and Cauchy distributions are related by the Fourier transform. 

Similarly, the average number in Eq.~\ref{eq:average total number} of the coherent state is $\bar N_\omega = A^2/(2\gamma)$. And, the inner product in Eq.~\ref{eq:inner product of states, real L2 inner product} is
\begin{align}
    \braket{h_{\omega}}{h_{\omega'}} = \exp\left(-\frac{A^2 (\omega-\omega')^2}{2\gamma  \left[4 \gamma ^2+(\omega-\omega')^2\right]}\right)
\end{align}
The QFI in Eq.~\ref{eq:QFI, general coherent} is also
\begin{align}
    \IQ(\omega) = \frac{A^2}{\gamma^3} = \frac{2\bar N_\omega}{\gamma^2}
\end{align}
which has the same scaling again given $\gamma\sim1/T$.

In comparison, the inner product between single-particle states in Eq.~\ref{eq:single-particle inner product} for $x_\omega(t)=\sqrt{\gamma}h_\omega(t)/A$ is
\begin{align}
    \braket{x_\omega}{x_{\omega'}} = \frac{4\gamma^2}{4\gamma^2 + (\omega-\omega')^2}
\end{align}
such that the QFI in Eq.~\ref{eq:single-particle QFI} is $\IQ(\omega) = \frac{2}{\gamma^2}$. This is the same as the QFI above for the coherent state with an average of one particle ($\bar N_\omega=1$). Meanwhile, the likelihood is $L(\theta|\theta')=\frac{2\gamma^3}{\pi[\gamma^2+(\theta-\theta')^2]^2}$ such that the MBMSE in Eq.~\ref{eq:MBMSE, improper prior} for the single-particle case is $\mathcal{V}=\gamma^2$ from the MSE. Alternatively, from the dual estimation problem, the prior Fisher information of $g(k)=\gamma e^{-2\gamma\abs{k}}$ with respect to $k$ is $\IP=4\gamma^2$ and the MBMSE is the quarter of this value. Whereas, the MBMSE for the coherent state case diverges since the coherent state inner product is asymptotically nonzero.

\section{Amplitude estimation}
\label{app:Amplitude estimation}
We can also consider instead estimating the amplitude $A$ for fixed frequency $\omega$ and phase $\phi$. In general, suppose that the classical waveform has the following form for a real scale parameter $A$ to estimate and some square-integrable complex function $g(t)$ defined on $\R$
\begin{align}\label{eq:general signal for amplitude estimation}
    h_A(t)=Ag(t) %
\end{align}
such that the inner product in Eq.~\ref{eq:inner product of states, real L2 inner product} is thus
\begin{align}
    \label{eq:inner product, amplitude estimation}
    \braket{h_{A}}{h_{A'}} = e^{-\frac{1}{2}\bar N_1(A-A')^2}
\end{align}
which is Toeplitz and an unnormalised Gaussian with variance $1/\bar N_1$ where $\bar N_1 = \frac{1}{2} \intginf{t} \abs{g(t)}^2$. The average number in Eq.~\ref{eq:average total number} is $\bar N_A = A^2 \bar N_1$. The QFI in Eq.~\ref{eq:QFI, general coherent} is
\begin{align}\label{eq:QFI, amplitude}
    \IQ(A) = 4 \bar N_1.
\end{align}
In the temporal mode associated with $g(t)$, the state is a coherent state with amplitude $\alpha_A=A\sqrt{\bar N_1}e^{i\varphi}$ for some $\varphi$ as shown in Appendix~\ref{app:inner product}. Without loss of generality, let $\varphi=\pi/2$ such that $\alpha_A=i\mu/\sqrt{2}$ where $\mu=A\sqrt{2\bar N_1}$. The MBMSE in Eq.~\ref{eq:MBMSE displacement} with respect to $\mu$ implies that the MBMSE with respect to $A$ is
\begin{align}
    \label{eq:MBMSE Gaussian prior, amplitude estimation}
    \Delta^2A = \frac{\sigma^2}{1+4\bar N_1\sigma^2} %
\end{align}
given a Gaussian prior on $A$ with variance $\sigma^2$. The MBMSE is attained by measuring the phase quadrature $\h p$ of the temporal mode associated with $g(t)$. If $4\bar N_1\sigma^2\gg1$, then the MBMSE is $\Delta^2A = 1/(4\bar N_1)$ which corresponds either to the wide-prior limit of $\sigma^2\gg1/\bar N_1$ for fixed $\bar N_1$ or the high-SNR limit of $\bar N_1\gg1/\sigma^2$ for fixed $\sigma$ such that the MBMSE achieves the QCRB. Conversely, if $4\bar N_1\sigma^2\ll1$, then the MBMSE remains at the prior variance $\Delta^2A = \sigma^2$ as the information from the prior dominates the information from the measurement. This corresponds to either the narrow-prior limit of $\sigma^2\ll1/\bar N_1$ for fixed $\bar N_1$ or the low-SNR limit of $\bar N_1\ll1/\sigma^2$ for fixed $\sigma$.

For example, in the sinusoidal case, the exact Gram matrix in Eq.~\ref{eq:inner product, amplitude estimation} is different for the signal defined on $(0, T)$ in Eq.~\ref{eq:signal windowed (0, T)} and $(-T/2, T/2)$ in Eq.~\ref{eq:signal windowed (-T/2, T/2)} since the average numbers are different. In the many-cycles limit of $T\gg1/\omega$, the matrix is equal for both cases to the following
\begin{align}
    \label{eq:Gram matrix, amplitude estimation}
    \braket{h_{A}}{h_{A'}} &\approx e^{-\frac{1}{8} (A-A')^2 T}
\end{align}
The average number in Eq.~\ref{eq:average total number} is $\bar N_A\approx A^2T/4$ with $\bar N_1\approx T/4$ in this limit such that the QFI in Eq.~\ref{eq:QFI, amplitude} is %
\begin{align}\label{eq:QFI, amplitude, sinusoid}
    \IQ(A) \approx T.
\end{align}
In the complex exponential case, with the signal defined on either $(0, T)$ in Eq.~\ref{eq:complex exponential signal, (0, T)} or $(-T/2, T/2)$ in Eq.~\ref{eq:complex exponential signal, (-T/2, T/2)}, the Gram matrix in Eq.~\ref{eq:inner product, amplitude estimation} is
\begin{align}
    \label{eq:Gram matrix, complex case, amplitude estimation}
    \braket{h_{A}}{h_{A'}} &= e^{-\frac{1}{4} (A-A')^2 T}
\end{align}
since the average number is $\bar N_A=A^2T/2$ with $\bar N_1=T/2$ such that the QFI is
\begin{align}\label{eq:QFI, amplitude, complex exponential}
    \IQ(A) = 2T
\end{align}
where the usual factor of two in comparison to the sinusoid case appears.

\section{Estimating the centre of a Gaussian pulse}
\label{sec:centre estimation}
The following is a different example which might have been expected to yield a Gaussian Gram matrix similarly to amplitude estimation in Appendix~\ref{app:Amplitude estimation} above. Instead, it behaves much more like frequency estimation of a complex sine-Gaussian in Eq.~\ref{eq:sine-Gaussian IP} or in the wide-prior limit in the main text. 

Suppose that we are estimating the centre of a Gaussian pulse that drives a coherent state. Let $x_\theta(t)$ be given in Eq.~\ref{eq:x(t) Gaussian} and $h_\theta(t)=-i\sqrt{2}\alpha x_\theta(t)$. The average number is $\bar N_\theta=\alpha^2$ and the Gram matrix is
\begin{align}\label{eq:Gram matrix, Gaussian pulse centre}
    G(\theta)=\exp\left(-\alpha^2 \left[1 - \exp\left(-\frac{\theta^2}{8 \sigma ^2}\right)\right]\right)
\end{align}
which is asymptotically $e^{-\alpha^2}$. This constant term leads to a divergence in $g(k)$ at DC and thus quadratic divergence of the posterior variance similarly to the frequency estimation case, except that the discontinuities from the first-order term of $G(\theta)$ seen in the frequency estimation case do not appear here since the variance of the Gaussian is finite unlike the sinc function. The MBMSE thus diverges. In comparison, the QCRB is $\IQ(\theta)^{-1}=\sigma^2/\alpha^2$ which is finite. If the average number of particles is one such that $\alpha=1$, then the QCRB is $\sigma^2$ which agrees with the value for the single-particle case. This is because the second term in Eq.~\ref{eq:single-particle QFI} vanishes since $x_\theta(t)$ is real.

\section{Phase estimation}
\label{app:Phase estimation}
Let us now estimate the phase $\phi$ for fixed amplitude $A$ and frequency $\omega$. In the sinusoidal case, the exact Gram matrix is not Toeplitz and is different for the signal defined on $(0, T)$ in Eq.~\ref{eq:signal windowed (0, T)} and $(-T/2, T/2)$ in Eq.~\ref{eq:signal windowed (-T/2, T/2)}, but in the many-cycles limit of $T\gg1/\omega$ the matrix is Toeplitz and equal for both cases to the following
\begin{align}
    \label{eq:phase estimation, Gram matrix}
    \braket{h_{\phi}}{h_{\phi'}} &\approx e^{-\frac{1}{2} A^2 T \sin^2[(\phi-\phi')/2]}
    .
\end{align}
The average number is $\bar N_\phi\approx A^2T/4$ and the QFI is the following in this limit
\begin{align}
    \label{eq:QFI, sinusoidal phase}
    \IQ(\phi) \approx A^2 T.
\end{align}
In the complex exponential case, with the signal defined on either $(0, T)$ in Eq.~\ref{eq:complex exponential signal, (0, T)} or $(-T/2, T/2)$ in Eq.~\ref{eq:complex exponential signal, (-T/2, T/2)} but with $\phi\to-\phi$ in either case, the Gram matrix is
\begin{align}
    \label{eq:phase estimation, Gram matrix, complex case}
    \braket{h_{\phi}}{h_{\phi'}} = e^{-\frac{1}{2} A^2 T \left(1-e^{i (\phi-\phi')}\right)}
\end{align}
where the average number is $\bar N_\phi=A^2T/2$ and the QFI is
\begin{align}
    \label{eq:QFI, complex phase}
    \IQ(\phi) = 2 A^2 T.
\end{align}
without assuming the many-cycles limit. Similarly to Eq.~\ref{eq:complex exponential inner product of states, (0, T)}, this Gram matrix is Hermitian and positive semi-definite but not real. Unlike for frequency estimation, the Gram matrices in Eq.~\ref{eq:phase estimation, Gram matrix} and Eq.~\ref{eq:phase estimation, Gram matrix, complex case} are not only Toeplitz but also circulant since they are $2\pi$-periodic in $\phi-\phi'$. Matrices which are circulant are necessarily Toeplitz.

We took $\phi\to-\phi$ in the complex exponential case above in Eq.~\ref{eq:phase estimation, Gram matrix, complex case} so that the corresponding spectral measure in Eq.~\ref{eq:discrete spectral measure} would consist of non-negative frequencies, $S=\mathbb{N}_0$. Given a quarter of the SNR squared $\frac{1}{2} A^2 T$, $g_k=e^{-\frac{1}{2} A^2 T} \frac{1}{k!}(\frac{1}{2} A^2 T)^{k}$ is Poissonian  if $k\geq0$ and $g_k=0$ if $k<0$. This $\phi\to-\phi$ case corresponds to $\h U(\phi)=e^{-i\phi\h N}$ such that $\h H=\h N$ instead of $\h U(\phi)=e^{i\phi\h N}$ and $\h H=-\h N$.

Regarding the BMSE, the periodic symmetry of this circulant family of states $\{\ket{h_\phi}\}_{\phi\in(-\pi,\pi]}$ means that we should switch from using the MSE in Eq.~\ref{eq:MSE} to a periodic cost function. We discuss this in Appendix~\ref{sec:Covariant figure-of-merit}.

\section{Numerical calculation of the MBMSE}
\label{app:Numerics}
In general, the Gram matrix may not be Toeplitz, e.g.\ in Eq.~\ref{eq:cosine inner product of states, (0, T), exact}, such that the analytic results in the main text do not apply. Instead, we can calculate the MBMSE numerically as follows. We focus on frequency estimation here but the procedure is general. Suppose that we want to estimate a frequency within $(\omega_0-\Delta\omega/2,\omega_0+\Delta\omega/2)$ and use a finite numerical frequency resolution $\delta\omega$ such that there are $n=\Delta\omega/\delta\omega$ waveforms or frequency bins. For example, the Gram matrix $G_{nn}$ is $n$-by-$n$ dimensional where we make the dimension explicit here for clarity. We find a basis for the waveform subspace by taking the spectral decomposition of the Gram matrix $G_{nn}=U_{nn}S_{nn}U_{nn}^\dag$ where $U_{nn}$ is unitary and $S_{nn}$ is diagonal. Since $G$ is Hermitian and positive-semidefinite, its eigenvalues (the entries of $S_{nn}$) are real and non-negative. Let $A_{Dn}$ be the matrix whose $n$ columns are the waveform states $\{\ket{h_\omega}\}_{\omega\in\mathbb{R}}$ each of which is dimension $D=d^M$ such that the Gram matrix is $G_{nn}=A_{nD}^\dag A_{Dn}$. Here, we discretise the bath in time into $M$ oscillators with truncated dimension $d$ as in Appendix~\ref{app:Discretising the time domain}. The goal of working within the waveform subspace is to avoid representing these $D$-dimensional states directly since the Hilbert space is too large and grows exponentially. We can calculate the Gram matrix for the coherent states numerically by using the analytic result in Eq.~\ref{eq:inner product of coherent states} without needing to explicitly represent the states in the large $D$-dimensional space. The spectral decomposition of $G_{nn}$ is the singular value decomposition (SVD) of $A_{Dn}$. This decomposition is more numerically stable than, e.g., directly performing the Gram-Schmidt process to $A_{Dn}$.

We may be concerned that improving the numerical frequency resolution $\delta\omega$ will dramatically increase the number of waveforms $n=\Delta\omega/\delta\omega$. However, we observe numerically that the eigenvalues of the Gram matrix quickly decay. For a given numerical threshold, below which the eigenvalues are set to zero, the rank $r\ll n$ of $S_{nn}$ remains stable even as $n$ increases. We may then use the truncated SVD to approximate the Gram matrix as $G_{nn}\approx U_{nr}S_{rr}U_{rn}^\dag$ by keeping only the eigenvalues above the threshold. Here, $U_{nr}$ is now only semi-unitary since $U_{rn}^\dag U_{nr} = I_r$ but $U_{nr}U_{rn}^\dag = P_{nn}$ where $P_{nn}$ is the Hermitian projection onto the column space of $U_{nr}$. This decomposition is the optimal low-rank approximation to $G_{nn}$ in the sense of minimising the Frobenius norm by the Eckart-Young theorem. The rank $r$ remains finite because as the frequency resolution shrinks the neighbouring frequencies become less orthogonal for finite SNR, i.e.\ $\braket{h_\omega}{h_{\omega+\delta\omega}}\to1$ as $\delta\omega\to0$ in Eq.~\ref{eq:inner product, infinitesimal, cosine}. 

Given the truncated SVD, we can now find a low-rank representation of the waveform states. Let $B_{Dr}=A_{Dn} U_{nr}S^{-1/2}_{rr}$ such that the inverse transformation is $A_{Dn}=B_{Dr}S^{1/2}_{rr}U_{rn}^\dag$. The $r$ columns of $B_{Dr}$ are orthogonal since $B_{rD}^\dag B_{Dr} \approx I_r$ given that $A_{nD}^\dag A_{Dn}\approx U_{nr}S_{rr}U_{rn}^\dag$. Let the $k$th column of $B_{Dr}$ be $\ket{\varphi_k}$ for $1\leq k\leq r$. The transformation between the waveform states $\ket{h_j}$ and these states $\ket{\varphi_k}$ for $1\leq j\leq n$ and $1\leq k\leq r$ is
\begin{align}
    \ket{h_j} &= \sum_{l=1}^r (U_{nr})_{jl}^* (S_{rr})_l^{1/2} \ket{\varphi_l}
    \\\ket{\varphi_k} &= (S_{rr})_k^{-1/2} \sum_{m=1}^n (U_{nr})_{mk} \ket{h_m}.\nonumber
\end{align}
Here, the normalisation of $\ket{\varphi_k}$ includes $(S_{rr})_k^{-1/2}$ which is well-defined since we truncated the SVD to only include positive eigenvalues above the numerical noise floor. We can numerically calculate the matrix coefficients of the mixed state and its Bayesian derivative in Eq.~\ref{eq:mixed state} in this basis and thus find the MBMSE in Eq.~\ref{eq:MBMSE sol}. The BSLD in Eq.~\ref{eq:BSLD sol} is an $r$-by-$r$ matrix $L_{rr}$ such that we can determine the coefficients of the optimal measurement basis $\{\ket{x}\}_x$ in Eq.~\ref{eq:BSLD} in the basis. We can then represent a given $\ket{x}$ as a superposition of the original waveform states, i.e.\ $\ket{x}=\sum_{j=1}^nc_{jx}\ket{h_j}$.

In the Toeplitz case for an infinite continuous family of states, we showed that the Gram matrix is diagonalised by the quantum Fourier transform such that the eigenvalues are $g(k)$ and the eigenstates are the Fourier states $\ket{k}$ and, provided that the prior is uninformative and uniform, that the optimal measurement basis is the quantum whitened superposition of the original waveform states. For finite Toeplitz matrices, the situation is more complicated and requires Szeg\H{o}'s theorem to understand the asymptotic relation between the eigenvalue distribution and the Fourier transform~\cite{bottcher2000toeplitz}. This is because Toeplitz matrices are only in general asymptotically diagonalised by the Fourier transform unlike finite circulant matrices.

We have considered only families of pure states throughout this work, but most of the techniques will also work for families of mixed states. In particular, we may want to study the effect of marginalising over unknown parameters rather than assume that they are known. For example, we could estimate the frequency $\om$ of a sinusoidal signal encoded in a coherent state with unknown amplitude $A$ and phase $\phi$, where we uniformly marginalise the amplitude over $A\in(0, A_0)$ and phase over $\phi\in(-\pi, \pi)$. In general, suppose that the pure state is $\ket{h_{\theta,\vec\phi}}$ for a given parameter $\theta$ that we wish to estimate and a vector of unknown parameters $\vec\phi$. Our knowledge of the device is thus captured by the classical mixture $\h\rho(\theta)=\inth{\vec\phi}p(\vec\phi)\proj{h_{\theta,\vec\phi}}$ and we want to study the family of states $\{\h\rho(\theta)\}_{\theta\in\R}$. These density matrices are bounded operators on the waveform subspace spanned by the possible pure states $\ket{h_{\theta,\vec\phi}}$. This means that we can first apply the SVD approach to $\{\ket{h_{\theta,\vec\phi}}\}_{\theta,\vec\phi}$ to find a basis for the (hopefully low-dimensional) waveform subspace and then represent $\h\rho(\theta)$ in this basis numerically. The procedure to calculate the MBMSE or QFI is then the same as before.

\nocite{apsrev42Control}
\bibliographystyle{bibliography/style.bst}
\renewcommand{\selectlanguage}[1]{}
\bibliography{bibliography/bib}

\begin{thebibliography}{44}%
\makeatletter
\providecommand \@ifxundefined [1]{%
 \@ifx{#1\undefined}
}%
\providecommand \@ifnum [1]{%
 \ifnum #1\expandafter \@firstoftwo
 \else \expandafter \@secondoftwo
 \fi
}%
\providecommand \@ifx [1]{%
 \ifx #1\expandafter \@firstoftwo
 \else \expandafter \@secondoftwo
 \fi
}%
\providecommand \natexlab [1]{#1}%
\providecommand \enquote  [1]{``#1''}%
\providecommand \bibnamefont  [1]{#1}%
\providecommand \bibfnamefont [1]{#1}%
\providecommand \citenamefont [1]{#1}%
\providecommand \href@noop [0]{\@secondoftwo}%
\providecommand \href [0]{\begingroup \@sanitize@url \@href}%
\providecommand \@href[1]{\@@startlink{#1}\@@href}%
\providecommand \@@href[1]{\endgroup#1\@@endlink}%
\providecommand \@sanitize@url [0]{\catcode `\\12\catcode `\$12\catcode `\&12\catcode `\#12\catcode `\^12\catcode `\_12\catcode `\%12\relax}%
\providecommand \@@startlink[1]{}%
\providecommand \@@endlink[0]{}%
\providecommand \url  [0]{\begingroup\@sanitize@url \@url }%
\providecommand \@url [1]{\endgroup\@href {#1}{\urlprefix }}%
\providecommand \urlprefix  [0]{URL }%
\providecommand \Eprint [0]{\href }%
\providecommand \doibase [0]{https://doi.org/}%
\providecommand \selectlanguage [0]{\@gobble}%
\providecommand \bibinfo  [0]{\@secondoftwo}%
\providecommand \bibfield  [0]{\@secondoftwo}%
\providecommand \translation [1]{[#1]}%
\providecommand \BibitemOpen [0]{}%
\providecommand \bibitemStop [0]{}%
\providecommand \bibitemNoStop [0]{.\EOS\space}%
\providecommand \EOS [0]{\spacefactor3000\relax}%
\providecommand \BibitemShut  [1]{\csname bibitem#1\endcsname}%
\let\auto@bib@innerbib\@empty
\bibitem [{\citenamefont {{E. Payne et al. (in preparation)}}(2025)}]{ethanpaper}%
  \BibitemOpen
  \bibfield  {author} {\bibinfo {author} {\bibnamefont {{E. Payne et al. (in preparation)}}},\ }\href@noop {} {\bibinfo {title} {Prospects for post-merger remnant signal inference with photon-counting interferometry}} (\bibinfo {year} {2025})\BibitemShut {NoStop}%
\bibitem [{\citenamefont {Lasky}(2021)}]{10.1088/2514-3433/ac2256ch9}%
  \BibitemOpen
  \bibfield  {author} {\bibinfo {author} {\bibfnamefont {P.~D.}\ \bibnamefont {Lasky}},\ }\bibfield  {title} {\bibinfo {title} {Gravitational wave astronomy},\ }in\ \href {https://doi.org/10.1088/2514-3433/ac2256ch9} {\emph {\bibinfo {booktitle} {Multimessenger Astronomy in Practice}}},\ \bibinfo {series and number} {2514-3433}\ (\bibinfo  {publisher} {IOP Publishing},\ \bibinfo {year} {2021})\ pp.\ \bibinfo {pages} {9--1 to 9--72}\BibitemShut {NoStop}%
\bibitem [{\citenamefont {Sarin}\ and\ \citenamefont {Lasky}(2022)}]{sarin_lasky_2022}%
  \BibitemOpen
  \bibfield  {author} {\bibinfo {author} {\bibfnamefont {N.}~\bibnamefont {Sarin}}\ \bibnamefont {and}\ \bibinfo {author} {\bibfnamefont {P.~D.}\ \bibnamefont {Lasky}},\ }\bibfield  {title} {\bibinfo {title} {Multimessenger astronomy with a khz-band gravitational-wave observatory},\ }\href {https://doi.org/10.1017/pasa.2022.1} {\bibfield  {journal} {\bibinfo  {journal} {Publ. Astron. Soc. Aust.}\ }\textbf {\bibinfo {volume} {39}},\ \bibinfo {pages} {e007} (\bibinfo {year} {2022})}\BibitemShut {NoStop}%
\bibitem [{\citenamefont {Kim}\ and\ \citenamefont {Carosi}(2010)}]{kim2010axions}%
  \BibitemOpen
  \bibfield  {author} {\bibinfo {author} {\bibfnamefont {J.~E.}\ \bibnamefont {Kim}}\ \bibnamefont {and}\ \bibinfo {author} {\bibfnamefont {G.}~\bibnamefont {Carosi}},\ }\bibfield  {title} {\bibinfo {title} {Axions and the strong ${{C P}}$ problem},\ }\href {https://doi.org/10.1103/RevModPhys.82.557} {\bibfield  {journal} {\bibinfo  {journal} {Rev. Mod. Phys.}\ }\textbf {\bibinfo {volume} {82}},\ \bibinfo {pages} {557} (\bibinfo {year} {2010})}\BibitemShut {NoStop}%
\bibitem [{\citenamefont {Choi}\ \emph {et~al.}(2021)\citenamefont {Choi}, \citenamefont {Im},\ and\ \citenamefont {Shin}}]{choi2021recent}%
  \BibitemOpen
  \bibfield  {author} {\bibinfo {author} {\bibfnamefont {K.}~\bibnamefont {Choi}}, \bibinfo {author} {\bibfnamefont {S.~H.}\ \bibnamefont {Im}},\ \bibnamefont {and}\ \bibinfo {author} {\bibfnamefont {C.~S.}\ \bibnamefont {Shin}},\ }\bibfield  {title} {\bibinfo {title} {Recent progress in the physics of axions and axion-like particles},\ }\href {https://doi.org/10.1146/annurev-nucl-120720-031147} {\bibfield  {journal} {\bibinfo  {journal} {Annu. Rev. Nucl. Part. Sci.}\ }\textbf {\bibinfo {volume} {71}},\ \bibinfo {pages} {225} (\bibinfo {year} {2021})}\BibitemShut {NoStop}%
\bibitem [{\citenamefont {Rosenberg}\ and\ \citenamefont {Van~Bibber}(2000)}]{rosenberg2000searches}%
  \BibitemOpen
  \bibfield  {author} {\bibinfo {author} {\bibfnamefont {L.~J.}\ \bibnamefont {Rosenberg}}\ \bibnamefont {and}\ \bibinfo {author} {\bibfnamefont {K.~A.}\ \bibnamefont {Van~Bibber}},\ }\bibfield  {title} {\bibinfo {title} {Searches for invisible axions},\ }\href {https://doi.org/10.1016/S0370-1573(99)00045-9} {\bibfield  {journal} {\bibinfo  {journal} {Phys. Rep.}\ }\textbf {\bibinfo {volume} {325}},\ \bibinfo {pages} {1} (\bibinfo {year} {2000})}\BibitemShut {NoStop}%
\bibitem [{\citenamefont {Graham}\ \emph {et~al.}(2015)\citenamefont {Graham}, \citenamefont {Irastorza}, \citenamefont {Lamoreaux}, \citenamefont {Lindner},\ and\ \citenamefont {van Bibber}}]{graham2015experimental}%
  \BibitemOpen
  \bibfield  {author} {\bibinfo {author} {\bibfnamefont {P.~W.}\ \bibnamefont {Graham}}, \bibinfo {author} {\bibfnamefont {I.~G.}\ \bibnamefont {Irastorza}}, \bibinfo {author} {\bibfnamefont {S.~K.}\ \bibnamefont {Lamoreaux}}, \bibinfo {author} {\bibfnamefont {A.}~\bibnamefont {Lindner}},\ \bibnamefont {and}\ \bibinfo {author} {\bibfnamefont {K.~A.}\ \bibnamefont {van Bibber}},\ }\bibfield  {title} {\bibinfo {title} {Experimental searches for the axion and axion-like particles},\ }\href {https://doi.org/10.1146/annurev-nucl-102014-022120} {\bibfield  {journal} {\bibinfo  {journal} {Annu. Rev. Nucl. Part. Sci.}\ }\textbf {\bibinfo {volume} {65}},\ \bibinfo {pages} {485} (\bibinfo {year} {2015})}\BibitemShut {NoStop}%
\bibitem [{\citenamefont {Peccei}(2008)}]{Peccei2008}%
  \BibitemOpen
  \bibfield  {author} {\bibinfo {author} {\bibfnamefont {R.~D.}\ \bibnamefont {Peccei}},\ }\bibinfo {title} {The strong cp problem and axions},\ in\ \href {https://doi.org/10.1007/978-3-540-73518-2_1} {\emph {\bibinfo {booktitle} {Axions: Theory, Cosmology, and Experimental Searches}}},\ \bibinfo {editor} {edited by\ \bibinfo {editor} {\bibfnamefont {M.}~\bibnamefont {Kuster}}, \bibinfo {editor} {\bibfnamefont {G.}~\bibnamefont {Raffelt}},\ \bibnamefont {and}\ \bibinfo {editor} {\bibfnamefont {B.}~\bibnamefont {Beltr{\'a}n}}}\ (\bibinfo  {publisher} {Springer Berlin Heidelberg},\ \bibinfo {address} {Berlin, Heidelberg},\ \bibinfo {year} {2008})\ pp.\ \bibinfo {pages} {3--17}\BibitemShut {NoStop}%
\bibitem [{\citenamefont {Battaglieri}\ \emph {et~al.}(2017)\citenamefont {Battaglieri}, \citenamefont {Belloni}, \citenamefont {Chou}, \citenamefont {Cushman}, \citenamefont {Echenard}, \citenamefont {Essig}, \citenamefont {Estrada}, \citenamefont {Feng}, \citenamefont {Flaugher}, \citenamefont {Fox} \emph {et~al.}}]{battaglieri2017us}%
  \BibitemOpen
  \bibfield  {author} {\bibinfo {author} {\bibfnamefont {M.}~\bibnamefont {Battaglieri}}, \bibinfo {author} {\bibfnamefont {A.}~\bibnamefont {Belloni}}, \bibinfo {author} {\bibfnamefont {A.}~\bibnamefont {Chou}}, \bibinfo {author} {\bibfnamefont {P.}~\bibnamefont {Cushman}}, \bibinfo {author} {\bibfnamefont {B.}~\bibnamefont {Echenard}}, \bibinfo {author} {\bibfnamefont {R.}~\bibnamefont {Essig}}, \bibinfo {author} {\bibfnamefont {J.}~\bibnamefont {Estrada}}, \bibinfo {author} {\bibfnamefont {J.~L.}\ \bibnamefont {Feng}}, \bibinfo {author} {\bibfnamefont {B.}~\bibnamefont {Flaugher}}, \bibinfo {author} {\bibfnamefont {P.~J.}\ \bibnamefont {Fox}}, \bibnamefont {et~al.},\ }\bibfield  {title} {\bibinfo {title} {Us cosmic visions: new ideas in dark matter 2017: community report},\ }\href@noop {} {\bibfield  {journal} {\bibinfo  {journal} {arXiv preprint arXiv:1707.04591}\ } (\bibinfo {year} {2017})}\BibitemShut {NoStop}%
\bibitem [{\citenamefont {Kubo}(1966)}]{KuboRPP66FluctuationdissipationTheorem}%
  \BibitemOpen
  \bibfield  {author} {\bibinfo {author} {\bibfnamefont {R.}~\bibnamefont {Kubo}},\ }\bibfield  {title} {\bibinfo {title} {The fluctuation-dissipation theorem},\ }\href {https://doi.org/10.1088/0034-4885/29/1/306} {\bibfield  {journal} {\bibinfo  {journal} {Rep. Prog. Phys.}\ }\textbf {\bibinfo {volume} {29}},\ \bibinfo {pages} {255} (\bibinfo {year} {1966})}\BibitemShut {NoStop}%
\bibitem [{\citenamefont {Buonanno}\ and\ \citenamefont {Chen}(2002)}]{BuonannoPRD02SignalRecycled}%
  \BibitemOpen
  \bibfield  {author} {\bibinfo {author} {\bibfnamefont {A.}~\bibnamefont {Buonanno}}\ \bibnamefont {and}\ \bibinfo {author} {\bibfnamefont {Y.}~\bibnamefont {Chen}},\ }\bibfield  {title} {\bibinfo {title} {Signal recycled laser-interferometer gravitational-wave detectors as optical springs},\ }\href {https://doi.org/10.1103/PhysRevD.65.042001} {\bibfield  {journal} {\bibinfo  {journal} {Phys. Rev. D}\ }\textbf {\bibinfo {volume} {65}},\ \bibinfo {pages} {042001} (\bibinfo {year} {2002})}\BibitemShut {NoStop}%
\bibitem [{\citenamefont {Bond}\ \emph {et~al.}(2016)\citenamefont {Bond}, \citenamefont {Brown}, \citenamefont {Freise},\ and\ \citenamefont {Strain}}]{bond2016interferometer}%
  \BibitemOpen
  \bibfield  {author} {\bibinfo {author} {\bibfnamefont {C.}~\bibnamefont {Bond}}, \bibinfo {author} {\bibfnamefont {D.}~\bibnamefont {Brown}}, \bibinfo {author} {\bibfnamefont {A.}~\bibnamefont {Freise}},\ \bibnamefont {and}\ \bibinfo {author} {\bibfnamefont {K.~A.}\ \bibnamefont {Strain}},\ }\bibfield  {title} {\bibinfo {title} {Interferometer techniques for gravitational-wave detection},\ }\href {https://doi.org/10.1007/s41114-016-0002-8} {\bibfield  {journal} {\bibinfo  {journal} {Living Rev. Relativ.}\ }\textbf {\bibinfo {volume} {19}},\ \bibinfo {pages} {1} (\bibinfo {year} {2016})}\BibitemShut {NoStop}%
\bibitem [{\citenamefont {Macieszczak}\ \emph {et~al.}(2014)\citenamefont {Macieszczak}, \citenamefont {Fraas},\ and\ \citenamefont {Demkowicz-Dobrza{\'n}ski}}]{macieszczak2014bayesian}%
  \BibitemOpen
  \bibfield  {author} {\bibinfo {author} {\bibfnamefont {K.}~\bibnamefont {Macieszczak}}, \bibinfo {author} {\bibfnamefont {M.}~\bibnamefont {Fraas}},\ \bibnamefont {and}\ \bibinfo {author} {\bibfnamefont {R.}~\bibnamefont {Demkowicz-Dobrza{\'n}ski}},\ }\bibfield  {title} {\bibinfo {title} {Bayesian quantum frequency estimation in presence of collective dephasing},\ }\href@noop {} {\bibfield  {journal} {\bibinfo  {journal} {New Journal of Physics}\ }\textbf {\bibinfo {volume} {16}},\ \bibinfo {pages} {113002} (\bibinfo {year} {2014})}\BibitemShut {NoStop}%
\bibitem [{\citenamefont {Jarzyna}\ and\ \citenamefont {Demkowicz-Dobrza{\'n}ski}(2015)}]{jarzyna2015true}%
  \BibitemOpen
  \bibfield  {author} {\bibinfo {author} {\bibfnamefont {M.}~\bibnamefont {Jarzyna}}\ \bibnamefont {and}\ \bibinfo {author} {\bibfnamefont {R.}~\bibnamefont {Demkowicz-Dobrza{\'n}ski}},\ }\bibfield  {title} {\bibinfo {title} {True precision limits in quantum metrology},\ }\href@noop {} {\bibfield  {journal} {\bibinfo  {journal} {New Journal of Physics}\ }\textbf {\bibinfo {volume} {17}},\ \bibinfo {pages} {013010} (\bibinfo {year} {2015})}\BibitemShut {NoStop}%
\bibitem [{\citenamefont {Rubio}\ and\ \citenamefont {Dunningham}(2019)}]{rubio2019quantum}%
  \BibitemOpen
  \bibfield  {author} {\bibinfo {author} {\bibfnamefont {J.}~\bibnamefont {Rubio}}\ \bibnamefont {and}\ \bibinfo {author} {\bibfnamefont {J.}~\bibnamefont {Dunningham}},\ }\bibfield  {title} {\bibinfo {title} {Quantum metrology in the presence of limited data},\ }\href@noop {} {\bibfield  {journal} {\bibinfo  {journal} {New Journal of Physics}\ }\textbf {\bibinfo {volume} {21}},\ \bibinfo {pages} {043037} (\bibinfo {year} {2019})}\BibitemShut {NoStop}%
\bibitem [{\citenamefont {Kaubruegger}\ \emph {et~al.}(2021)\citenamefont {Kaubruegger}, \citenamefont {Vasilyev}, \citenamefont {Schulte}, \citenamefont {Hammerer},\ and\ \citenamefont {Zoller}}]{kaubruegger2021quantum}%
  \BibitemOpen
  \bibfield  {author} {\bibinfo {author} {\bibfnamefont {R.}~\bibnamefont {Kaubruegger}}, \bibinfo {author} {\bibfnamefont {D.~V.}\ \bibnamefont {Vasilyev}}, \bibinfo {author} {\bibfnamefont {M.}~\bibnamefont {Schulte}}, \bibinfo {author} {\bibfnamefont {K.}~\bibnamefont {Hammerer}},\ \bibnamefont {and}\ \bibinfo {author} {\bibfnamefont {P.}~\bibnamefont {Zoller}},\ }\bibfield  {title} {\bibinfo {title} {Quantum variational optimization of ramsey interferometry and atomic clocks},\ }\href@noop {} {\bibfield  {journal} {\bibinfo  {journal} {Physical review X}\ }\textbf {\bibinfo {volume} {11}},\ \bibinfo {pages} {041045} (\bibinfo {year} {2021})}\BibitemShut {NoStop}%
\bibitem [{\citenamefont {Marciniak}\ \emph {et~al.}(2022)\citenamefont {Marciniak}, \citenamefont {Feldker}, \citenamefont {Pogorelov}, \citenamefont {Kaubruegger}, \citenamefont {Vasilyev}, \citenamefont {van Bijnen}, \citenamefont {Schindler}, \citenamefont {Zoller}, \citenamefont {Blatt},\ and\ \citenamefont {Monz}}]{marciniak2022optimal}%
  \BibitemOpen
  \bibfield  {author} {\bibinfo {author} {\bibfnamefont {C.~D.}\ \bibnamefont {Marciniak}}, \bibinfo {author} {\bibfnamefont {T.}~\bibnamefont {Feldker}}, \bibinfo {author} {\bibfnamefont {I.}~\bibnamefont {Pogorelov}}, \bibinfo {author} {\bibfnamefont {R.}~\bibnamefont {Kaubruegger}}, \bibinfo {author} {\bibfnamefont {D.~V.}\ \bibnamefont {Vasilyev}}, \bibinfo {author} {\bibfnamefont {R.}~\bibnamefont {van Bijnen}}, \bibinfo {author} {\bibfnamefont {P.}~\bibnamefont {Schindler}}, \bibinfo {author} {\bibfnamefont {P.}~\bibnamefont {Zoller}}, \bibinfo {author} {\bibfnamefont {R.}~\bibnamefont {Blatt}},\ \bibnamefont {and}\ \bibinfo {author} {\bibfnamefont {T.}~\bibnamefont {Monz}},\ }\bibfield  {title} {\bibinfo {title} {Optimal metrology with programmable quantum sensors},\ }\href@noop {} {\bibfield  {journal} {\bibinfo  {journal} {Nature}\ }\textbf {\bibinfo {volume} {603}},\ \bibinfo {pages} {604} (\bibinfo {year} {2022})}\BibitemShut {NoStop}%
\bibitem [{\citenamefont {Direkci}\ \emph {et~al.}(2024)\citenamefont {Direkci}, \citenamefont {Finkelstein}, \citenamefont {Endres},\ and\ \citenamefont {Gefen}}]{direkci2024heisenberg}%
  \BibitemOpen
  \bibfield  {author} {\bibinfo {author} {\bibfnamefont {S.}~\bibnamefont {Direkci}}, \bibinfo {author} {\bibfnamefont {R.}~\bibnamefont {Finkelstein}}, \bibinfo {author} {\bibfnamefont {M.}~\bibnamefont {Endres}},\ \bibnamefont {and}\ \bibinfo {author} {\bibfnamefont {T.}~\bibnamefont {Gefen}},\ }\bibfield  {title} {\bibinfo {title} {Heisenberg-limited bayesian phase estimation with low-depth digital quantum circuits},\ }\href@noop {} {\bibfield  {journal} {\bibinfo  {journal} {arXiv preprint arXiv:2407.06006}\ } (\bibinfo {year} {2024})}\BibitemShut {NoStop}%
\bibitem [{\citenamefont {Tsang}\ \emph {et~al.}(2011)\citenamefont {Tsang}, \citenamefont {Wiseman},\ and\ \citenamefont {Caves}}]{TsangPRL11FundamentalQuantum}%
  \BibitemOpen
  \bibfield  {author} {\bibinfo {author} {\bibfnamefont {M.}~\bibnamefont {Tsang}}, \bibinfo {author} {\bibfnamefont {H.~M.}\ \bibnamefont {Wiseman}},\ \bibnamefont {and}\ \bibinfo {author} {\bibfnamefont {C.~M.}\ \bibnamefont {Caves}},\ }\bibfield  {title} {\bibinfo {title} {Fundamental {{Quantum Limit}} to {{Waveform Estimation}}},\ }\href {https://doi.org/10.1103/PhysRevLett.106.090401} {\bibfield  {journal} {\bibinfo  {journal} {Phys. Rev. Lett.}\ }\textbf {\bibinfo {volume} {106}},\ \bibinfo {pages} {090401} (\bibinfo {year} {2011})}\BibitemShut {NoStop}%
\bibitem [{\citenamefont {Miao}\ \emph {et~al.}(2017)\citenamefont {Miao}, \citenamefont {Adhikari}, \citenamefont {Ma}, \citenamefont {Pang},\ and\ \citenamefont {Chen}}]{Miao+2017}%
  \BibitemOpen
  \bibfield  {author} {\bibinfo {author} {\bibfnamefont {H.}~\bibnamefont {Miao}}, \bibinfo {author} {\bibfnamefont {R.~X.}\ \bibnamefont {Adhikari}}, \bibinfo {author} {\bibfnamefont {Y.}~\bibnamefont {Ma}}, \bibinfo {author} {\bibfnamefont {B.}~\bibnamefont {Pang}},\ \bibnamefont {and}\ \bibinfo {author} {\bibfnamefont {Y.}~\bibnamefont {Chen}},\ }\bibfield  {title} {\bibinfo {title} {Towards the fundamental quantum limit of linear measurements of classical signals},\ }\href {https://doi.org/10.1103/PhysRevLett.119.050801} {\bibfield  {journal} {\bibinfo  {journal} {Phys. Rev. Lett.}\ }\textbf {\bibinfo {volume} {119}},\ \bibinfo {pages} {050801} (\bibinfo {year} {2017})}\BibitemShut {NoStop}%
\bibitem [{\citenamefont {Gardner}\ \emph {et~al.}(2024)\citenamefont {Gardner}, \citenamefont {Gefen}, \citenamefont {Haine}, \citenamefont {Hope},\ and\ \citenamefont {Chen}}]{gardner2024achieving}%
  \BibitemOpen
  \bibfield  {author} {\bibinfo {author} {\bibfnamefont {J.~W.}\ \bibnamefont {Gardner}}, \bibinfo {author} {\bibfnamefont {T.}~\bibnamefont {Gefen}}, \bibinfo {author} {\bibfnamefont {S.~A.}\ \bibnamefont {Haine}}, \bibinfo {author} {\bibfnamefont {J.~J.}\ \bibnamefont {Hope}},\ \bibnamefont {and}\ \bibinfo {author} {\bibfnamefont {Y.}~\bibnamefont {Chen}},\ }\bibfield  {title} {\bibinfo {title} {Achieving the fundamental quantum limit of linear waveform estimation},\ }\href {https://doi.org/10.1103/PhysRevLett.132.130801} {\bibfield  {journal} {\bibinfo  {journal} {Phys. Rev. Lett.}\ }\textbf {\bibinfo {volume} {132}},\ \bibinfo {pages} {130801} (\bibinfo {year} {2024})}\BibitemShut {NoStop}%
\bibitem [{\citenamefont {Holevo}(2011)}]{Holevo2011book}%
  \BibitemOpen
  \bibfield  {author} {\bibinfo {author} {\bibfnamefont {A.~S.}\ \bibnamefont {Holevo}},\ }\href {https://doi.org/10.1007/978-88-7642-378-9} {\emph {\bibinfo {title} {Probabilistic and Statistical Aspects of Quantum Theory}}}\ (\bibinfo  {publisher} {Springer Science \& Business Media},\ \bibinfo {year} {2011})\BibitemShut {NoStop}%
\bibitem [{\citenamefont {Vasilyev}\ \emph {et~al.}(2024)\citenamefont {Vasilyev}, \citenamefont {Shankar}, \citenamefont {Kaubruegger},\ and\ \citenamefont {Zoller}}]{vasilyev2024optimal}%
  \BibitemOpen
  \bibfield  {author} {\bibinfo {author} {\bibfnamefont {D.~V.}\ \bibnamefont {Vasilyev}}, \bibinfo {author} {\bibfnamefont {A.}~\bibnamefont {Shankar}}, \bibinfo {author} {\bibfnamefont {R.}~\bibnamefont {Kaubruegger}},\ \bibnamefont {and}\ \bibinfo {author} {\bibfnamefont {P.}~\bibnamefont {Zoller}},\ }\bibfield  {title} {\bibinfo {title} {Optimal multiparameter metrology: The quantum compass solution},\ }\href@noop {} {\bibfield  {journal} {\bibinfo  {journal} {arXiv preprint arXiv:2404.14194}\ } (\bibinfo {year} {2024})}\BibitemShut {NoStop}%
\bibitem [{\citenamefont {{A. Monras, Phase space formalism for quantum estimation of Gaussian states (2013)}}()}]{monras2013phase}%
  \BibitemOpen
  \bibfield  {author} {\bibinfo {author} {\bibnamefont {{A. Monras, Phase space formalism for quantum estimation of Gaussian states (2013)}}},\ }\href {https://doi.org/10.48550/arXiv.1303.3682} {\bibinfo {title} {arxiv:1303.3682 [quant-ph]}}\BibitemShut {NoStop}%
\bibitem [{\citenamefont {Van~Trees}(2002)}]{van2002detection}%
  \BibitemOpen
  \bibfield  {author} {\bibinfo {author} {\bibfnamefont {H.~L.}\ \bibnamefont {Van~Trees}},\ }\href {https://doi.org/10.1002/0471221104} {\emph {\bibinfo {title} {{Detection, Estimation, and Modulation Theory: Part IV}}}}\ (\bibinfo  {publisher} {John Wiley \& Sons, Incorporated},\ \bibinfo {year} {2002})\BibitemShut {NoStop}%
\bibitem [{\citenamefont {Gill}\ and\ \citenamefont {Levit}(1995)}]{gill1995applications}%
  \BibitemOpen
  \bibfield  {author} {\bibinfo {author} {\bibfnamefont {R.~D.}\ \bibnamefont {Gill}}\ \bibnamefont {and}\ \bibinfo {author} {\bibfnamefont {B.~Y.}\ \bibnamefont {Levit}},\ }\bibfield  {title} {\bibinfo {title} {{Applications of the van Trees inequality: a Bayesian Cram{\'e}r-Rao bound}},\ }\href {https://doi.org/10.2307/3318681} {\bibfield  {journal} {\bibinfo  {journal} {Bernoulli}\ ,\ \bibinfo {pages} {59}} (\bibinfo {year} {1995})}\BibitemShut {NoStop}%
\bibitem [{\citenamefont {Genoni}\ \emph {et~al.}(2013)\citenamefont {Genoni}, \citenamefont {Paris}, \citenamefont {Adesso}, \citenamefont {Nha}, \citenamefont {Knight},\ and\ \citenamefont {Kim}}]{genoni2013optimal}%
  \BibitemOpen
  \bibfield  {author} {\bibinfo {author} {\bibfnamefont {M.~G.}\ \bibnamefont {Genoni}}, \bibinfo {author} {\bibfnamefont {M.~G.}\ \bibnamefont {Paris}}, \bibinfo {author} {\bibfnamefont {G.}~\bibnamefont {Adesso}}, \bibinfo {author} {\bibfnamefont {H.}~\bibnamefont {Nha}}, \bibinfo {author} {\bibfnamefont {P.~L.}\ \bibnamefont {Knight}},\ \bibnamefont {and}\ \bibinfo {author} {\bibfnamefont {M.}~\bibnamefont {Kim}},\ }\bibfield  {title} {\bibinfo {title} {Optimal estimation of joint parameters in phase space},\ }\href@noop {} {\bibfield  {journal} {\bibinfo  {journal} {Physical Review A—Atomic, Molecular, and Optical Physics}\ }\textbf {\bibinfo {volume} {87}},\ \bibinfo {pages} {012107} (\bibinfo {year} {2013})}\BibitemShut {NoStop}%
\bibitem [{\citenamefont {Gardner}\ \emph {et~al.}(2025{\natexlab{a}})\citenamefont {Gardner}, \citenamefont {Gefen}, \citenamefont {Haine}, \citenamefont {Hope}, \citenamefont {Preskill}, \citenamefont {Chen},\ and\ \citenamefont {McCuller}}]{gardner2024stochastic}%
  \BibitemOpen
  \bibfield  {author} {\bibinfo {author} {\bibfnamefont {J.~W.}\ \bibnamefont {Gardner}}, \bibinfo {author} {\bibfnamefont {T.}~\bibnamefont {Gefen}}, \bibinfo {author} {\bibfnamefont {S.~A.}\ \bibnamefont {Haine}}, \bibinfo {author} {\bibfnamefont {J.~J.}\ \bibnamefont {Hope}}, \bibinfo {author} {\bibfnamefont {J.}~\bibnamefont {Preskill}}, \bibinfo {author} {\bibfnamefont {Y.}~\bibnamefont {Chen}},\ \bibnamefont {and}\ \bibinfo {author} {\bibfnamefont {L.}~\bibnamefont {McCuller}},\ }\bibfield  {title} {\bibinfo {title} {Stochastic waveform estimation at the fundamental quantum limit},\ }\href {https://doi.org/10.1103/h91r-4ws9} {\bibfield  {journal} {\bibinfo  {journal} {PRX Quantum}\ }\textbf {\bibinfo {volume} {6}},\ \bibinfo {pages} {030311} (\bibinfo {year} {2025}{\natexlab{a}})}\BibitemShut {NoStop}%
\bibitem [{\citenamefont {Gardner}\ \emph {et~al.}(2025{\natexlab{b}})\citenamefont {Gardner}, \citenamefont {Haine}, \citenamefont {Hope}, \citenamefont {Chen},\ and\ \citenamefont {Gefen}}]{gardner2025lindblad}%
  \BibitemOpen
  \bibfield  {author} {\bibinfo {author} {\bibfnamefont {J.~W.}\ \bibnamefont {Gardner}}, \bibinfo {author} {\bibfnamefont {S.~A.}\ \bibnamefont {Haine}}, \bibinfo {author} {\bibfnamefont {J.~J.}\ \bibnamefont {Hope}}, \bibinfo {author} {\bibfnamefont {Y.}~\bibnamefont {Chen}},\ \bibnamefont {and}\ \bibinfo {author} {\bibfnamefont {T.}~\bibnamefont {Gefen}},\ }\bibfield  {title} {\bibinfo {title} {Lindblad estimation with fast and precise quantum control},\ }\href@noop {} {\bibfield  {journal} {\bibinfo  {journal} {arXiv preprint arXiv:2501.03364}\ } (\bibinfo {year} {2025}{\natexlab{b}})}\BibitemShut {NoStop}%
\bibitem [{\citenamefont {Rife}\ and\ \citenamefont {Boorstyn}(1974)}]{RifeITIT74SingleTone}%
  \BibitemOpen
  \bibfield  {author} {\bibinfo {author} {\bibfnamefont {D.}~\bibnamefont {Rife}}\ \bibnamefont {and}\ \bibinfo {author} {\bibfnamefont {R.}~\bibnamefont {Boorstyn}},\ }\bibfield  {title} {\bibinfo {title} {Single tone parameter estimation from discrete-time observations},\ }\href {https://doi.org/10.1109/TIT.1974.1055282} {\bibfield  {journal} {\bibinfo  {journal} {IEEE Trans. Inf. Theory}\ }\textbf {\bibinfo {volume} {20}},\ \bibinfo {pages} {591} (\bibinfo {year} {1974})}\BibitemShut {NoStop}%
\bibitem [{\citenamefont {Steinhardt}\ and\ \citenamefont {Bretherton}(1985)}]{SteinhardtI8IICASSP85ThresholdsFrequency}%
  \BibitemOpen
  \bibfield  {author} {\bibinfo {author} {\bibfnamefont {A.}~\bibnamefont {Steinhardt}}\ \bibnamefont {and}\ \bibinfo {author} {\bibfnamefont {C.}~\bibnamefont {Bretherton}},\ }\bibfield  {title} {\bibinfo {title} {Thresholds in frequency estimation},\ }in\ \href {https://doi.org/10.1109/ICASSP.1985.1168170} {\emph {\bibinfo {booktitle} {{{ICASSP}} 85 {{IEEE Int}}. {{Conf}}. {{Acoust}}. {{Speech Signal Process}}.}}},\ Vol.~\bibinfo {volume} {10}\ (\bibinfo {year} {1985})\ pp.\ \bibinfo {pages} {1273--1276}\BibitemShut {NoStop}%
\bibitem [{\citenamefont {James}\ \emph {et~al.}(2002)\citenamefont {James}, \citenamefont {Anderson},\ and\ \citenamefont {Williamson}}]{james2002characterization}%
  \BibitemOpen
  \bibfield  {author} {\bibinfo {author} {\bibfnamefont {B.}~\bibnamefont {James}}, \bibinfo {author} {\bibfnamefont {B.~D.}\ \bibnamefont {Anderson}},\ \bibnamefont {and}\ \bibinfo {author} {\bibfnamefont {R.~C.}\ \bibnamefont {Williamson}},\ }\bibfield  {title} {\bibinfo {title} {Characterization of threshold for single tone maximum likelihood frequency estimation},\ }\href@noop {} {\bibfield  {journal} {\bibinfo  {journal} {IEEE transactions on signal processing}\ }\textbf {\bibinfo {volume} {43}},\ \bibinfo {pages} {817} (\bibinfo {year} {2002})}\BibitemShut {NoStop}%
\bibitem [{\citenamefont {Knockaert}(1997)}]{KnockaertITSP97BarankinBound}%
  \BibitemOpen
  \bibfield  {author} {\bibinfo {author} {\bibfnamefont {L.}~\bibnamefont {Knockaert}},\ }\bibfield  {title} {\bibinfo {title} {The {{Barankin}} bound and threshold behavior in frequency estimation},\ }\href {https://doi.org/10.1109/78.622965} {\bibfield  {journal} {\bibinfo  {journal} {IEEE Trans. Signal Process.}\ }\textbf {\bibinfo {volume} {45}},\ \bibinfo {pages} {2398} (\bibinfo {year} {1997})}\BibitemShut {NoStop}%
\bibitem [{\citenamefont {Schmitt}\ \emph {et~al.}(2017)\citenamefont {Schmitt}, \citenamefont {Gefen}, \citenamefont {St{\"u}rner}, \citenamefont {Unden}, \citenamefont {Wolff}, \citenamefont {M{\"u}ller}, \citenamefont {Scheuer}, \citenamefont {Naydenov}, \citenamefont {Markham}, \citenamefont {Pezzagna} \emph {et~al.}}]{schmitt2017submillihertz}%
  \BibitemOpen
  \bibfield  {author} {\bibinfo {author} {\bibfnamefont {S.}~\bibnamefont {Schmitt}}, \bibinfo {author} {\bibfnamefont {T.}~\bibnamefont {Gefen}}, \bibinfo {author} {\bibfnamefont {F.~M.}\ \bibnamefont {St{\"u}rner}}, \bibinfo {author} {\bibfnamefont {T.}~\bibnamefont {Unden}}, \bibinfo {author} {\bibfnamefont {G.}~\bibnamefont {Wolff}}, \bibinfo {author} {\bibfnamefont {C.}~\bibnamefont {M{\"u}ller}}, \bibinfo {author} {\bibfnamefont {J.}~\bibnamefont {Scheuer}}, \bibinfo {author} {\bibfnamefont {B.}~\bibnamefont {Naydenov}}, \bibinfo {author} {\bibfnamefont {M.}~\bibnamefont {Markham}}, \bibinfo {author} {\bibfnamefont {S.}~\bibnamefont {Pezzagna}}, \bibnamefont {et~al.},\ }\bibfield  {title} {\bibinfo {title} {Submillihertz magnetic spectroscopy performed with a nanoscale quantum sensor},\ }\href@noop {} {\bibfield  {journal} {\bibinfo  {journal} {Science}\ }\textbf {\bibinfo {volume} {356}},\ \bibinfo {pages} {832} (\bibinfo {year} {2017})}\BibitemShut {NoStop}%
\bibitem [{\citenamefont {Schmitt}\ \emph {et~al.}(2021)\citenamefont {Schmitt}, \citenamefont {Gefen}, \citenamefont {Louzon}, \citenamefont {Osterkamp}, \citenamefont {Staudenmaier}, \citenamefont {Lang}, \citenamefont {Markham}, \citenamefont {Retzker}, \citenamefont {McGuinness},\ and\ \citenamefont {Jelezko}}]{schmitt2021optimal}%
  \BibitemOpen
  \bibfield  {author} {\bibinfo {author} {\bibfnamefont {S.}~\bibnamefont {Schmitt}}, \bibinfo {author} {\bibfnamefont {T.}~\bibnamefont {Gefen}}, \bibinfo {author} {\bibfnamefont {D.}~\bibnamefont {Louzon}}, \bibinfo {author} {\bibfnamefont {C.}~\bibnamefont {Osterkamp}}, \bibinfo {author} {\bibfnamefont {N.}~\bibnamefont {Staudenmaier}}, \bibinfo {author} {\bibfnamefont {J.}~\bibnamefont {Lang}}, \bibinfo {author} {\bibfnamefont {M.}~\bibnamefont {Markham}}, \bibinfo {author} {\bibfnamefont {A.}~\bibnamefont {Retzker}}, \bibinfo {author} {\bibfnamefont {L.~P.}\ \bibnamefont {McGuinness}},\ \bibnamefont {and}\ \bibinfo {author} {\bibfnamefont {F.}~\bibnamefont {Jelezko}},\ }\bibfield  {title} {\bibinfo {title} {Optimal frequency measurements with quantum probes},\ }\href@noop {} {\bibfield  {journal} {\bibinfo  {journal} {npj Quantum Information}\ }\textbf {\bibinfo {volume} {7}},\ \bibinfo {pages} {55} (\bibinfo {year} {2021})}\BibitemShut {NoStop}%
\bibitem [{\citenamefont {Holevo}(1978)}]{holevo1978estimation}%
  \BibitemOpen
  \bibfield  {author} {\bibinfo {author} {\bibfnamefont {A.}~\bibnamefont {Holevo}},\ }\bibfield  {title} {\bibinfo {title} {Estimation of shift parameters of a quantum state},\ }\href@noop {} {\bibfield  {journal} {\bibinfo  {journal} {Reports on Mathematical Physics}\ }\textbf {\bibinfo {volume} {13}},\ \bibinfo {pages} {379} (\bibinfo {year} {1978})}\BibitemShut {NoStop}%
\bibitem [{\citenamefont {Chiribella}\ \emph {et~al.}(2005)\citenamefont {Chiribella}, \citenamefont {D’ariano},\ and\ \citenamefont {Sacchi}}]{chiribella2005optimal}%
  \BibitemOpen
  \bibfield  {author} {\bibinfo {author} {\bibfnamefont {G.}~\bibnamefont {Chiribella}}, \bibinfo {author} {\bibfnamefont {G.}~\bibnamefont {D’ariano}},\ \bibnamefont {and}\ \bibinfo {author} {\bibfnamefont {M.~F.}\ \bibnamefont {Sacchi}},\ }\bibfield  {title} {\bibinfo {title} {Optimal estimation of group transformations using entanglement},\ }\href@noop {} {\bibfield  {journal} {\bibinfo  {journal} {Physical Review A—Atomic, Molecular, and Optical Physics}\ }\textbf {\bibinfo {volume} {72}},\ \bibinfo {pages} {042338} (\bibinfo {year} {2005})}\BibitemShut {NoStop}%
\bibitem [{\citenamefont {Wiseman}\ and\ \citenamefont {Milburn}(2009)}]{WisemanMilburn2009book}%
  \BibitemOpen
  \bibfield  {author} {\bibinfo {author} {\bibfnamefont {H.~M.}\ \bibnamefont {Wiseman}}\ \bibnamefont {and}\ \bibinfo {author} {\bibfnamefont {G.~J.}\ \bibnamefont {Milburn}},\ }\href@noop {} {\emph {\bibinfo {title} {Quantum Measurement and Control}}}\ (\bibinfo  {publisher} {Cambridge University Press},\ \bibinfo {year} {2009})\BibitemShut {NoStop}%
\bibitem [{\citenamefont {{S. Direkci et al. (in preparation)}}(2025)}]{supaper}%
  \BibitemOpen
  \bibfield  {author} {\bibinfo {author} {\bibnamefont {{S. Direkci et al. (in preparation)}}},\ }\href@noop {} {\bibinfo {title} {Extending the dynamic range in atomic clocks with weak measurements}} (\bibinfo {year} {2025})\BibitemShut {NoStop}%
\bibitem [{rep()}]{repo}%
  \BibitemOpen
  \href@noop {} {\bibinfo {title} {{J.~W.~Gardner.} \emph{boldBat}. 2025. \url{https://git.ligo.org/jameswalter.gardner/boldbat}}}\BibitemShut {NoStop}%
\bibitem [{\citenamefont {Demkowicz-Dobrza{\'n}ski}\ \emph {et~al.}(2015)\citenamefont {Demkowicz-Dobrza{\'n}ski}, \citenamefont {Jarzyna},\ and\ \citenamefont {Ko{\l}ody{\'n}ski}}]{demkowicz2015quantum}%
  \BibitemOpen
  \bibfield  {author} {\bibinfo {author} {\bibfnamefont {R.}~\bibnamefont {Demkowicz-Dobrza{\'n}ski}}, \bibinfo {author} {\bibfnamefont {M.}~\bibnamefont {Jarzyna}},\ \bibnamefont {and}\ \bibinfo {author} {\bibfnamefont {J.}~\bibnamefont {Ko{\l}ody{\'n}ski}},\ }\bibfield  {title} {\bibinfo {title} {Quantum limits in optical interferometry},\ }\href@noop {} {\bibfield  {journal} {\bibinfo  {journal} {Progress in Optics}\ }\textbf {\bibinfo {volume} {60}},\ \bibinfo {pages} {345} (\bibinfo {year} {2015})}\BibitemShut {NoStop}%
\bibitem [{\citenamefont {Bartlett}\ \emph {et~al.}(2007)\citenamefont {Bartlett}, \citenamefont {Rudolph},\ and\ \citenamefont {Spekkens}}]{bartlett2007reference}%
  \BibitemOpen
  \bibfield  {author} {\bibinfo {author} {\bibfnamefont {S.~D.}\ \bibnamefont {Bartlett}}, \bibinfo {author} {\bibfnamefont {T.}~\bibnamefont {Rudolph}},\ \bibnamefont {and}\ \bibinfo {author} {\bibfnamefont {R.~W.}\ \bibnamefont {Spekkens}},\ }\bibfield  {title} {\bibinfo {title} {Reference frames, superselection rules, and quantum information},\ }\href@noop {} {\bibfield  {journal} {\bibinfo  {journal} {Reviews of Modern Physics}\ }\textbf {\bibinfo {volume} {79}},\ \bibinfo {pages} {555} (\bibinfo {year} {2007})}\BibitemShut {NoStop}%
\bibitem [{\citenamefont {Jeffreys}(1946)}]{jeffreys1946invariant}%
  \BibitemOpen
  \bibfield  {author} {\bibinfo {author} {\bibfnamefont {H.}~\bibnamefont {Jeffreys}},\ }\bibfield  {title} {\bibinfo {title} {An invariant form for the prior probability in estimation problems},\ }\href@noop {} {\bibfield  {journal} {\bibinfo  {journal} {Proceedings of the Royal Society of London. Series A. Mathematical and Physical Sciences}\ }\textbf {\bibinfo {volume} {186}},\ \bibinfo {pages} {453} (\bibinfo {year} {1946})}\BibitemShut {NoStop}%
\bibitem [{\citenamefont {B{\"o}ttcher}\ and\ \citenamefont {Grudsky}(2000)}]{bottcher2000toeplitz}%
  \BibitemOpen
  \bibfield  {author} {\bibinfo {author} {\bibfnamefont {A.}~\bibnamefont {B{\"o}ttcher}}\ \bibnamefont {and}\ \bibinfo {author} {\bibfnamefont {S.~M.}\ \bibnamefont {Grudsky}},\ }\href@noop {} {\emph {\bibinfo {title} {Toeplitz matrices, asymptotic linear algebra and functional analysis}}},\ Vol.~\bibinfo {volume} {67}\ (\bibinfo  {publisher} {Springer},\ \bibinfo {year} {2000})\BibitemShut {NoStop}%
\end{thebibliography}%

\end{document}